\documentclass[]{pasj02}

\usepackage{threeparttable}
\usepackage{natbib} 
\usepackage{caption}
\usepackage{adjustbox}
\usepackage{array}

\jyear{2026}
\Received{2026/02/25}
\Accepted{2026/06/01}

\graphicspath{{./}{figures/}} 

\begin{document} 

\title{Near- to Far-Infrared Spectral Energy Distribution Analysis of Interacting Galaxies in Hickson Compact Groups\,56 and Stephan\textquoteright s Quintet (HCG\,92)\thanks[\dagger]{This is a pre-copyedited, author-produced version of an article accepted for publication in ``Publications of the Astronomical Society of Japan'' following peer review. The version of record (PASJ, 2026, Vol. ??, p. ???) is available online at [https://doi.org/10.1093/pasj/?????].}}

\author{Ayato \textsc{ikeuchi}\altaffilmark{1,2}}\email{ikeuchi@astron.s.u-tokyo.ac.jp} \orcid{0009-0004-8015-1614}
\author{Itsuki \textsc{sakon}\altaffilmark{2}} \orcid{0000-0001-7641-5497}
\author{Takashi \textsc{onaka}\altaffilmark{1}} \orcid{0000-0002-8234-6747}
\author{Fr\'{e}d\'{e}ric \textsc{galliano}\altaffilmark{3}} \orcid{0000-0002-4414-3367}
\author{Ronin \textsc{wu}\altaffilmark{4}} \orcid{0000-0002-3736-5391}
\altaffiltext{1}{Department of Astronomy, Graduate School of Science, The University of Tokyo, 7-3-1 Hongo, Bunkyo-ku, Tokyo, 113-0033, Japan}
\altaffiltext{2}{Institute of Astronomy, Graduate School of Science, The University of Tokyo, 2-21-1, Osawa, Mitaka, Tokyo, 181-0015, Japan}
\altaffiltext{3}{Université Paris-Saclay, Université Paris Cité, CEA, CNRS, AIM, 91191, Gif-sur-Yvette, France}
\altaffiltext{4}{QunaSys Europe, Blegdamsvej 17, Building K, 2100 K\o benhavn \O, Denmark}


\maketitle

\begin{abstract}
We investigate star formation activity in galaxies belonging to two Hickson Compact Groups (HCGs), HCG\,56 and HCG\,92 (Stephan\textquoteright s Quintet), both of which show clear evidence of interactions, using spectral energy distribution (SED) analysis across the near- to far-infrared range. By combining data from the Infrared Satellite AKARI, the Spitzer Space Telescope, and the Herschel Space Observatory, we examine how galactic interactions influence the physical conditions and the evolution of group members. The observed SEDs of member galaxies are compared with model SEDs representing both star-forming galaxies and active galactic nuclei (AGN). Star formation rates (SFRs) are estimated using two independent methods: (i) the strength of mid-infrared polycyclic aromatic hydrocarbon (PAH) bands and (ii) far-infrared luminosities attributed to star formation, as derived from the models. Although both methods yield generally consistent results, SFRs based on PAH features are systematically lower, possibly due to the PAH destruction in some interacting galaxies. When plotted against the stellar mass, all member galaxies are found below the main sequence of star-forming galaxies in the SDSS field, suggesting that interaction-induced starbursts are not seen in HCG\,56 and HCG\,92.
\end{abstract}


\section{Introduction}
Galaxy interactions---gravitational encounters between galaxies, including collisions and mergers---are widely recognized as key drivers of galaxy formation and evolution. In the star formation rate (SFR) versus stellar mass ($M_{\star}$) diagram, normal star-forming galaxies follow a tight correlation known as the star formation main sequence \citep{key-Brinchmann2004,key-Peng2010}. Interacting galaxies, however, often deviate from this sequence, exhibiting either enhanced SFRs (e.g., starbursts and LIRGs) or suppressed SFRs (e.g., Active Galactic Nucleus (AGN)-hosting or quiescent galaxies). During interactions, gas and dust within galaxies can become compressed and turbulent, leading to the formation of dense regions that trigger intense bursts of star formation \citep{key-Melnick1990,key-Mihos1996}. Conversely, gas and dust funneled toward the galactic center may fuel nuclear activity and potentially enhance AGN activity. The subsequent release of hot gas from the AGNs can heat or expel cold ISM material, thereby suppressing star formation \citep{key-Barnes1992,key-Ellison2008}. Interstellar dust plays a pivotal role in regulating a variety of physical processes in interacting systems, including the absorption and re-radiation of starlight, shielding of molecular gas from dissociating radiation, and the cooling of dense gas that can lead to star formation. It also influences the redistribution of mass through dust-driven outflows and by tracing gas inflows toward galactic nuclei. In turn, dust itself undergoes significant evolution in such environments, including production in stellar ejecta, redistribution via tidal interactions, accumulation in dense regions, and condensation into complex grain structures.

Hickson Compact Groups (HCGs) offer valuable environments for studying the state of interstellar dust under ongoing galaxy interactions in the local Universe. \citet{key-Hickson1982} compiled a catalog of compact groups using photographic plates from the Palomar Observatory Sky Survey, based on the galaxy population, isolation, and compactness. According to \citet{key-Hickson1982}, the fraction of spiral galaxies in HCGs is approximately 43\%, which is significantly lower than the $\sim$75\% observed in the general field \citep{key-Gisler1980}. Additionally, 43\% of galaxies in HCGs exhibit tidal structures such as bridges or tails \citep{key-Toomre1972,key-Mendes1994}. Star formation quenching in HCGs---likely driven by the depletion of interstellar gas---has been linked to a rapid morphological transformation toward elliptical galaxies \citep{key-Verdes2001,key-Alatalo2015}. These characteristics underscore the impact of galaxy interactions, making HCGs ideal laboratories for investigating the evolutionary consequences of such processes. 

Despite recent advances, infrared studies of HCGs remain limited. Previous infrared studies of HCG member galaxies have provided valuable insights into the state of the interstellar medium (ISM). Using the spectroscopy obtained with the Infrared Spectrograph (IRS; \citealt{key-Houck2004}) on board the Spitzer Space Telescope (hereafter Spitzer/IRS), \citet{key-Cluver2013} investigated warm molecular hydrogen emission and found evidence for shock excitation in several HCGs. They also reported suppressed PAH emission relative to H$_2$, suggesting that dynamical interactions and associated shocks can destroy or alter PAH carriers in these environments. Meanwhile, \citet{key-Bitsakis2011,key-Bitsakis2014} performed analyses of broad-band SEDs across a large sample of HCG galaxies, deriving global properties such as stellar mass, SFR, and dust attenuation using population synthesis modeling. However, broad-band photometry alone often leaves degeneracies in the mid-infrared between PAH emission, stellar continuum, and hot dust, especially in AGN-hosting systems.  Therefore, it is important to complement near- to far-infrared photometry with near- to mid-infrared spectroscopy, which anchors the PAH band strengths and the mid-infrared (MIR) continuum shape, enabling more robust starburst/AGN decomposition. 

For HCG\,92, commonly known as Stephan\textquoteright s Quintet, \citet{key-Natale2010} carried out a detailed study based on MIR to far-infrared (FIR) observations with Spitzer and showed that a substantial fraction of the ongoing star formation is associated with intergalactic structures rather than being confined to the main stellar disks. Their study provided an important view of the dust-emitting components in the system, including NGC 7319 (HCG 92c), SQ-A, SQ-B, the shock region, and the extended emission. 

In addition to those results, \citet{key-Appleton2023} presented multi-wavelength observations of Stephan\textquoteright s Quintet, combining images from the James Webb Space Telescope (JWST) and the Hubble Space Telescope (HST), along with CO(2--1) spectroscopy from the Atacama Large Millimeter/submillimeter Array (ALMA), to study the intergalactic medium (IGM) on spatial scales of 25--150\,pc. They reported that the Mid-Infrared Instrument (MIRI) on JWST F1000W and F770W bands are dominated by emission from the $0$--$0$ S(3) H$_2$ line and from PAH and $0$--$0$ S(5) H$_2$, respectively. Their results suggested that the strong mid-infrared H$_2$ emission throughout Stephan\textquoteright s Quintet originates from a fog of warm gas produced by fragmentation of dense, cold molecular clouds and subsequent mixing and recycling in the post-shock medium following the collision with the intruder galaxy NGC~7318b. 

While significant progress has been made in characterizing interstellar dust in interacting systems, many aspects of its physical state, chemical composition, and response to dynamical and radiative processes remain uncertain. In this study, we focus on detailed SED modeling of two prototypical interacting groups, HCG\,56 and HCG\,92, by combining near- to far-infrared photometry from the Infrared Satellite AKARI \citep{key-Murakami2007}, the Spitzer Space Telescope \citep{key-Werner2004}, and the Herschel Space Observatory\footnote{Herschel is an ESA space observatory with science instruments provided by European-led Principal Investigator consortia and with important participation from NASA.} \citep{key-Pilbratt2010}, together with near- to mid-infrared spectroscopy from the Infrared Camera (IRC) on board AKARI \citep{key-Onaka2007}. Building on the work of \citet{key-Natale2010}, we present an analysis of aperture-matched, galaxy-scale SEDs with improved FIR coverage and AKARI/IRC spectroscopy. This approach helps constrain PAH strengths, MIR continuum shapes, and AGN/starburst decomposition. We also incorporate modern dust composition models \citep{key-Galliano2011,key-Jones2013} and AGN dust templates \citep{key-Siebenmorgen2015} to disentangle star-forming dust and AGN components. By leveraging broad spectral coverage and physically motivated modeling, we explore the interplay between dust, star formation, and AGN activity in HCG\,56 and HCG\,92 in detail.

This paper is organized as follows. Section 2 describes the infrared imaging and spectroscopic data obtained with AKARI, Spitzer, and Herschel, along with the sample selection of HCG\,56 and HCG\,92. In Section 3, we present the spectral energy distributions (SEDs) of the member galaxies and identify key spectral features, including PAH bands and mid-infrared continuum properties. Section 4 details the estimation of star formation rates using both PAH and FIR indicators, and discusses their systematic discrepancies. Section 5 summarizes our main findings. Additional information on the spectral decomposition and SED modeling is provided in the Appendices.

\section{Observations}
We used near- to mid-infrared spectroscopic data collected with AKARI/IRC \citep{key-Ohyama2007,key-Onaka2007} in the framework of the open time program SHARP (PI: Itsuki Sakon) and the near- to far infrared imaging data taken with AKARI/IRC and the Far-Infrared Surveyor (FIS) \citep{key-Kawada2007} on board AKARI (hereafter AKARI/FIS) in the mission program ISMGN (PI: Hidehiro Kaneda). We also employed the archival photometric data of galaxies in HCG\,56 and HCG\,92 collected with Spitzer \citep{key-Werner2004} and Herschel \citep{key-Pilbratt2010}. 

\subsection{Targets}
\subsubsection{HCG\,56}
HCG\,56 consists of five member galaxies (HCG\,56a--e), which have already been classified as interacting galaxies (VV\,150, Arp\,322, and I\,Zw\,027) prior to Hickson\textquoteright s formal definition. \citet{key-Hickson1982} later redefined this group systematically. The distances to HCG\,56b, 56c, 56d, and 56e are approximately 113\,Mpc, whereas HCG\,56a lies slightly farther away at around 145\,Mpc. However, the difference in radial velocities among these galaxies is not considered significant. HCG\,56b is classified as a Seyfert 1 galaxy \citep{key-Veron2006} and shows strong mid-infrared continuum emission \citep{key-Bitsakis2014}, indicating the presence of an AGN. A bridge-like structure is visible between HCG\,56b and 56c in optical images, suggesting ongoing interaction, which further supports its classification as an interacting system.
\subsubsection{HCG\,92}
HCG\,92 comprises five member galaxies (HCG\,92a--e) and is commonly known as ``Stephan\textquoteright s Quintet (SQ),'' named after \'{E}douard Stephan, who first classified it in 1877 as a prototype compact group. The group also appears in other catalogs of interacting galaxies, as VV\,288 and Arp\,319. Due to the large radial velocity difference between HCG\,92a ($\sim$780\,km\,s$^{-1}$) and the other members ($\sim$6500\,km\,s$^{-1}$; see NED\footnote{The NASA/IPAC Extragalactic Database (NED) is operated by the Jet Propulsion Laboratory, California Institute of Technology, under contract with NASA.}), HCG\,92a is considered a foreground galaxy along the line of sight. In contrast, NGC~7320C has a redshift similar to that of HCG\,92b--e, and tidal tails extend from HCG\,92c toward NGC~7320C. Therefore, HCG\,92b--e and NGC~7320C are regarded as physically interacting members. HCG\,92c is classified as a Seyfert 2 galaxy \citep{key-Veron2006}, while HCG\,92b is currently passing through the group\textquoteright s center, producing shock-heated X-ray gas between 92b and 92c \citep{key-Hwang2012}. In addition, the ISO/ISOCAM infrared sources SQ-A and SQ-B, located within the HCG\,92 system but lacking obvious single-galaxy optical counterparts, have been detected in VLA maps at 1.40 and 4.86\,GHz \citep{key-Xu2003,key-Nikiel-Wroczynski2013}. Both regions are associated with the tidal and disturbed intergalactic structures identified in Stephan\textquoteright s Quintet \citep{key-Xu1999}. 
However, their physical interpretations are not identical. Recent studies have emphasized SQ-A as an intragroup starburst region whose activity was triggered by a high-speed collision between distinct gas systems in the intragroup medium \citep{key-Xu2025}, and detailed JWST-based analyses also support its starburst nature within the shocked intergalactic environment of Stephan\textquoteright s Quintet \citep{key-Appleton2023}. By contrast, SQ-B is widely discussed as a tidal star-forming knot in the tidal debris, and has often been regarded as a tidal dwarf galaxy candidate based on its location, gas content, and star formation activity \citep{key-Lisenfeld2004}.

\subsection{Data}
\subsubsection{AKARI}
Near- and mid-infrared imaging observations of HCG\,56 (pointing IDs: 1402230.1 and 1402231.1) and HCG\,92 (pointing ID: 1402236.1) were conducted using AKARI/IRC \citep{key-Onaka2007}. These datasets were calibrated by the AKARI data science team, details of which are provided in \citet{key-Egusa2016}. In addition to the standard data reduction procedures, which included the background subtraction described by \citet{key-Egusa2016}, we performed additional sky subtraction on the S7, S11, L15, and L24 band data for HCG\,56. This was done using the two independent pointings (1402230.1 and 1402231.1) to correct for residual zodiacal light and Galactic diffuse emission. However, this procedure could not be applied to the near-infrared data, where the dominant sky background originates from stellar light. In this wavelength regime, the presence of stars introduces localized bright regions across the field, while adjacent areas may remain relatively faint, resulting in a highly irregular background. As a result, reliable background subtraction is not feasible. 

Slitless spectroscopy in the near- and mid-infrared was also performed for HCG\,56 (pointing ID: 320004.1) and HCG\,92 (pointing ID: 3220003.1) using AKARI/IRC. All member galaxies of HCG\,56 and HCG\,92 were within a $\sim 10^{\prime} \times 10^{\prime}$ field of view of the slitless mode. Near-infrared spectra were obtained in prism mode (NP), covering wavelengths from 1.8 to 5.5\,$\mu$m with a resolution of $R\sim19$ at 3.5\,$\mu$m \citep{key-Ohyama2007}. Mid-infrared spectra were acquired in the grism mode (SG1 and SG2), covering 4.6--9.2\,$\mu$m with $R\sim50$ for SG1 and 7.2--13.4\,$\mu$m with $R\sim50$ for SG2.

To mitigate contamination from HCG\,92a and spectral blending with HCG\,92b--e, we conducted additional slit spectroscopy observations of HCG\,92a (pointing ID: 3221001.1). Near-infrared spectra were taken in the NG mode, covering 2.55--4.9\,$\mu$m with $R\sim120$ at 3.6\,$\mu$m \citep{key-Onaka2007,key-Ohyama2007}, and mid-infrared spectra were again obtained using the SG1 and SG2 grism modes. Owing to the IRC aperture-mask geometry (slits plus slitless apertures within the same NIR field), HCG\,92c was also recorded in this pointing; we therefore extract the NG (and the corresponding SG1/SG2) spectra of HCG\,92c from the same dataset and use them in Section~3.2.3 and Figure~\ref{92c_fig.2}.

For the AKARI/IRC spectroscopic data, we employ our own reduction tools that follow the official pipeline but are optimized for extended sources, ensuring reliable extraction for each target. We adopted narrow extraction windows in the cross-dispersion direction, typically corresponding to only 2--3 detector pixels per source (approximately 3--4 arcsec for the NIR/NP and NG data and 5--7 arcsec for the MIR-S/SG1 and SG2 data), in order to maximize the signal-to-noise ratio and to minimize contamination from neighboring sources and background structures. Because these extraction widths do not exactly match the effective apertures used for the imaging photometry, the extracted spectra were scaled by a single constant factor so as to match the corresponding broad-band photometric points over the overlapping wavelength range. The data reduction procedures for the AKARI spectroscopic data are described in Appendix~1.

Far-infrared imaging of HCG\,56 (pointing ID: 1402232.1) and HCG\,92 (pointing ID: 1402238.1) was carried out with AKARI/FIS \citep{key-Kawada2007} in the N60, Wide-S, Wide-L, and N160 bands. However, due to the low spatial resolution at far-infrared wavelengths, photometric decomposition of individual group members is not possible. Consequently, we rely on Herschel data for photometry and SED analysis at far-infrared wavelengths.

\subsubsection{Spitzer}
Spitzer Space Telescope Infrared Array Camera (IRAC; hereafter Spitzer/IRAC) \citep{key-Fazio2004} and the Multiband Imaging Photometer for Spitzer (MIPS; hereafter Spitzer/MIPS) \citep{key-Rieke2004} datasets were obtained from the Spitzer Heritage Archive (SHA). The individual data frames, calibrated using the Spitzer pipeline, are classified as Level 1 products (Basic Calibrated Data, BCD), while combined frames are referred to as Level 2 products (Post-BCD, PBCD) in SHA. All datasets used in this study are Level 2 (PBCD) products. For HCG\,56, IRAC observations in channels 1--4 and MIPS channel 1 were carried out under the MIC\_HCG program (PI: Emeric Le Floc\textquoteright h). The AORKEY for the IRAC channels 1--4 is 23026944; these observations were performed in IRAC Map mode. The MIPS channel 1 observation (AORKEY: 23031296) was conducted in MIPS Photometry mode. The FITS files were generated using Spitzer pipeline version S18.25.0 for the IRAC bands and S18.12.0 for the MIPS band. For HCG\,92, IRAC channels 1--4 were observed as part of the SJH\_TDGS program (PI: James R. Houck), with AORKEY 6011392, also in IRAC Map mode. MIPS channel 1 was observed under the STEPHANS program (PI: Philip N. Appleton), with AORKEY 22383616, using the MIPS Photometry mode. 

The FITS files for these datasets were processed using pipeline version S18.25.0 for the IRAC bands and S18.13.0 for the MIPS band. We include Spitzer data to cross-check the photometry between AKARI and Spitzer. For reference, the Spitzer/IRS low-resolution slit widths are $\sim$3.6$^{\prime\prime}$ for SL and $\sim$10.5$^{\prime\prime}$ for LL \citep[e.g.,][]{key-Cluver2013}. For HCG\,56, IRAC and MIPS flux densities have already been reported by \citet{key-Bitsakis2010}. Our independently measured IRAC/MIPS fluxes for HCG\,56b agree with their values within 5--20\%, and those for HCG\,56a, 56c+56d, and 56e differ by at most $\sim$30--40\%, which is comparable to the expected systematics caused by different aperture definitions and background subtraction. We therefore conclude that our Spitzer photometry is on the same absolute scale as previous work and adopt our own measurements in order to use a consistently defined set of source/background regions from the NIR through the FIR in the SED fitting.
\subsubsection{Herschel}
Herschel Space Observatory Photodetector Array Camera and Spectrometer (PACS; hereafter Herschel/PACS) \citep{key-Poglitsch2010} and the Spectral and Photometric Imaging Receiver (SPIRE; hereafter Herschel/SPIRE) \citep{key-Griffin2010} datasets were obtained from the Herschel Science Archive (HSA). In this study, we utilized photometric data from both PACS and SPIRE instruments. Data reduction was carried out using SPG pipeline versions v14.1.0 and v14.2.0. Since the scan map products---Level 2.5 or higher for the PACS photometer and Level 2 or higher for the SPIRE photometer---are already calibrated and suitable for scientific analysis, we use these products to examine the dust properties of the target galaxies.

For HCG\,56, PACS imaging was performed as part of the OT2\_vcharman\_2 program (PI: Vassilis Charmandaris) in the PACSPhoto mode, with observation ID 1342255878. The blue (70\,$\mu$m) and red (160\,$\mu$m) channels were selected for this observation, and we use the Jscanam maps (HPPJSMAP[B/R]) for our analysis. SPIRE imaging of HCG\,56 was conducted under the OT1\_mcluver\_2 program (PI: Michelle Cluver), with observation ID 1342256634, using the SpirePhotoSmallScan mode. The PSW (250\,$\mu$m), PMW (350\,$\mu$m), and PLW (500\,$\mu$m) filters were used for SPIRE, and we analyze the Level 2 extended source maps. All member galaxies of HCG\,56 are covered in both the PACS and SPIRE datasets. 

For HCG\,92, PACS imaging was carried out as part of the OT2\_pguillar\_5 program (PI: Pierre Guillard) in the PACSPhoto mode, with observation ID 1342245660. In this observation, the green (100\,$\mu$m) and red channels were employed. We use the Jscanam maps (HPPJSMAP[B/R]) for the analysis. SPIRE imaging was conducted under the OT1\_pguillar\_1 program (PI: Pierre Guillard), with observation ID 1342220635, using the SpirePhotoSmallScan mode. Observations were performed using the PSW, PMW, and PLW filters, and the Level 2 extended source maps are used in our analysis. All member galaxies of HCG\,92 were observed in both the PACS and SPIRE instruments. For the far-infrared regime, \citet{key-Bitsakis2014} include Herschel PACS and SPIRE photometry in their MAGPHYS SED fitting for HCG\,56 and HCG\,92, but individual PACS/SPIRE flux densities for the separate group members are not tabulated in their paper. We have therefore re-analyzed the archival PACS and SPIRE maps and performed our own aperture photometry, using apertures and background annuli that are consistent with those adopted for the Spitzer and AKARI data. To our knowledge, no previous study has explicitly published consistently defined-aperture PACS/SPIRE flux densities  for the individual components of HCG\,56 and HCG\,92; the values reported in this work are the first such measurements. For HCG\,56a, the PACS 160\,$\mu$m (RED) flux density appears significantly higher than that expected from the shorter-wavelength PACS and the SPIRE photometry.

\section{Results}
We construct spectral energy distributions (SEDs) for each galaxy using the spectroscopic and photometric datasets described in Section~2. 

\subsection{HCG\,56}
\begin{figure}[htbp!]
\begin{center}
\includegraphics[width=\linewidth]{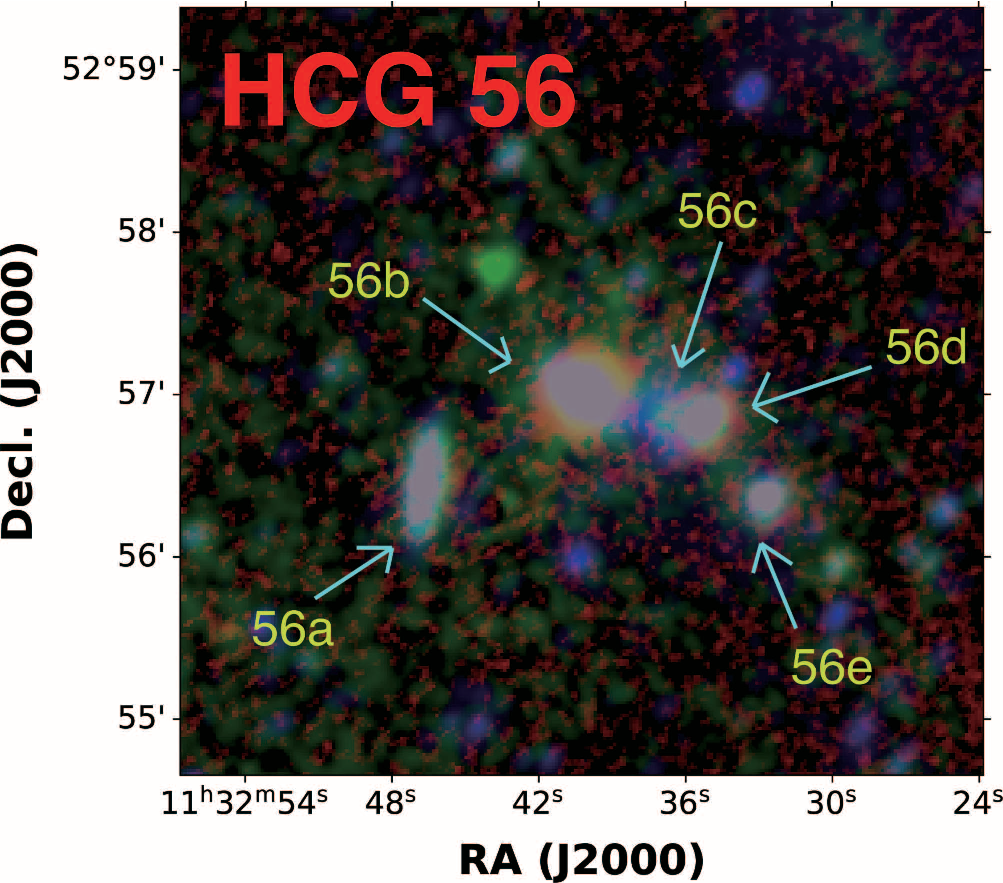}
\end{center}
\captionof{figure}{An AKARI/IRC composite image of HCG\,56 made with N3(3.2\,$\mu$m, blue), S11(11\,$\mu$m, green) and L24 (24\,$\mu$m, red) bands. All images were convolved to a common resolution corresponding to the beam size of the L24 band ($6.7^{\prime\prime}$; \citealt{key-Arimatsu2011}). {Alt text: A composite infrared image of Hickson Compact Group 56. The image combines three infrared bands.}}
\label{56_fig.1}
\end{figure}

Figure~\ref{56_fig.1} shows a false-color composite image of HCG\,56, constructed from AKARI/IRC N3 (3.2\,$\mu$m; blue), S11 (11\,$\mu$m; green), and L24 (24\,$\mu$m; red) bands. The reddish appearance of HCG\,56b, compared to the other members, indicates enhanced L24 emission, likely due to strong mid-infrared continuum from an AGN.  The results of photometry of the member galaxies (HCG\,56a, b, c+d, and e) are summarized in Table~\ref{56a_table}.

\subsubsection{HCG\,56a}
\begin{figure}[htbp!]
\begin{center}
\includegraphics[width=\linewidth]{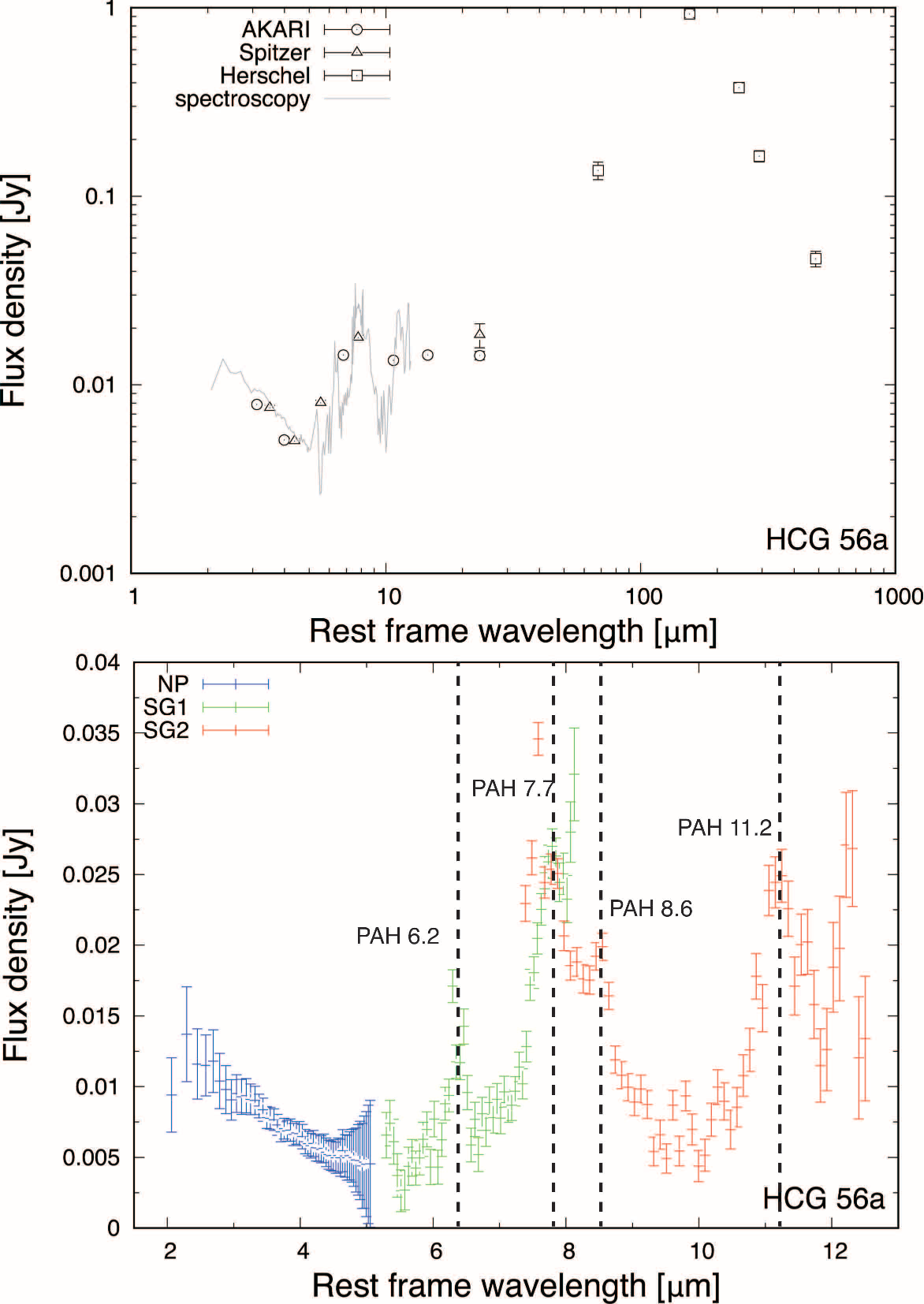}
 \end{center}
\captionof{figure}{(Top): SED of HCG\,56a in rest frame, constructed from AKARI, Spitzer, and Herschel data as detailed in Section~3.1.1. Photometric values are taken from Table~\ref{56a_table}. Open circle, triangle, and square represent AKARI, Spitzer, and Herschel bands, respectively. The AKARI slitless spectrum is overlaid in gray.  (Bottom): Near- to mid-infrared spectrum from AKARI/IRC NP, SG1, and SG2, shown in blue, green, and red, respectively. {Alt text: Two panel figure for HCG 56a. The top panel is a spectral energy distribution with photometric points across near to far infrared wavelengths with an overlaid spectrum. The bottom panel shows the rest frame spectrum.}}
\label{56a_fig.1} \label{56a_fig.2}
\end{figure}

We perform aperture photometry on the calibrated images using square apertures tailored to each source. Since the SPIRE/PLW band, which has the largest beam size among the datasets used here ($36.6^{\prime\prime}$; \citealt{key-Griffin2010}), provides the lowest spatial resolution, the aperture size is generally determined from the PLW image unless the galaxies could be clearly separated in that band. The actual physical coordinates of the pixels of each instrument are not exactly the same, and the photometry area differs slightly depending on the instrument and filter. No image convolution or PSF matching was applied for the aperture photometry itself. Instead, we defined the source apertures and background regions with reference to the lowest-resolution SPIRE/PLW image and measured the fluxes in the corresponding sky regions on each band image. Because the pixel grids and WCS sampling differ slightly among instruments and filters, the exact sampled area is not perfectly identical from band to band. The background levels are estimated from nearby regions free of detectable sources and subtracted from each calibrated image, except for the AKARI/IRC N3 and N4 bands and Spitzer/IRAC channel 1 and channel 2. For these NIR images, we do not apply additional background subtraction because the dominant residual structures are the spatially varying stellar light rather than the uniform sky background, preventing reliable subtraction. The top panel of Figure~\ref{56a_fig.1} presents the SED constructed from multiple bands: AKARI/IRC N3 and N4; S7 and S11; L15 and L24; Spitzer/IRAC channels 1--4; Spitzer/MIPS channel 1; Herschel/PACS blue and red; and Herschel/SPIRE PSW, PMW, and PLW. The bottom panel of Figure~\ref{56a_fig.2} shows the near- to mid-infrared spectrum of HCG\,56a, obtained from AKARI/IRC slitless spectroscopy. Prominent PAH features are detected at 6.2\,$\mu$m, 7.7\,$\mu$m, 8.6\,$\mu$m, and 11.2\,$\mu$m. At the IRC spectral resolution, the 11.2\,$\mu$m PAH feature is well separated from [Ne\,\textsc{ii}] 12.8\,$\mu$m and is not blended with this line. These SED characteristics, together with the PAH features, match those of normal star-forming galaxies.

\begin{table*}[htbp!]
\centering
\caption{Near- to far-infrared photometry for HCG\,56 members.}
\label{56a_table}\label{56b_table}\label{56cd_table}\label{56e_table}
\begin{threeparttable}
\begin{adjustbox}{max width=\textwidth, max totalheight=\textheight, keepaspectratio}
\begin{tabular}{l l c c c}
\hline
Source & Band & $\lambda$ [$\mu$m] & Flux $\pm$ Error [mJy] & Area [$\times 10^{3}$ arcsec$^{2}$] \\
\hline
HCG\,56a & AKARI/IRC N3 & 3.2  & $7.87 \pm 0.20$ & 2.65 \\
        & AKARI/IRC N4 & 4.1  & $5.10 \pm 0.17$ & 2.25 \\
        & AKARI/IRC S7 & 7.0  & $14.4 \pm 0.3$ & 2.96 \\
        & AKARI/IRC S11 & 11.0 & $13.5 \pm 0.3$ & 2.53 \\
        & AKARI/IRC L15 & 15.0 & $14.44 \pm 0.4$ & 2.52 \\
        & AKARI/IRC L24 & 24.0 & $14.3 \pm 0.7$ & 2.40 \\
        & Spitzer/IRAC ch.1 & 3.6 & $7.54 \pm 0.23$ & 2.85 \\
        & Spitzer/IRAC ch.2 & 4.5 & $5.03 \pm 0.15$ & 2.29 \\
        & Spitzer/IRAC ch.3 & 5.8 & $8.02 \pm 0.24$ & 2.21 \\
        & Spitzer/IRAC ch.4 & 7.9 & $17.8 \pm 0.5$ & 2.40 \\
        & Spitzer/MIPS ch.1 & 24.0 & $18.4 \pm 2.7$ & 2.97 \\
        & Herschel/PACS blue & 70 & $(1.37 \pm 0.15)\times 10^{2}$ & 2.60 \\
        & Herschel/PACS red  & 160 & $(9.27 \pm 0.43)\times 10^{2}$\tnote{$\dag$} & 2.72 \\
        & Herschel/SPIRE PSW & 250 & $(3.77 \pm 0.23)\times 10^{2}$ & 2.80 \\
        & Herschel/SPIRE PMW & 350 & $(1.63 \pm 0.11)\times 10^{2}$ & 2.40 \\
        & Herschel/SPIRE PLW & 500 & $46.6 \pm 4.4$ & 2.35 \\
\hline
HCG\,56b & AKARI/IRC N3 & 3.2  & $22.7 \pm 0.6$ & 0.950 \\
        & AKARI/IRC N4 & 4.1  & $24.9 \pm 0.8$ & 1.01 \\
        & AKARI/IRC S7 & 7.0  & $45.0 \pm 1.0$ & 1.22 \\
        & AKARI/IRC S11 & 11.0 & $89.2 \pm 2.1$ & 1.23 \\
        & AKARI/IRC L15 & 15.0 & $(1.60 \pm 0.05)\times 10^{2}$ & 1.30 \\
        & AKARI/IRC L24 & 24.0 & $(2.00 \pm 0.09)\times 10^{2}$ & 1.27 \\
        & Spitzer/IRAC ch.1 & 3.6 & $26.3 \pm 0.8$ & 0.977 \\
        & Spitzer/IRAC ch.2 & 4.5 & $29.1 \pm 0.9$ & 1.04 \\
        & Spitzer/IRAC ch.3 & 5.8 & $40.5 \pm 1.2$ & 1.08 \\
        & Spitzer/IRAC ch.4 & 7.9 & $63.8 \pm 2.0$ & 1.19 \\
        & Spitzer/MIPS ch.1 & 24.0 & $(1.80 \pm 0.07)\times 10^{2}$ & 1.35 \\
        & Herschel/PACS blue & 70 & $(3.19 \pm 0.13)\times 10^{2}$ & 1.18 \\
        & Herschel/PACS red  & 160 & $(1.73 \pm 0.14)\times 10^{2}$ & 1.24 \\
        & Herschel/SPIRE PSW & 250 & --\tnote{$\ddag$} & --\tnote{$\ddag$} \\
        & Herschel/SPIRE PMW & 350 & --\tnote{$\ddag$} & --\tnote{$\ddag$} \\
        & Herschel/SPIRE PLW & 500 & --\tnote{$\ddag$} & --\tnote{$\ddag$} \\
\hline
HCG\,56c+d & AKARI/IRC N3 & 3.2  & $12.8 \pm 0.3$ & 1.13 \\
          & AKARI/IRC N4 & 4.1  & $7.92 \pm 0.27$ & 1.04 \\
          & AKARI/IRC S7 & 7.0  & $16.6 \pm 0.4$ & 1.15 \\
          & AKARI/IRC S11 & 11.0 & $15.0 \pm 0.4$ & 0.992 \\
          & AKARI/IRC L15 & 15.0 & $17.6 \pm 0.5$ & 1.09 \\
          & AKARI/IRC L24 & 24.0 & $25.0 \pm 1.2$ & 1.29 \\
          & Spitzer/IRAC ch.1 & 3.6 & $11.7 \pm 0.4$ & 1.03 \\
          & Spitzer/IRAC ch.2 & 4.5 & $7.55 \pm 0.23$ & 1.02 \\
          & Spitzer/IRAC ch.3 & 5.8 & $9.79 \pm 0.29$ & 0.936 \\
          & Spitzer/IRAC ch.4 & 7.9 & $20.5 \pm 0.6$ & 1.09 \\
          & Spitzer/MIPS ch.1 & 24.0 & $22.8 \pm 1.9$ & 1.17 \\
          & Herschel/PACS blue & 70 & $(2.78 \pm 0.12)\times 10^{2}$ & 1.08 \\
          & Herschel/PACS red  & 160 & $(4.58 \pm 0.23)\times 10^{2}$ & 1.35 \\
          & Herschel/SPIRE PSW & 250 & --\tnote{$\ddag$} & --\tnote{$\ddag$} \\
          & Herschel/SPIRE PMW & 350 & --\tnote{$\ddag$} & --\tnote{$\ddag$} \\
          & Herschel/SPIRE PLW & 500 & --\tnote{$\ddag$} & --\tnote{$\ddag$} \\
\hline
HCG\,56e & AKARI/IRC N3 & 3.2  & $2.63 \pm 0.07$ & 0.683 \\
        & AKARI/IRC N4 & 4.1  & $1.69 \pm 0.06$ & 0.801 \\
        & AKARI/IRC S7 & 7.0  & $4.64 \pm 0.11$ & 0.799 \\
        & AKARI/IRC S11 & 11.0 & $4.09 \pm 0.11$ & 0.723 \\
        & AKARI/IRC L15 & 15.0 & $4.67 \pm 0.16$ & 0.724 \\
        & AKARI/IRC L24 & 24.0 & $8.28 \pm 0.42$ & 0.759 \\
        & Spitzer/IRAC ch.1 & 3.6 & $2.50 \pm 0.08$ & 0.759 \\
        & Spitzer/IRAC ch.2 & 4.5 & $1.62 \pm 0.05$ & 0.696 \\
        & Spitzer/IRAC ch.3 & 5.8 & $2.32 \pm 0.07$ & 0.729 \\
        & Spitzer/IRAC ch.4 & 7.9 & $5.04 \pm 0.15$ & 0.504 \\
        & Spitzer/MIPS ch.1 & 24.0 & $7.58 \pm 1.36$ & 0.792 \\
        & Herschel/PACS blue & 70 & $(1.20 \pm 0.07)\times 10^{2}$ & 0.538 \\
        & Herschel/PACS red  & 160 & $(1.16 \pm 0.05)\times 10^{2}$ & 0.737 \\
        & Herschel/SPIRE PSW & 250 & --\tnote{$\ddag$} & --\tnote{$\ddag$} \\
        & Herschel/SPIRE PMW & 350 & --\tnote{$\ddag$} & --\tnote{$\ddag$} \\
        & Herschel/SPIRE PLW & 500 & --\tnote{$\ddag$} & --\tnote{$\ddag$} \\
\hline
\end{tabular}
\end{adjustbox}
\begin{tablenotes}
\item[$\dag$] The HCG\,56a PACS 160\,$\mu$m flux is likely contaminated.
\item[$\ddag$] These sources are blended. Please see Section~\ref{56_SPIREbands}.
\end{tablenotes}
\end{threeparttable}
\end{table*}

\subsubsection{HCG\,56b}
\begin{figure}[htbp!]
\begin{center}
\includegraphics[width=\linewidth]{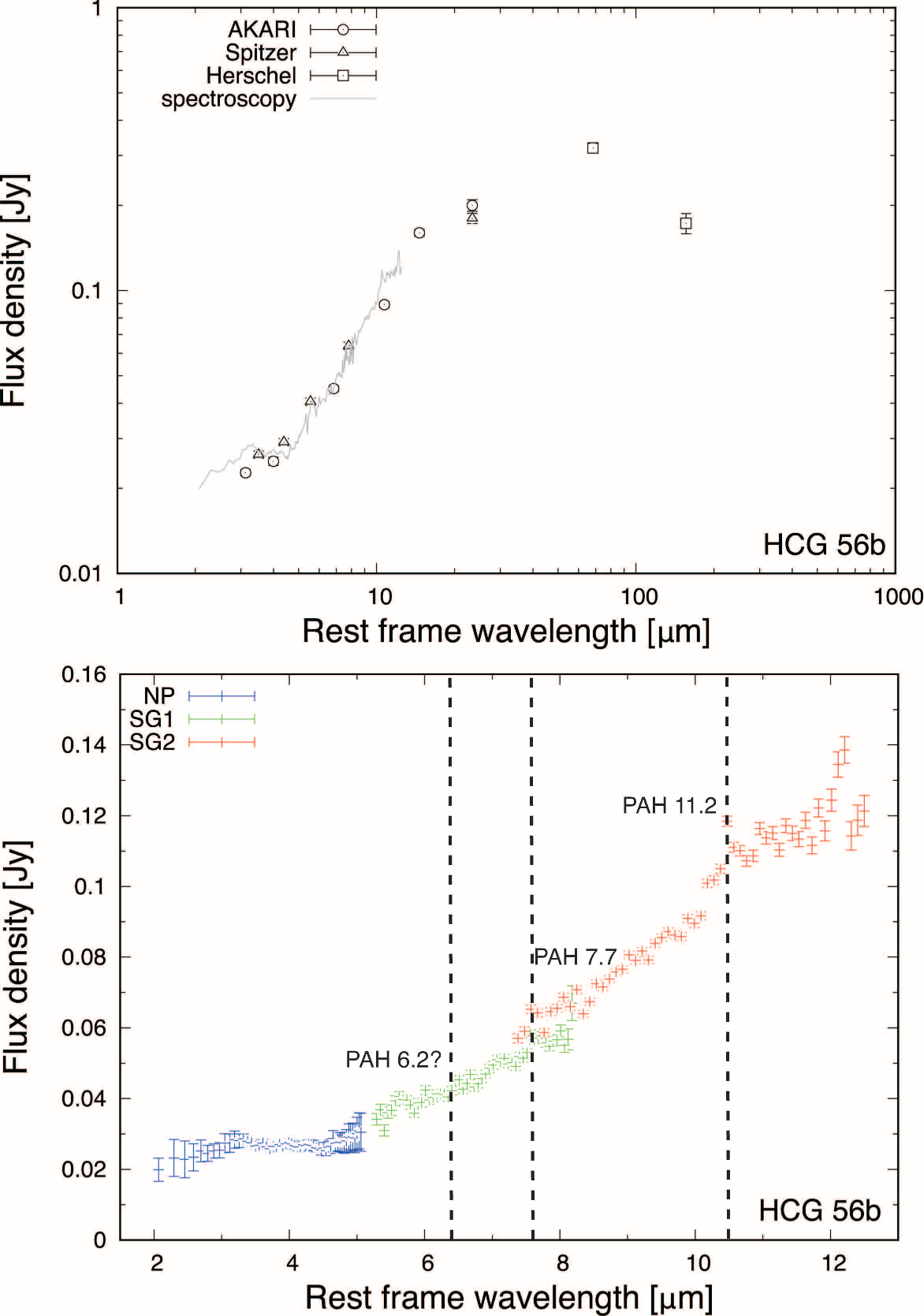}
 \end{center}
 \captionof{figure}{(Top): Same as Figure~\ref{56a_fig.1} (top), but for HCG\,56b. Photometric values are from Table~\ref{56b_table}. (Bottom): Same as Figure~\ref{56a_fig.1} (bottom), but for HCG\,56b. {Alt text: Two panel figure for HCG 56b. The top panel is a spectral energy distribution with photometric points across near to far infrared wavelengths with an overlaid spectrum. The bottom panel shows the rest frame spectrum.}}
\label{56b_fig.1} \label{56b_fig.2}
\end{figure}

The top panel of Figure~\ref{56b_fig.2} presents the SED of HCG\,56b. Due to the close proximity of HCG\,56b, 56c, 56d, and 56e, photometric decomposition in the Herschel/SPIRE bands is not feasible. The treatment of photometry for these unresolved sources is described in Section~\ref{56_SPIREbands}.

The AKARI IRC spectrum of HCG\,56b shows a small gap between the SG1 and SG2 modules around $\sim$7.5--8 $\mu$m: the IRC SG1 segment is fainter than IRS by $\sim$5--10\% in the 6--8 $\mu$m range, while SG2 is brighter by $\sim$5--10\% between 8 and 13 $\mu$m. This level of discrepancy is within the uncertainties of the absolute calibration of IRC and IRS and may be attributed partly to a combination of different aperture sizes, background subtraction, and  extended nature of the PAH emission. The larger IRC aperture likely includes additional diffuse PAH emission in SG2, whereas our background annulus may remove extended continuum emission in SG1. We therefore do not attempt to rescale the spectrum segments. The modest SG1--SG2 gap does not affect the conclusions on the strength of the PAH emission.

HCG\,56b shows, in both AKARI/IRC (the bottom panel of Figure~\ref{56b_fig.2}) and Spitzer/IRS data \citep{key-Cluver2013}, general suppression of the 6.2, 7.7, and 11.2~$\mu$m PAH bands, with the 7.7~$\mu$m feature in particular being buried in the continuum. The continuum rises steeply from 3 to 8~$\mu$m and becomes increasingly red toward the mid-infrared, indicative of a significant contribution from hot dust associated with the AGN torus. The presence of red continuum without silicate absorption at 9.7\,$\mu$m is characteristic of Seyfert 1 galaxies \citep{key-Veron2006}. In the IRS spectrum \citep{key-Cluver2013}, the [S~IV]~10.51~$\mu$m line is prominent, and high-excitation lines such as [Ne~V]~14.32~$\mu$m, characteristic of AGN, are also present. Due to its lower spectral resolution, AKARI/IRC cannot separate these emission lines clearly, but the reddening of the continuum is readily apparent. The larger AKARI aperture may capture a small contribution of PAH emission from surrounding star-forming regions, but the overall suppression of PAHs remains evident. The bridge structure observed between HCG\,56b and 56c \citep{key-Rubin1991} suggests that gas inflow from HCG\,56c has triggered or enhanced the AGN activity in HCG\,56b.

\subsubsection{HCG\,56c and 56d}
HCG\,56c is located near HCG\,56d and exhibits significantly fainter mid- and far-infrared emission. As a result, photometric decomposition between HCG\,56c and 56d is not feasible. The top panel of Figure~\ref{56cd_fig.1} shows the combined SED of both galaxies. Using spectral decomposition techniques (see Appendix~1), we independently extract mid-infrared spectra for HCG\,56c and HCG\,56d, although the signal-to-noise ratio for HCG\,56c is only 2--3$\sigma$. The bottom panels of Figure~\ref{56c_fig.1} show the mid-infrared spectra of HCG\,56c and HCG\,56d, respectively. Both AKARI/IRC and Spitzer/IRS spectra \citep{key-Cluver2013} of HCG\,56c show the presence of the PAH features at 6.2, 7.7, and 8.6~$\mu$m, with the 11.2~$\mu$m band being the most prominent, although they are not particularly strong. The 9.7~$\mu$m silicate absorption is weak to moderate, the results of both instruments being consistent with each other. In the IRS data, the continuum baseline declines gradually from $\sim$5 to 20~$\mu$m, rather than showing the rising trend typically associated with warm dust from vigorous star formation. The larger extracting aperture of the IRC may include faint PAH emission from surrounding star-forming regions. 

Overall, the relative weakness of the PAHs, the lack of a strongly rising mid-IR continuum, and the relatively dust-poor nature of the source suggest that HCG\,56c is not a pure star-forming galaxy, but instead may represent a composite system where star formation is present but not dominant, with possible contributions from a weak AGN or shock-heated dust. This picture is consistent with the SED-based analysis of \citet{key-Bitsakis2011}, who argue that the traditional early-type classification of HCG\,56c is unlikely to be correct and that the galaxy hosts a significant young stellar component in addition to the old stellar population. 

Both AKARI/IRC and Spitzer/IRS spectra \citep{key-Cluver2013} of HCG\,56d clearly detect the PAH emission features at 6.2, 7.7, 8.6, and 11.2~$\mu$m, with well-defined peak-to-trough contrast characteristic of substantial star formation. The 9.7~$\mu$m silicate absorption is weak, and the overall mid-infrared spectrum is not continuum-dominated. In the wavelength range where the IRC (SG2) and IRS (SL) spectra overlap, the IRC flux density is systematically higher than that of IRS by about 20\% in the continuum at around 10--12~$\mu$m. This modest offset is consistent with the combination of (i) different observation geometries (slitless IRC versus slit IRS), (ii) the larger effective extraction area in the IRC data, which may include low-level, diffuse PAH emission from the surroundings, and (iii) absolute calibration uncertainties. Importantly, despite this normalization difference, the two instruments show broadly consistent relative band-to-continuum ratios and overall spectral shapes for HCG\,56d.

Taken at face value, the mid-infrared spectrum of HCG\,56d is more naturally interpreted as star-formation-dominated than as AGN-dominated. This view is supported by the strong PAH emission and the absence of a steeply rising hot-dust continuum such as that seen in clearly AGN-dominated systems like HCG\,56b. Consistently, in the galaxy-scale SED fitting of the combined HCG\,56c+d system (Section~3.1.6), we did not find it necessary to introduce an explicit AGN component, unlike the case of HCG\,56b. Although this does not by itself exclude a weak nuclear contribution in HCG\,56d, it indicates that an AGN is not required to explain the infrared emission of the unresolved HCG\,56c+d system at the level considered in the present analysis. HCG\,56d has historically been classified as a Seyfert~2 galaxy by \citet{key-Khachikian1974}, and the Herschel photometry does not allow us to separate HCG\,56c and HCG\,56d individually in the far-infrared. We therefore regard HCG\,56d as a star-forming-dominated system with some residual uncertainty in its nuclear classification, rather than as a purely normal star-forming galaxy established without caveat.

A similar level of cross-instrument normalization uncertainty should be borne in mind when comparing IRC and IRS spectra of HCG\,56c, although its mid-infrared emission is intrinsically less PAH-dominated than HCG\,56d. The relatively weak PAH band-to-continuum contrast, together with the faint mid- and far-infrared emission of HCG\,56c, may be consistent with a scenario in which the interaction has redistributed the cold ISM (e.g., gas transfer toward HCG\,56b and/or shock processing), thereby lowering the current star-formation efficiency and/or modifying the dust properties in HCG\,56c.

\begin{figure}[htbp!]
\begin{center}
\includegraphics[width=\linewidth]{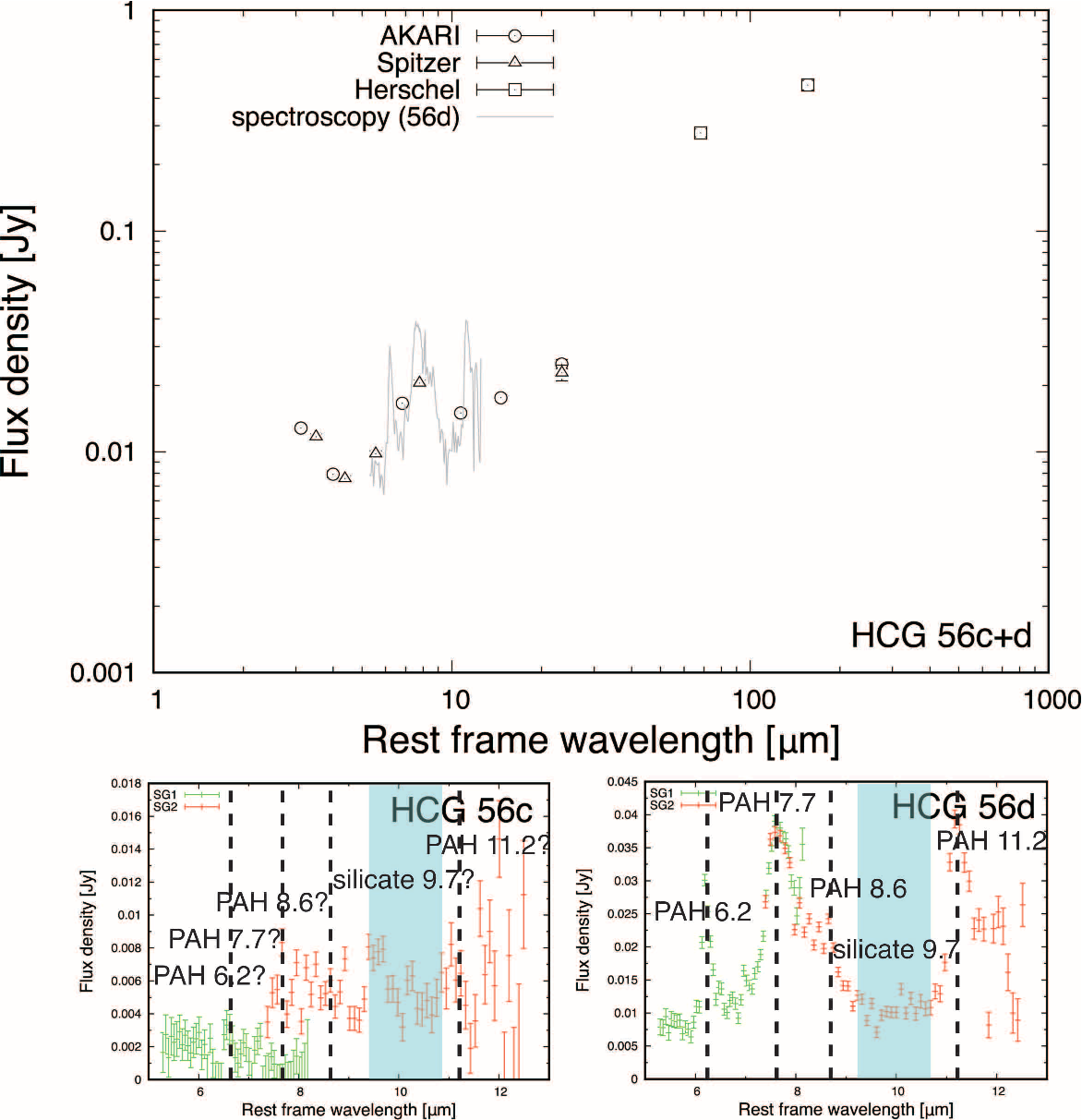}
\end{center}
\captionof{figure}{(Top): Combined SED of HCG\,56c and 56d. The symbols are the same as in Figure~\ref{56a_fig.1}.  (Bottom left and right): AKARI/IRC mid-infrared spectra of HCG\,56c and HCG\,56d, respectively. See Appendix~1 for spectral decomposition procedure. {Alt text: Three panel figure for HCG 56c and HCG 56d. The top panel shows the combined spectral energy distribution from near infrared to far infrared wavelengths. The bottom left panel shows the rest frame spectrum for HCG 56c. The bottom right panel shows the rest frame spectrum for HCG 56d.}}
\label{56cd_fig.1}\label{56c_fig.1}\label{56d_fig.1}
\end{figure}

\subsubsection{HCG\,56e}
\begin{figure}[htbp!]
\begin{center}
\includegraphics[width=\linewidth]{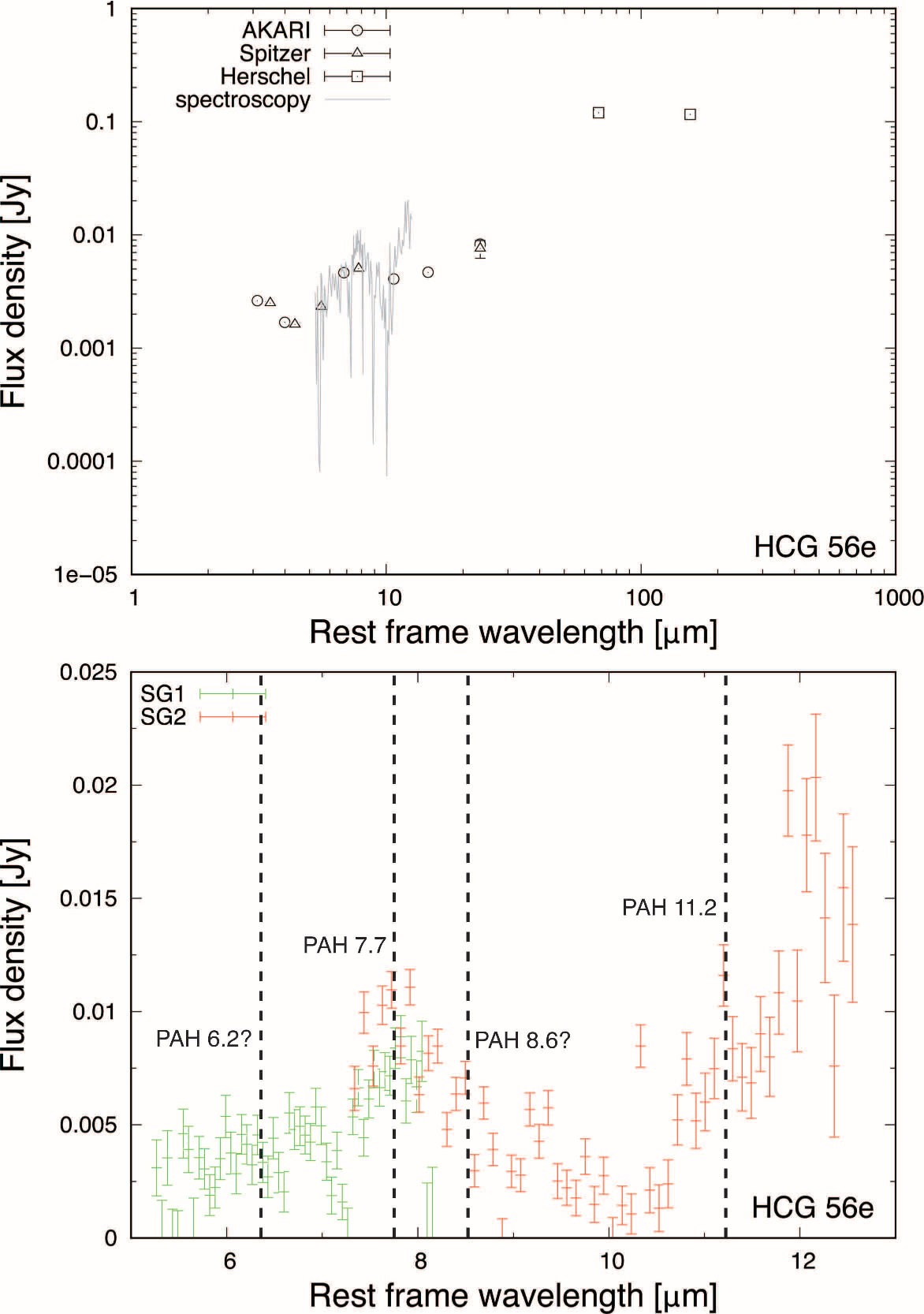}
\end{center}
\captionof{figure}{(Top): Same as Figure~\ref{56a_fig.1} (top), but for HCG\,56e.  (Bottom): AKARI/IRC mid-infrared spectrum of HCG\,56e. The symbols are the same as in Figure~\ref{56a_fig.1}. {Alt text: Two panel figure for HCG 56e. The top panel is a spectral energy distribution with photometric points across near to far infrared wavelengths with an overlaid spectrum. The bottom panel shows the rest frame spectrum.}}
\label{56e_fig.1}\label{56e_fig.2}
\end{figure}

The top panel of Figure~\ref{56e_fig.1} shows the SED of HCG\,56e. The bottom panel of Figure~\ref{56e_fig.2} presents the IRC mid-infrared spectrum, extracted using spectral decomposition techniques (see Appendix~1). HCG\,56e shows, in the Spitzer/IRS spectrum, clearly the PAH features at 6.2, 7.7, 8.6, and 11.2~$\mu$m, whereas in the lower-resolution AKARI/IRC spectrum the PAHs appear weaker and are more embedded in the continuum. The mid-infrared continuum is moderately red, rising from $\sim$3 to 8~$\mu$m, consistent with the emission from warm to hot dust possibly associated with an AGN torus or shock heating, but the slope is not as steep as in typical AGN-dominated systems. The 9.7~$\mu$m silicate absorption is weak or absent. The IRS spectrum also shows low- to intermediate-excitation fine-structure lines such as [Ar~III] and [S~III], while the high-excitation [S~IV] line is not prominent. Compared to the nuclear Spitzer/IRS spectrum of \citet{key-Cluver2013}, the AKARI/IRC spectrum is only moderately brighter, by $\sim$20--30\% (i.e., a factor of $\sim$1.2--1.3) in the overlapping 6--8~$\mu$m range. This is plausibly due to the larger effective extraction area of the slitless IRC observation, which may include diffuse PAH emission from the surroundings, whereas the IRS spectrum primarily traces the nuclear region. As also noted by \citet{key-Cluver2013}, the mid-infrared colors and nuclear classification of HCG\,56e are consistent with a star-forming system, despite its peculiar morphology and the presence of a moderately red continuum. This interpretation is supported by the UV-to-IR SED modeling of \citet{key-Bitsakis2011}, who argue that the conventional early-type classification is not adequate and that HCG\,56e hosts a non-negligible young stellar component (i.e., residual/recent star formation). Combined with our IRC data, these characteristics suggest that HCG\,56e is best described as a composite system in which star formation is significant but may be accompanied by contributions from an AGN or other heating mechanisms.

\subsubsection{Blended Source of Herschel/SPIRE Band in HCG\,56}
\label{56_SPIREbands}
Due to the limited spatial resolution of the Herschel/SPIRE band images, the galaxies HCG\,56b, 56c, 56d, and 56e could not be individually separated. Therefore, photometry at the SPIRE PSW, PMW, and PLW bands was performed by summing the flux densities from these unresolved sources. The results are presented in Table~\ref{56bcde_table} and were used to set upper limits on the flux densities of each individual galaxy at these wavelengths during the SED model fitting in Section~\ref{56_sedmodel}.

\begin{table*}[htbp!]
\centering
\caption{Far-infrared photometry for HCG\,56 blended sources.}
\label{56bcde_table}
\begin{adjustbox}{max width=\textwidth, max totalheight=\textheight, keepaspectratio}
\begin{tabular}{l l c c c}
\hline
Source & Band & $\lambda$ [$\mu$m] & Flux $\pm$ Error [mJy] & Area [$\times 10^{3}$ arcsec$^{2}$] \\
\hline
HCG\,56b+c+d+e & Herschel/SPIRE PSW & 250 & (4.81$\pm$0.30)$\times$10$^{2}$ & 8.42 \\
        & Herschel/SPIRE PMW & 350 & (2.27$\pm$0.19)$\times$10$^{2}$ & 13.2\\
        & Herschel/SPIRE PLW & 500 & 81.8$\pm$8.5 & 9.60  \\
\hline
\end{tabular}
\end{adjustbox}
\end{table*}

\subsubsection{Dust SED Analysis of HCG\,56 Galaxies}
\label{56_sedmodel}
The near- to far-infrared SEDs of the member galaxies in HCG\,56 are analyzed using two different dust models: the AC (amorphous carbon) composition model \citep{key-Galliano2011} and the THEMIS (The Heterogeneous dust Evolution Model for Interstellar Solids) composition model \citep{key-Jones2013}. These two models present differences. The AC model of \citet{key-Galliano2011} is a modification of the model developed in \citet{key-Zubko2004}, replacing graphite by amorphous carbon in order to account for the Herschel observations of the Large Magellanic Cloud. In addition to amorphous carbon, it also contains silicates and PAHs. On the other hand, the THEMIS model's mixture is made of amorphous carbon and silicate grains, both coated with carbon mantles. The aromatic features are not accounted for by PAHs but by small amorphous carbon grains. Despite these differences, both models have a similar far-IR-to-submm opacity (Fig. 4 of \citealt{key-Galliano2018}). The comparison of the two models therefore provides an assessment of the robustness of our results. Details of the dust model and SED fitting procedures are provided in Appendix~2. We adopt a non-uniformly illuminated dust model for SED fitting. The non-uniformly illuminated dust model accounts for the fact that interstellar or circumnuclear dust is exposed to a wide range of radiation field intensities, rather than a single, uniform illumination. In such models, dust grains located at different distances from the heating source --- or shielded by varying amounts of intervening material --- receive different levels of radiation, resulting in a distribution of dust temperatures. This approach reproduces the broad mid- to far-infrared SEDs observed in AGN and star-forming regions more realistically than uniformly illuminated dust models. The application of these SED models is summarized in Appendix~2. 

The dust SED models used in this paper are primarily designed to fit star-forming galaxies. Therefore, as \citet{key-Cluver2013} stated, the contribution of AGN-heated dust must be taken into account in HCG\,56b. Accordingly, its observed SED is fitted including additional AGN radiative transfer emission models from \citet{key-Siebenmorgen2015}.

In this study, the parameter $U$ represents the scaling factor of the interstellar radiation field (ISRF) defined by \citet{key-Mathis1983}, where $U = 1$ corresponds to the local ISRF in the solar neighborhood (integrated flux $2.2 \times 10^{-5}$\,Wm$^{-2}$ between 0.0912 and 8\,$\mu$m; see Appendix~2).

In the cases of HCG\,56a, 56c+d, and 56e, the dust mass $M_{\rm dust}$, the lower cut-off of the power-law distribution of the strength of radiation field $U_{\rm min}$, the upper cut-off of the power-law distribution of the strength of radiation field $U_{\rm max}$, the index of the power-law distribution of the strength of radiation field $\alpha$, the PAH-to-dust mass ratio $q_{\rm PAH}$, the charged PAH fraction $f_{+}$, and the mass of the old stellar population $M_{\star}$ are free parameters in the fits. $U_{\rm max}$, $U_{\rm min}$, and $\alpha$ are defined in Equation (A1). The best-fit results are presented in Table~\ref{56_sed_table}.

For HCG\,56b, AGN-related parameters---including the central source luminosity $L^{\rm AGN}_{\star}$, the inner radius $R^{\rm AGN}$, the clump volume filling factor $V^{\rm AGN}_{c}$, the clump $V$-band optical depth $A^{\rm AGN}_{c}$, the disk $V$-band optical depth $A^{\rm AGN}_{d}$, and the viewing angle $\theta ^{\rm AGN}$---are also free parameters (for details, see Appendix~2.2 and \citealt{key-Siebenmorgen2014}). The corresponding best-fit values are likewise shown in Table~\ref{56_sed_table}.

\begin{figure}[htbp!]
\begin{center}
\includegraphics[width=\linewidth]{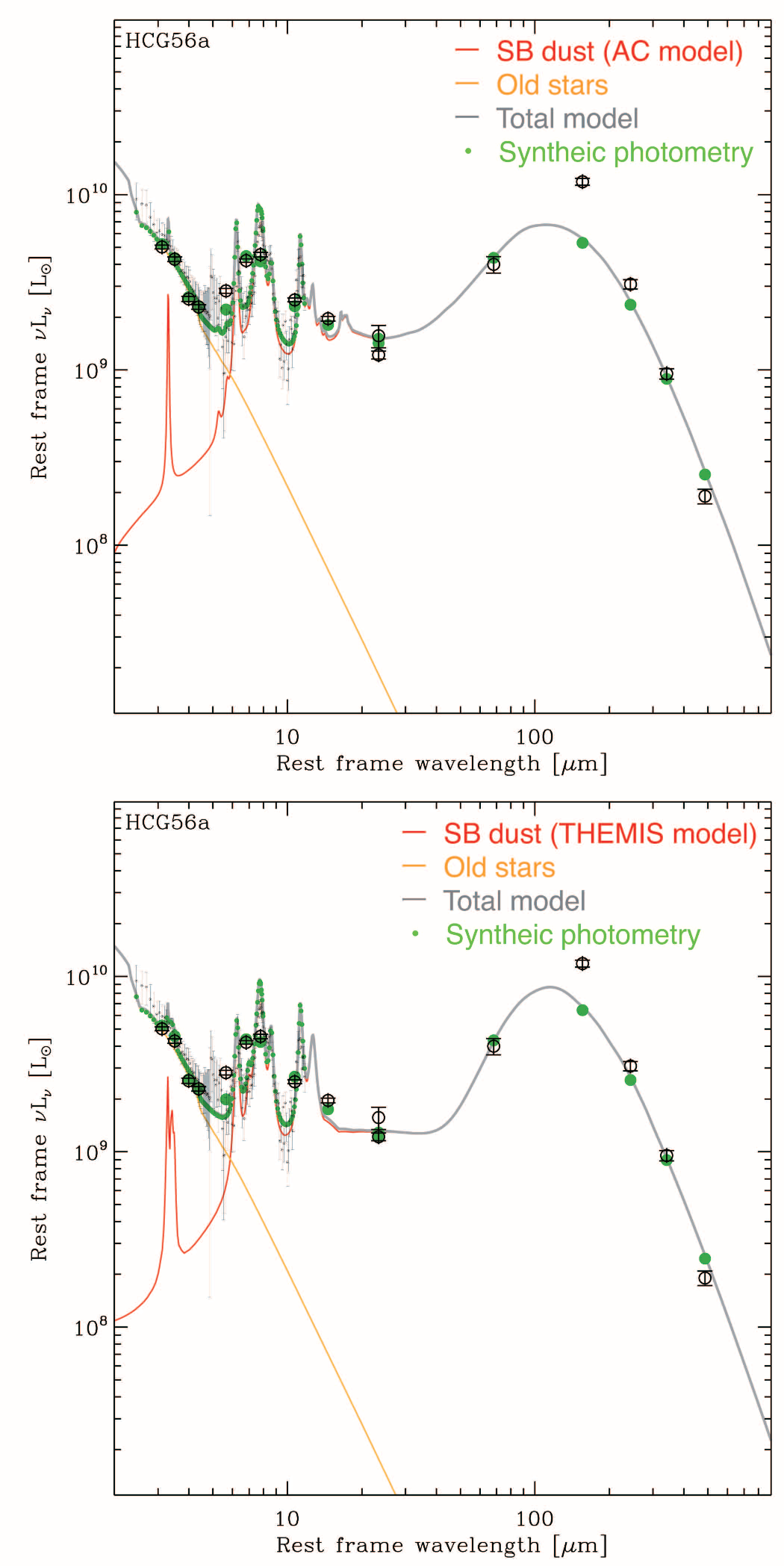}
\end{center}
\captionof{figure}{(Top): Best-fit model SED of HCG\,56a using the amorphous carbon (AC) dust model \citep{key-Galliano2011}.\\(Bottom): Same as top, but using the THEMIS dust model \citep{key-Jones2013}. 
The orange, red, blue, and gray lines represent the stellar, dust, AGN, and total model components, respectively. The observed photometric data are shown by the open circles, and the spectroscopic points by the filled circles. The green circles indicate synthetic flux densities derived from the best-fit model.
Photometric value of PACS red (160 $\mu$m) seems to be an artifact. {Alt text: Two panel model fit figure for HCG 56a. Each panel shows observed infrared data points and the best fit spectral energy distribution model. Separate curves show stellar emission, dust emission, and the total model, for two different dust models.}}
\label{56a_sed_fig.1}\label{56a_sed_fig.2}
\end{figure}

\begin{figure*}[htbp!]
\begin{adjustbox}{max width=\textwidth, max totalheight=\textheight, keepaspectratio, center}
\includegraphics{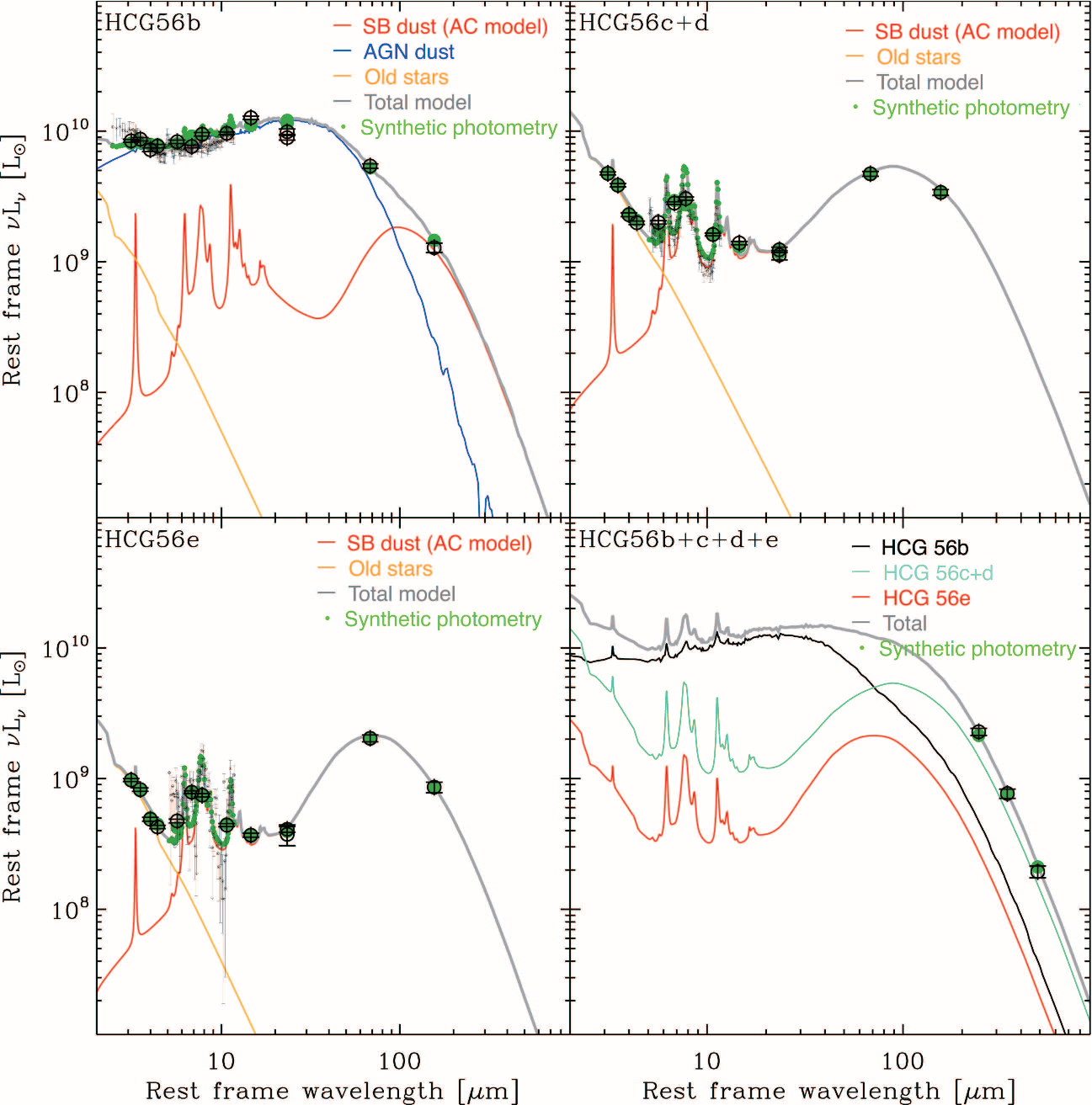}
\end{adjustbox}
\caption{(Top left): SED fitting of HCG\,56b using the AC dust model with AGN templates from \citet{key-Siebenmorgen2015}. 
(Top right): Combined fit for HCG\,56c+d with the AC model; these sources are treated as unresolved in FIR. 
(Bottom left): SED fit of HCG\,56e using the AC model. 
(Bottom right): Total model reconstruction for the blended SPIRE flux density of HCG\,56b, 56c+d, and 56e. 
Notation follows Figure~\ref{56a_sed_fig.1}. {Alt text: Four panel model fit figure for HCG 56 members. Panels show best fit spectral energy distribution models and observed points for HCG 56b, the combined HCG 56c and HCG 56d, and HCG 56e. The last panel shows the blended far infrared model recovery.}}
\label{56bcde_sed_fig.1}
\end{figure*}

\begin{figure*}[htbp!]
\begin{adjustbox}{max width=\textwidth, max totalheight=\textheight, keepaspectratio, center}
\includegraphics{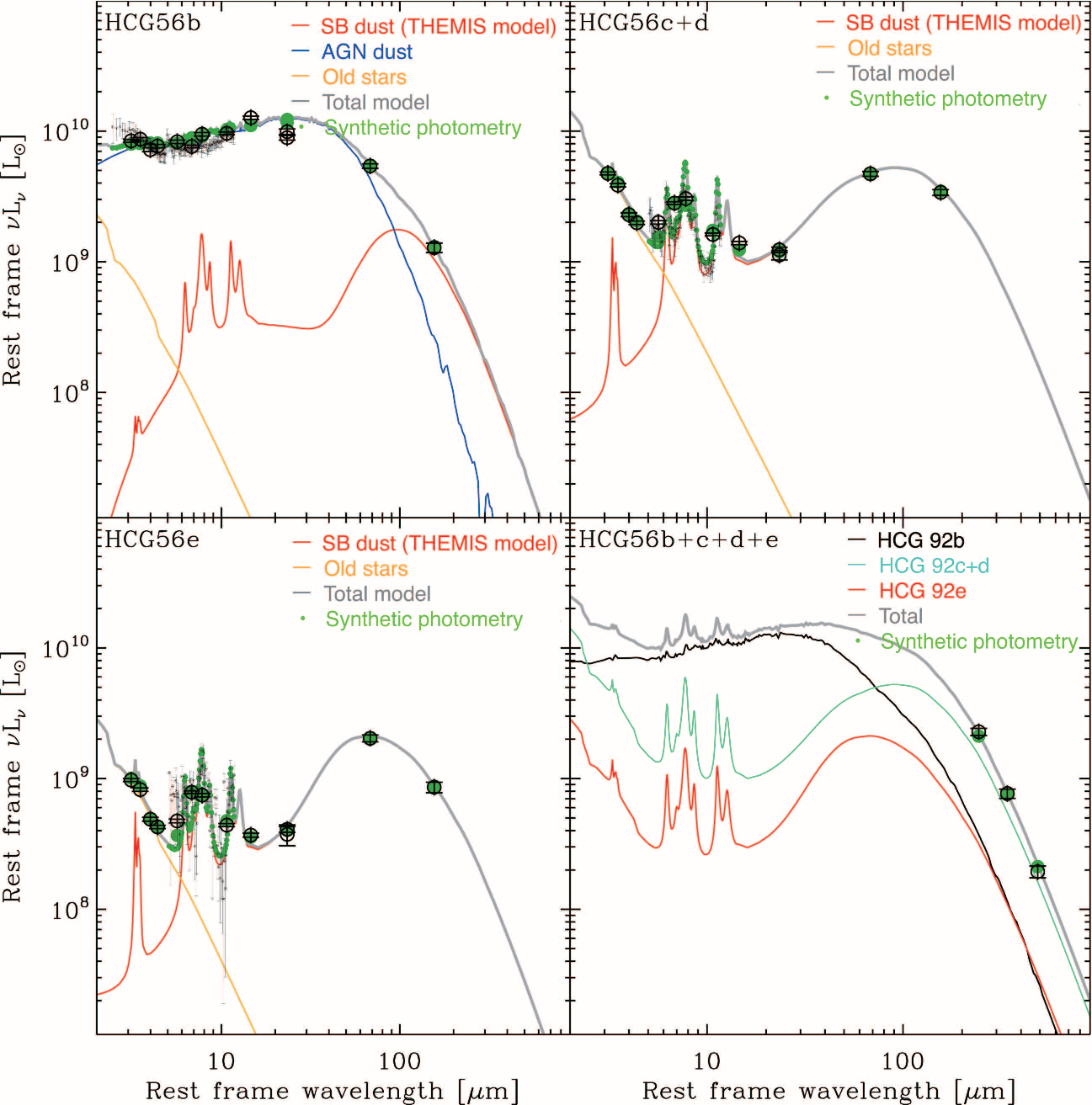}
\end{adjustbox}
\captionsetup{justification=raggedright,singlelinecheck=false}
\caption{Same as Figure~\ref{56bcde_sed_fig.1}, but using the THEMIS dust model instead of AC. {Alt text: Four panel model fit figure for HCG 56 members but using the alternative dust model. Panels show best fit spectral energy distribution models and observed points for HCG 56b, the combined HCG 56c and HCG 56d, and HCG 56e. The last panel shows the blended far infrared model recovery.}}
\label{56bcde_sed_fig.2}
\end{figure*}

As illustrated in Figures~\ref{56a_sed_fig.1},~\ref{56bcde_sed_fig.1}, and~\ref{56bcde_sed_fig.2}, the near- to mid-infrared SEDs are well reproduced by both AC and THEMIS models, and the differences in the best-fit parameter values between the two dust models are generally minor.

\begin{table*}[htbp!]
\begin{adjustbox}{max width=\textwidth, max totalheight=\textheight, keepaspectratio, center}
\begin{threeparttable}
\caption{Best fit parameters of HCG\,56 galaxies.\tnote{$\dag$}}
\label{56_sed_table}
\begin{tabular}{ccc}
\hline\hline
&AC COMPOSITION&THEMIS COMPOSITION\\ \hline
\multicolumn{3}{c}{\bf {HCG\,56a}}\\ \hline
ln ($M_{\rm dust}$ [$M_{\odot}$])&17.3$\pm$0.190&16.9$\pm$0.0700\\
ln ($U_{\rm min}$ [2.2$\times$10$^{-5}$Wm$^{-2}$])&-0.434$\pm$0.708&0.618$\pm$0.111\\
ln ($U_{\rm max}-U_{\rm min}$ [2.2$\times$10$^{-5}$Wm$^{-2}$])&13.8$\pm$6.94&0\\
$\alpha$&2.50&2.50\\
$q_{\rm PAH}$&($8.20\pm0.670$)$\times$10$^{-2}$&($1.53\pm0.0790$)$\times$10$^{-1}$\\
$f_{+}$&($4.79\pm0.390$)$\times$10$^{-1}$&($3.42\pm0.410$)$\times$10$^{-1}$\\
$M_{\star}$ [$M_{\odot}$]&(1.18$\pm$0.0200)$\times$10$^{11}$&(1.14$\pm$0.0190)$\times$10$^{11}$\\
ln ($\langle U\rangle$ [2.2$\times$10$^{-5}$Wm$^{-2}$])&0.664$\pm$0.248&0.818$\pm$0.0940\\ \hline
\multicolumn{3}{c}{\bf {HCG\,56b}}\\ \hline
ln ($M_{\rm dust}$ [$M_{\odot}$])&15.2$\pm$0.460&14.4$\pm$0.600\\
ln ($U_{\rm min}$ [2.2$\times$10$^{-5}$Wm$^{-2}$])&0$\pm$0.308&1.61$\pm$0.801\\
ln ($U_{\rm max}-U_{\rm min}$ [2.2$\times$10$^{-5}$Wm$^{-2}$])&2.30$\pm$0.622&0.611$^{+0.967}_{-0.611}$\\
$\alpha$&1.00$^{+0.455}_{-0}$&2.50$\pm$0.638\\
$q_{\rm PAH}$&($1.62\pm0.290$)$\times$10$^{-1}$&($1.47\pm0.384$)$\times$10$^{-1}$\\
$f_{+}$&0$^{+0.0720}_{-0}$&0\\
$L^{\rm AGN}_{\star}$ [$L_{\odot}$]&(2.50$\pm$0.188)$\times$10$^{11}$&(2.62$\pm$0.150)$\times$10$^{11}$\\
$L^{\rm AGN}_{\rm IR}$ [$L_{\odot}$]&(1.71$\pm$0.136)$\times$10$^{9}$&(1.79$\pm$0.112)$\times$10$^{9}$\\
$R^{\rm AGN}$ [$\mathit pc$]&3.00$\times$10$^{2}$&($3.00^{+0.0180}_{-0}$)$\times$10$^{2}$\\
$V^{\rm AGN}_{c}$ [\%]&77.7$\pm$3.80&77.7$\pm$4.60\\
$A^{\rm AGN}_{c}$&10.9$\pm$2.64&10.2$\pm$2.48\\
$A^{\rm AGN}_{d}$&($1.00\pm0.0740$)$\times$10$^{3}$&($1.00\pm0.0510$)$\times$10$^{3}$\\
$\theta ^{\rm AGN}$ [degrees]&19.0$^{+1.23}_{-0}$&19.0$^{+0.820}_{-0}$\\
$M_{\star}$ [$M_{\odot}$]&(2.72$\pm$0.987)$\times$10$^{10}$&($1.76^{+1.61}_{-0.764}$)$\times$10$^{10}$\\
ln ($\langle U\rangle$ [2.2$\times$10$^{-5}$Wm$^{-2}$])&1.43$\pm$0.426&1.76$\pm$0.654\\ \hline
\multicolumn{3}{c}{\bf {HCG\,56c+d}}\\ \hline
ln ($M_{\rm dust}$ [$M_{\odot}$])&16.8$\pm$0.640&16.7$\pm$0.700\\
ln ($U_{\rm min}$ [2.2$\times$10$^{-5}$Wm$^{-2}$])&-0.829$\pm$0.444&-1.06$\pm$0.694\\
ln ($U_{\rm max}-U_{\rm min}$ [2.2$\times$10$^{-5}$Wm$^{-2}$])&6.42$\pm$1.92&6.84$\pm$1.78\\
$\alpha$&2.05$\pm$0.157&2.08$\pm$0.172\\
$q_{\rm PAH}$&($6.09\pm1.86$)$\times$10$^{-2}$&($1.15\pm0.189$)$\times$10$^{-1}$\\
$f_{+}$&($3.99\pm0.740$)$\times$10$^{-1}$&($3.29\pm0.560$)$\times$10$^{-1}$\\
$M_{\star}$ [$M_{\odot}$]&(1.07$\pm$0.0500)$\times$10$^{11}$&(1.09$\pm$0.0180)$\times$10$^{11}$\\
ln ($\langle U\rangle$ [2.2$\times$10$^{-5}$Wm$^{-2}$])&1.03$\pm$0.359&0.797$\pm$0.600\\ \hline
\multicolumn{3}{c}{\bf {HCG\,56e}}\\ \hline
ln ($M_{\rm dust}$ [$M_{\odot}$])&14.5$\pm$0.850&15.0$\pm$0.950\\
ln ($U_{\rm min}$ [2.2$\times$10$^{-5}$Wm$^{-2}$])&-0.0777$\pm$1.13&-1.47$\pm$1.06\\
ln ($U_{\rm max}-U_{\rm min}$ [2.2$\times$10$^{-5}$Wm$^{-2}$])&5.76$\pm$0.770&5.84$\pm$1.06\\
$\alpha$&1.73$\pm$0.300&1.69$\pm$0.175\\
$q_{\rm PAH}$&($4.42\pm1.87$)$\times$10$^{-2}$&($9.43\pm2.33$)$\times$10$^{-2}$\\
$f_{+}$&($5.24\pm0.750$)$\times$10$^{-1}$&($4.27\pm0.520$)$\times$10$^{-1}$\\
$M_{\star}$ [$M_{\odot}$]&(2.16$\pm$0.0999)$\times$10$^{10}$&(2.18$\pm$0.0410)$\times$10$^{10}$\\
ln ($\langle U\rangle$ [2.2$\times$10$^{-5}$Wm$^{-2}$])&2.28$\pm$0.616&1.48$\pm$0.834\\ \hline
\end{tabular}
\begin{tablenotes}
\item[$\dag$]The values in the second and third columns are the results obtained from the AC composition model \citep{key-Galliano2011} and the THEMIS composition model \citep{key-Jones2013}, respectively. The $L^{\rm AGN}_{\rm IR}$ and $\langle U\rangle$ are calculated using the results of the best fit parameters.
\end{tablenotes}
\end{threeparttable}
\end{adjustbox}
\end{table*}

\subsection{Stephan\textquoteright s Quintet (HCG\,92)}
\begin{figure}[htbp!]
\begin{center}
\includegraphics[width=\linewidth]{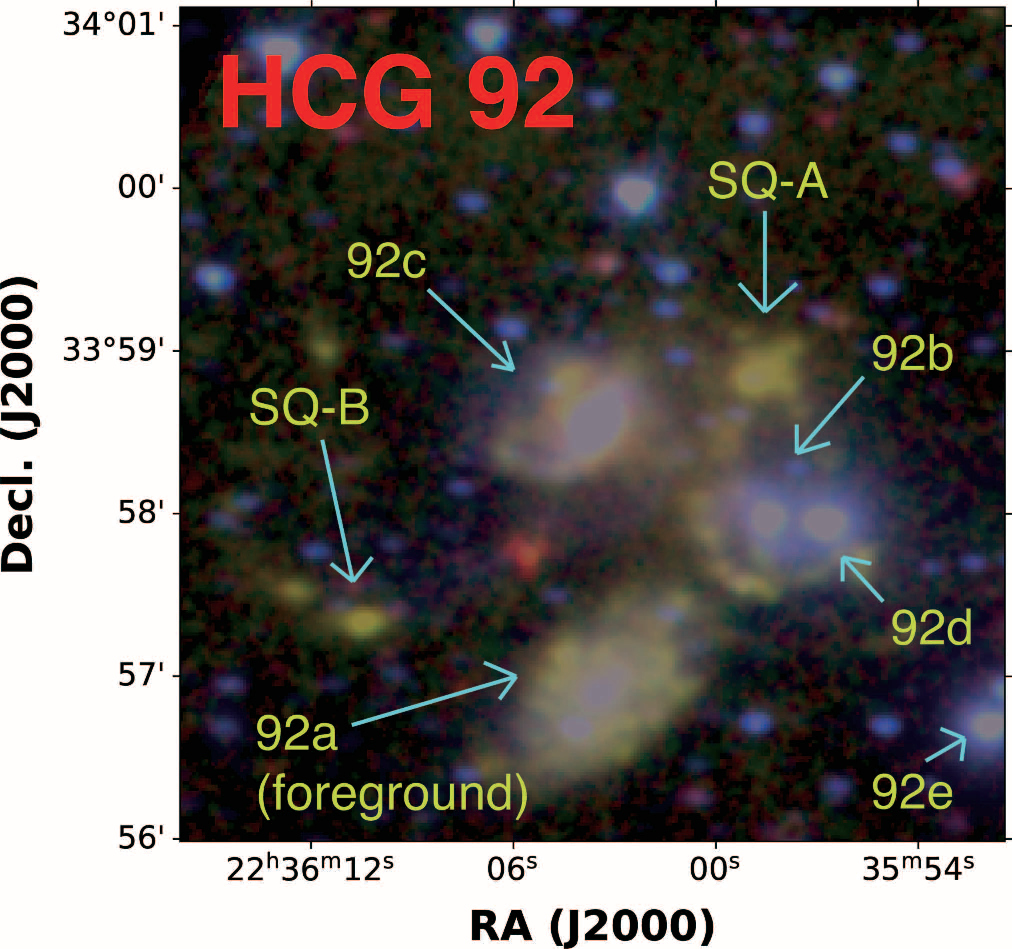}
\end{center}
\captionof{figure}{An AKARI/IRC composite image of HCG 92 made with N3 (3.2\,$\mu$m, blue), S7 (7\,$\mu$m, green), and S11 (11\,$\mu$m, red) bands. The beam sizes of all images have been degraded to match that of S11. The labels indicate HCG\,92a (NGC\,7320), HCG\,92b (NGC\,7318b), HCG\,92c (NGC\,7319), HCG\,92d (NGC\,7318a), and HCG\,92e (NGC\,7317), together with the intergalactic star-forming regions SQ-A and SQ-B. {Alt text: A composite infrared image of Stephan\textquoteright s Quintet (Hickson Compact Group 92). The image combines three infrared bands.}}
\label{92_fig.1}
\end{figure}

Figure~\ref{92_fig.1} shows a false-color image of Stephan\textquoteright s Quintet, created using the AKARI/IRC N3 (3.2\,$\mu$m) band for blue, the S7 (7\,$\mu$m) band for green, and the S11 (11\,$\mu$m) band for red.  Table~\ref{92a_table} summarizes the photometry results of the member galaxies (HCG\,92a, b, c, d, and e).

\subsubsection{HCG\,92a (NGC\,7320)}
\begin{figure}[htbp!]
\begin{center}
\includegraphics[width=\linewidth]{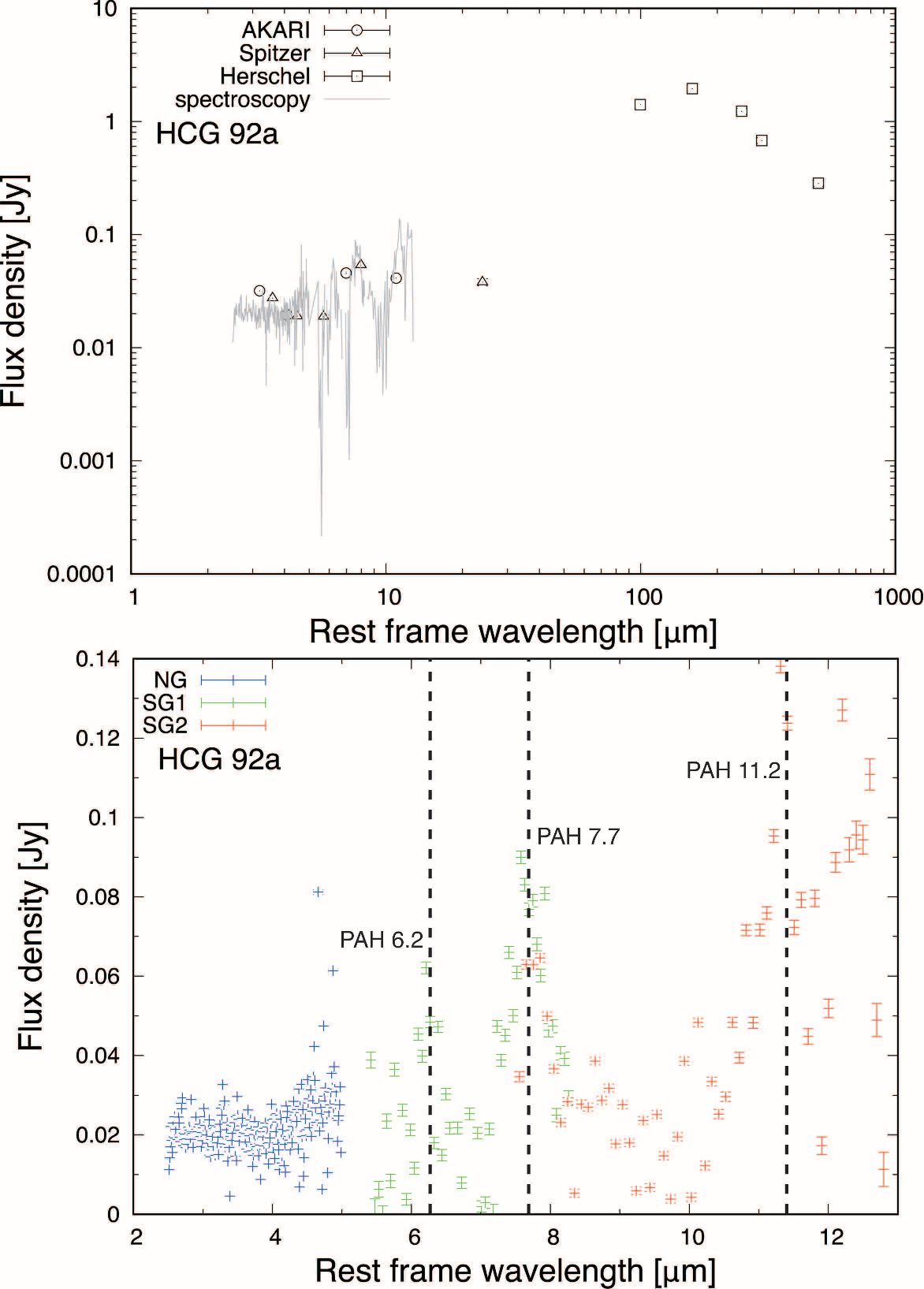}
\end{center}
\captionof{figure}{(Top): SED of HCG\,92a in the rest frame, constructed from AKARI, Spitzer, and Herschel data (Table~\ref{92a_table}). Photometric values are plotted as open circles (AKARI), triangles (Spitzer), and squares (Herschel). The gray line indicates the slit spectrum scaled to the photometry data.  (Bottom): Near- to mid-infrared spectrum from AKARI/IRC NG, SG1, and SG2, shown in blue, green, and red, respectively. {Alt text: Two panel figure for HCG 92a. The top panel is a spectral energy distribution with photometric points across near to far infrared wavelengths with an overlaid spectrum. The bottom panel shows the rest frame spectrum.}}
\label{92a_fig.1}\label{92a_fig.2}
\end{figure}

The top panel of Figure~\ref{92a_fig.1} shows the SED of HCG\,92a, constructed from the AKARI/IRC N3 and N4 bands; S7 and S11 bands; Spitzer/IRAC channels 1--4; Spitzer/MIPS channel 1; Herschel/PACS green and red bands; and Herschel/SPIRE PSW, PMW, and PLW bands. The bottom panel of Figure~\ref{92a_fig.2} presents the near- to mid-infrared spectrum of HCG\,92a, obtained with AKARI/IRC slit spectroscopy and scaled to match the photometric data points. PAH band features are detected at 6.2, 7.7, and 11.2\,$\mu$m clearly. The overall SED and spectrum indicate that HCG\,92a is a foreground, normal star-forming galaxy, and no further detailed comparison with other HCG\,92 members is warranted.

\begin{table*}[htbp!]
\centering
\caption{Near- to far-infrared photometry for HCG\,92 members.}
\label{92a_table}\label{92b_table}\label{92c_table}\label{92d_table}\label{92e_table}
\begin{threeparttable}
\begin{adjustbox}{max width=\textwidth, max totalheight=\textheight, keepaspectratio}
\begin{tabular}{l l c c c}
\hline
Source & Band & $\lambda$ [$\mu$m] & Flux $\pm$ Error [mJy] & Area [$\times 10^{3}$ arcsec$^{2}$] \\
\hline
HCG\,92a & AKARI/IRC N3 & 3.2  & $31.8 \pm 0.8$ & 6.32 \\
        & AKARI/IRC N4 & 4.1  & $19.6 \pm 0.7$ & 5.60 \\
        & AKARI/IRC S7 & 7.0  & $45.7 \pm 1.1$ & 6.92 \\
        & AKARI/IRC S11 & 11.0 & $41.3 \pm 1.0$ & 6.02 \\
        & Spitzer/IRAC ch.1 & 3.6 & $27.4 \pm 0.8$ & 5.93 \\
        & Spitzer/IRAC ch.2 & 4.5 & $19.0 \pm 0.6$ & 6.29 \\
        & Spitzer/IRAC ch.3 & 5.8 & $18.8 \pm 0.6$ & 5.81 \\
        & Spitzer/IRAC ch.4 & 7.9 & $53.8 \pm 1.6$ & 6.22 \\
        & Spitzer/MIPS ch.1 & 24.0 & $38.0 \pm 2.8$ & 6.30 \\
        & Herschel/PACS green & 100 & $(1.41 \pm 0.04)\times 10^{3}$ & 6.38 \\
        & Herschel/PACS red   & 160 & $(1.95 \pm 0.08)\times 10^{3}$ & 6.53 \\
        & Herschel/SPIRE PSW  & 250 & $(1.23 \pm 0.07)\times 10^{3}$ & 6.05 \\
        & Herschel/SPIRE PMW  & 350 & $(6.78 \pm 0.04)\times 10^{2}$ & 6.40 \\
        & Herschel/SPIRE PLW  & 500 & $(2.84 \pm 0.18)\times 10^{2}$ & 6.86 \\
\hline
HCG\,92b & AKARI/IRC N3 & 3.2  & $33.4 \pm 0.9$ & 3.03 \\
        & AKARI/IRC N4 & 4.1  & $19.5 \pm 0.7$ & 2.63 \\
        & AKARI/IRC S7 & 7.0  & $21.0 \pm 0.5$ & 2.94 \\
        & AKARI/IRC S11 & 11.0 & $20.5 \pm 0.5$ & 2.94 \\
        & Spitzer/IRAC ch.1 & 3.6 & $29.2 \pm 0.9$ & 3.06 \\
        & Spitzer/IRAC ch.2 & 4.5 & $17.2 \pm 0.5$ & 2.75 \\
        & Spitzer/IRAC ch.3 & 5.8 & $17.1 \pm 0.5$ & 2.99 \\
        & Spitzer/IRAC ch.4 & 7.9 & $18.0 \pm 0.5$ & 2.60 \\
        & Spitzer/MIPS ch.1 & 24.0 & $12.4 \pm 1.6$ & 2.76 \\
        & Herschel/PACS green & 100 & $(3.44 \pm 0.10)\times 10^{2}$ & 2.62 \\
        & Herschel/PACS red   & 160 & $(6.80 \pm 0.28)\times 10^{2}$ & 2.53 \\
        & Herschel/SPIRE PSW  & 250 & $(4.92 \pm 0.29)\times 10^{2}$ & 2.88 \\
        & Herschel/SPIRE PMW  & 350 & $(2.02 \pm 0.13)\times 10^{2}$ & 2.50 \\
        & Herschel/SPIRE PLW  & 500 & $(1.09 \pm 0.07)\times 10^{2}$ & 3.14 \\
\hline
HCG\,92c & AKARI/IRC N3 & 3.2  & $43.9 \pm 1.1$ & 3.41 \\
        & AKARI/IRC N4 & 4.1  & $41.1 \pm 1.4$ & 3.74 \\
        & AKARI/IRC S7 & 7.0  & $73.7 \pm 1.7$ & 3.54 \\
        & AKARI/IRC S11 & 11.0 & $98.0 \pm 2.3$ & 3.77 \\
        & Spitzer/IRAC ch.1 & 3.6 & $43.9 \pm 1.3$ & 3.88 \\
        & Spitzer/IRAC ch.2 & 4.5 & $42.6 \pm 1.3$ & 3.90 \\
        & Spitzer/IRAC ch.3 & 5.8 & $55.2 \pm 1.7$ & 3.73 \\
        & Spitzer/IRAC ch.4 & 7.9 & $85.4 \pm 2.6$ & 3.76 \\
        & Spitzer/MIPS ch.1 & 24.0 & $(1.88 \pm 0.08)\times 10^{2}$ & 3.46 \\
        & Herschel/PACS green & 100 & $(8.93 \pm 0.25)\times 10^{2}$ & 3.63 \\
        & Herschel/PACS red   & 160 & $(1.31 \pm 0.05)\times 10^{3}$ & 3.89 \\
        & Herschel/SPIRE PSW  & 250 & $(8.04 \pm 0.48)\times 10^{2}$ & 3.56 \\
        & Herschel/SPIRE PMW  & 350 & $(3.71 \pm 0.23)\times 10^{2}$ & 3.60 \\
        & Herschel/SPIRE PLW  & 500 & $(1.49 \pm 0.10)\times 10^{2}$ & 3.92 \\
\hline
HCG\,92d & AKARI/IRC N3 & 3.2  & $36.4 \pm 1.0$ & 2.61 \\
        & AKARI/IRC N4 & 4.1  & $22.6 \pm 0.8$ & 2.58 \\
        & AKARI/IRC S7 & 7.0  & $16.5 \pm 0.4$ & 2.71 \\
        & AKARI/IRC S11 & 11.0 & $12.7 \pm 0.3$ & 2.54 \\
        & Spitzer/IRAC ch.1 & 3.6 & $33.7 \pm 1.0$ & 2.90 \\
        & Spitzer/IRAC ch.2 & 4.5 & $20.6 \pm 0.6$ & 2.74 \\
        & Spitzer/IRAC ch.3 & 5.8 & $18.4 \pm 0.6$ & 2.53 \\
        & Spitzer/IRAC ch.4 & 7.9 & $18.6 \pm 0.6$ & 2.75 \\
        & Spitzer/MIPS ch.1 & 24.0 & $5.18 \pm 1.55$ & 2.65 \\
        & Herschel/PACS green & 100 & $(1.20 \pm 0.04)\times 10^{2}$ & 2.13 \\
        & Herschel/PACS red   & 160 & $(2.32 \pm 0.10)\times 10^{2}$ & 2.70 \\
        & Herschel/SPIRE PSW  & 250 & $(1.65 \pm 0.11)\times 10^{2}$ & 2.81 \\
        & Herschel/SPIRE PMW  & 350 & $80.2 \pm 6.6$ & 2.80 \\
        & Herschel/SPIRE PLW  & 500 & $47.5 \pm 4.5$ & 2.94 \\
\hline
HCG\,92e & AKARI/IRC N3 & 3.2  & $30.8 \pm 0.8$ & 2.30 \\
        & AKARI/IRC N4 & 4.1  & $18.0 \pm 0.6$ & 2.22 \\
        & AKARI/IRC S7 & 7.0  & $8.18 \pm 0.19$ & 2.62 \\
        & AKARI/IRC S11 & 11.0 & $5.23 \pm 0.13$ & 2.43 \\
        & Spitzer/IRAC ch.1 & 3.6 & $27.0 \pm 0.8$ & 2.18 \\
        & Spitzer/IRAC ch.2 & 4.5 & $16.8 \pm 0.5$ & 2.13 \\
        & Spitzer/IRAC ch.3 & 5.8 & $13.9 \pm 0.4$ & 2.16 \\
        & Spitzer/IRAC ch.4 & 7.9 & $6.84 \pm 0.21$ & 1.91 \\
        & Spitzer/MIPS ch.1 & 24.0 & --\tnote{$\dag$} & 2.28 \\
        & Herschel/PACS green & 100 & --\tnote{$\dag$} & 2.08 \\
        & Herschel/PACS red   & 160 & $12.5 \pm 2.9$ & 2.30 \\
        & Herschel/SPIRE PSW  & 250 & $13.5 \pm 4.4$ & 2.30 \\
        & Herschel/SPIRE PMW  & 350 & $6.25 \pm 3.94$ & 2.00 \\
        & Herschel/SPIRE PLW  & 500 & $6.97 \pm 3.20$ & 2.35 \\
\hline
\end{tabular}
\end{adjustbox}
\begin{tablenotes}
\item[$\dag$] Below the detection limit.
\end{tablenotes}
\end{threeparttable}
\end{table*}

\subsubsection{HCG\,92b (NGC\,7318b)}
\begin{figure*}[htbp!]
\begin{adjustbox}{max width=\textwidth, max totalheight=\textheight, keepaspectratio, center}
\includegraphics{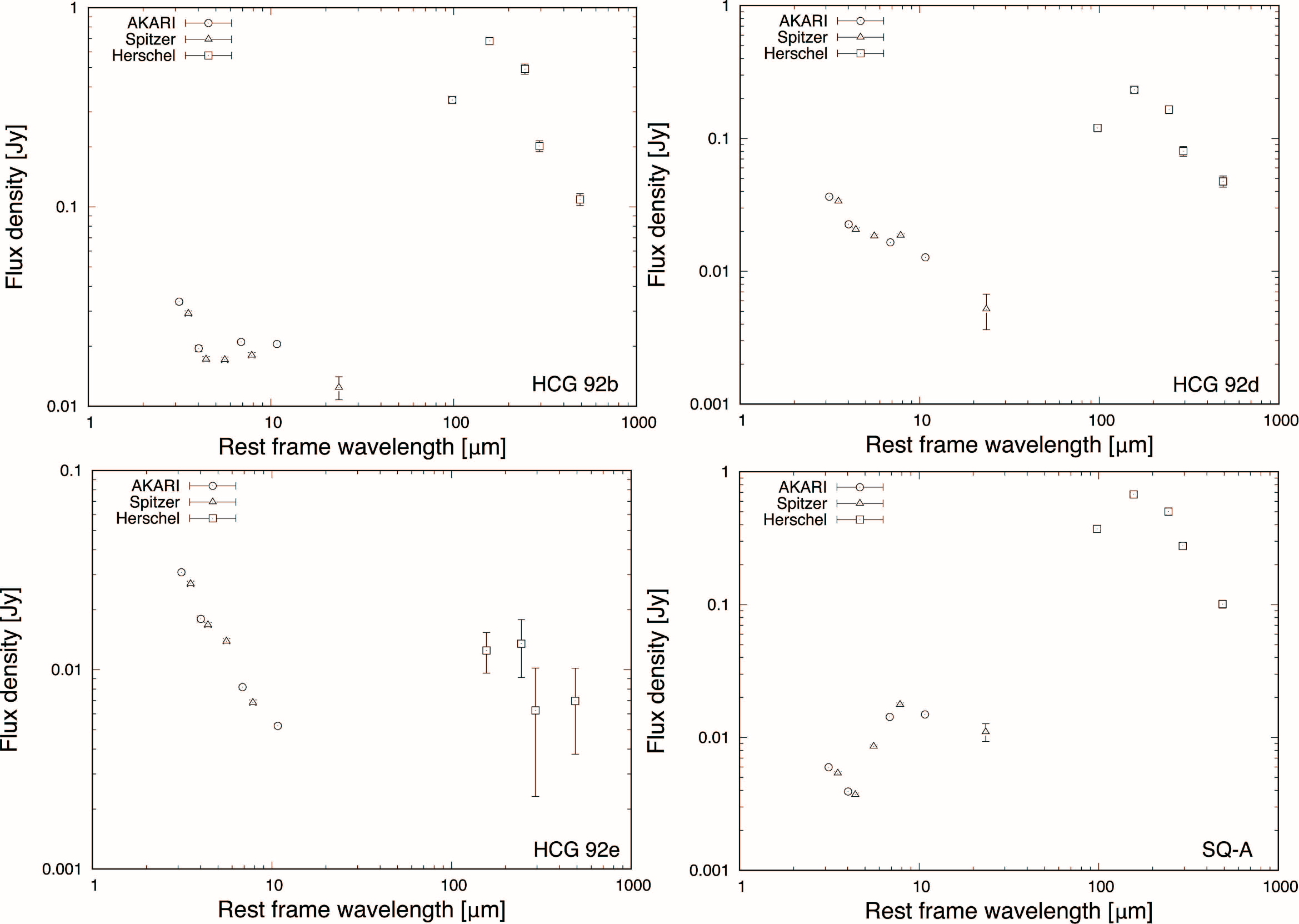}
\end{adjustbox}
\captionsetup{justification=raggedright, singlelinecheck=false, width=\textwidth}
\caption{Photometric SEDs in rest frame for HCG\,92b, 92d, 92e, and SQ-A. Notation follows Figure~\ref{92a_fig.1} (top). {Alt text: Four panel figure showing photometric spectral energy distributions in the rest frame. The panels correspond to HCG 92b, HCG 92d, HCG 92e, and SQ-A. Each panel plots infrared photometric points across near to far infrared wavelengths.}}
\label{92b_fig.1}\label{92d_fig.1}\label{92e_fig.1}\label{92sqa_fig.1}
\end{figure*}

The top left panel of Figure~\ref{92b_fig.1} shows the SED of HCG\,92b. For HCG\,92b and 92d, the spectra overlapped along the dispersion direction and could not be clearly separated; therefore, only photometric data were used for the analysis. The FIR emission dominated in the SED, with relatively weak mid-infrared emission, suggesting a substantial long-wavelength dust component, although the present photometric data alone do not distinguish whether its heating is dominated by star formation within the galaxy body or by dust associated with the disturbed group environment. No obvious excess is seen at wavelengths shorter than $\sim 10~\mu$m, indicating limited warm dust or AGN-heated components. Only photometric data are available for HCG\,92b, and thus spectral diagnostics such as PAH feature strengths or ionic line ratios cannot be applied. Further classification relies on the broadband SED shape alone.

\subsubsection{HCG\,92c (NGC\,7319)}
\begin{figure}[htbp!]
\begin{center}
\includegraphics[width=\linewidth]{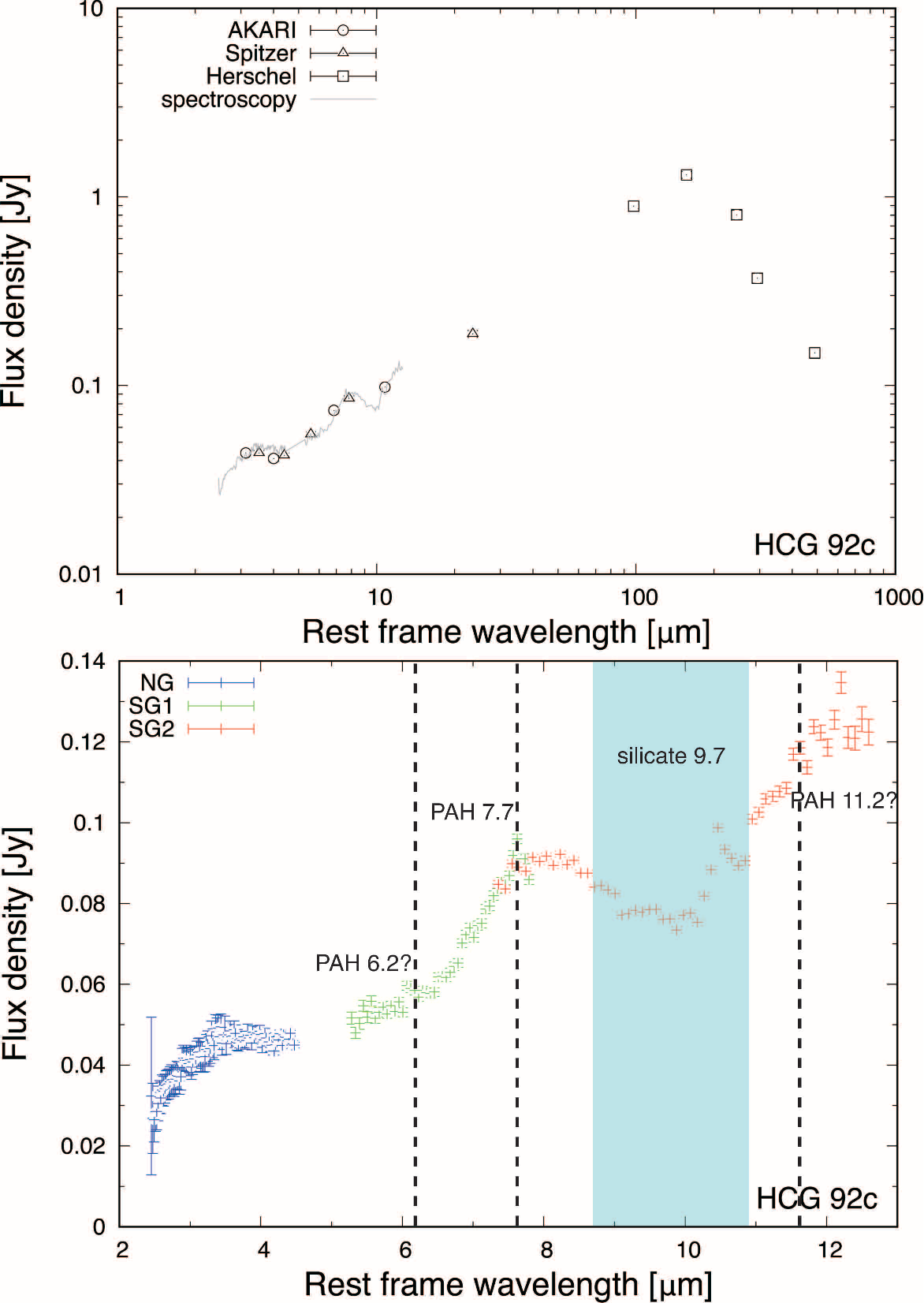}
\end{center}
\captionof{figure}{(Top): Same as Figure~\ref{92a_fig.1} (top), but for HCG\,92c (Table~\ref{92c_table}).\\(Bottom): Near- to mid-infrared spectrum of HCG\,92c, of the AKARI/IRC NG, SG1, and SG2 in rest frame, scaled to photometry. Color coding follows Figure~\ref{92a_fig.1} (bottom). {Alt text: Two panel figure for HCG 92c. The top panel is a spectral energy distribution with photometric points across near to far infrared wavelengths with an overlaid spectrum. The bottom panel shows the rest frame spectrum.}}
\label{92c_fig.1}\label{92c_fig.2}
\end{figure}

The SED of HCG\,92c is shown in the top panel of Figure~\ref{92c_fig.1}. The near- to mid-infrared spectrum of HCG\,92c shown in the bottom panel of Figure~\ref{92c_fig.2} was extracted from the AKARI/IRC pointing ID 3221001.1 (nominally a slit observation of HCG\,92a), in which HCG\,92c was serendipitously recorded in the slitless aperture. We apply a small multiplicative scaling ($<10\%$) to the spectrum of HCG\,92c, which is attributable to a minor mismatch in effective extraction apertures. This galaxy is classified as a Seyfert~2 galaxy \citep{key-Veron2006}. Its spectrum is characterized by a steeply rising red continuum and a prominent silicate absorption feature at 9.7\,$\mu$m. PAH features at 6.2\,$\mu$m, 7.7\,$\mu$m, and 11.2\,$\mu$m are extremely weak, with equivalent widths well below those typically found in star-forming galaxies, indicating strong suppression of PAH emission. The overall mid-infrared shape suggests that the emission is dominated by AGN-heated dust, with little contamination from surrounding star-forming regions in the AKARI aperture.

\begin{figure}[htbp!]
\begin{center}
\includegraphics[width=\linewidth]{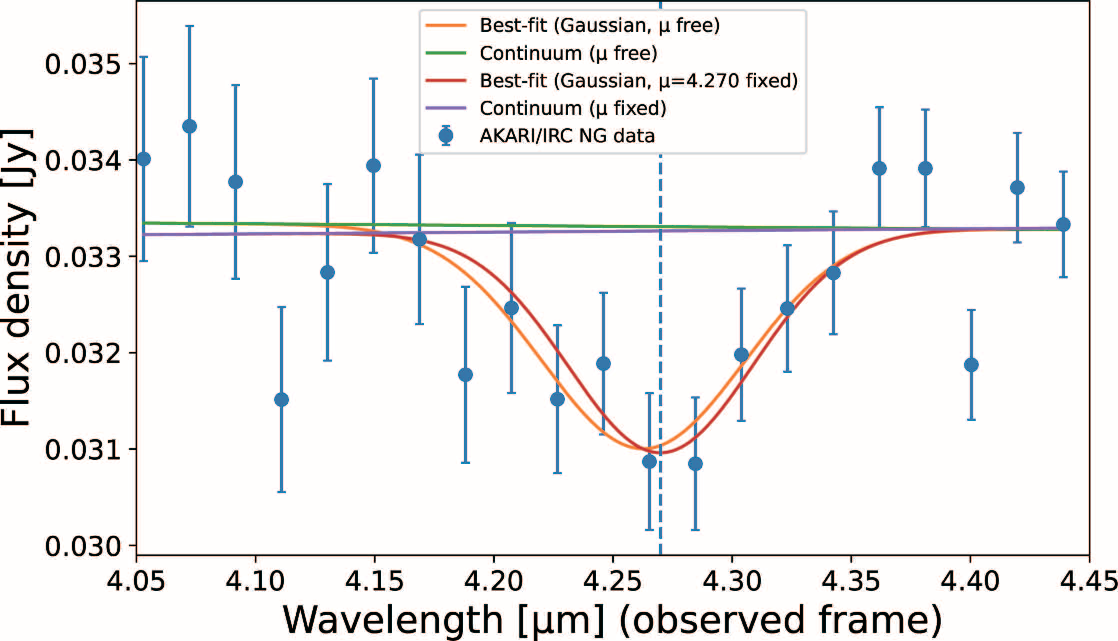}
\end{center}
\captionof{figure}{Zoom-in of the AKARI/IRC/NG spectrum of HCG\,92c around the 4.27\,$\mu$m absorption feature (observer frame; no redshift correction). Blue points show the observed fluxes with 1$\sigma$ uncertainties. The spectrum is locally fitted over 4.05--4.45\,$\mu$m using a linear continuum plus a Gaussian absorption component. Solid curves indicate the best-fit models for two cases: the Gaussian centroid treated as a free parameter and fixed at $\mu=4.270\,\mu$m; the corresponding linear continua are also shown. The vertical dashed line marks 4.270\,$\mu$m. {Alt text: The observed frame infrared spectrum of HCG 92c from 4.05 to 4.45 micrometer. Data points include one sigma uncertainties. Two best fit models are employed with a linear continuum and a Gaussian absorption feature near 4.27 micrometer.}}
\label{92c_fig.3}
\end{figure}

In the AKARI/IRC/NG spectrum of HCG\,92c, we found an absorption feature around 4.27\,$\mu$m in the observer frame. We performed a local fit over 4.05--4.45\,$\mu$m with a linear continuum plus a Gaussian absorption component (Figure~\ref{92c_fig.3}) and evaluated the detection significance using the best-fit amplitude $A$ of the Gaussian absorption component (negative for absorption) and its formal 1$\sigma$ uncertainty $\sigma_A$. The fit yields a statistically significant absorption with $|A|/\sigma_A=4.8$ when the centroid wavelength $\mu$ is treated as a free parameter ($\mu=4.263\pm0.011\,\mu$m; $\chi^2_\nu=1.13$), and $|A|/\sigma_A=4.7$ when the centroid is fixed to $\mu=4.270\,\mu$m ($\chi^2_\nu=1.09$). Both fits provide an acceptable goodness of fit in terms of the reduced chi-square, $\chi^2_\nu \simeq 1.1$. If this feature is interpreted as CO$_2$ ice absorption, its observed central wavelength suggests that it is not associated with the redshifted host galaxy HCG\,92c, but would instead point to a foreground component at approximately $z=0$, possibly in the Galaxy or in nearby foreground material along the line of sight. However, its physical origin is uncertain at present, and we do not discuss it further in this paper.

\subsubsection{HCG\,92d (NGC\,7318a)}
The SED of HCG\,92d is shown in the top right panel of Figure~\ref{92d_fig.1}. Although no mid-infrared spectroscopic data are available, the SED exhibits a broad peak around 100\,$\mu$m and a relatively flat 3--8\,$\mu$m continuum, consistent with the thermal emission from a moderately warm dust component heated by recent star formation. Compared with the SED of HCG\,92b, both galaxies show the dominance of FIR emission, peaking at similar wavelengths, indicative of comparable cold dust temperatures. However, HCG\,92b displays a slightly enhanced mid-infrared continuum relative to its FIR emission, which suggests a larger fraction of warm dust or a possible AGN contribution. In contrast, HCG\,92d appears more consistent with galaxy-scale dust heating by stellar radiation, although a weak nuclear AGN contribution cannot be excluded.

Recent CO mapping of Stephan\textquoteright s Quintet by \citet{key-Emonts2025} showed that no significant CO(2--1) is detected from the inner regions of NGC\,7318a and NGC\,7318b (HCG\,92d and HCG\,92b, respectively), while the strongest molecular gas associated with the NGC\,7318a/b pair is instead found along the southern arm of NGC\,7318b and in the adjacent shocked ridge. They argued that much of the gas in the main bodies of these galaxies has likely been stripped into the intergalactic medium by past collisions. At the same time, \citet{key-Aromal2025} showed that very young star clusters are abundant in the southern parts of the system and closely trace the CO-bright southern arm and Left Arc, but that active recent star formation is also present in regions with weak or undetected CO emission. Taken together, these results suggest that the strong continuum emission longer than 100\,$\mu$m in HCG\,92b and HCG\,92d does not necessarily imply the presence of gas-rich nuclear star formation. A more plausible interpretation is that the far-infrared emission arises from dust heated by recent but spatially distributed star formation, including extra-nuclear young stellar populations, with a possible additional contribution from dust in tidally disturbed material within the group environment. A contribution from AGN-heated dust cannot be completely excluded, especially for HCG\,92b, but the present photometric data alone do not allow us to isolate such a component robustly. Furthermore, the angular resolution of the far-infrared data is not sufficient to determine whether the emission at wavelengths longer than 100\,$\mu$m is spatially coincident with the young star clusters, the galaxy nuclei, or more diffuse tidal material. A contribution from dust heated by the diffuse radiation field of evolved stellar populations may also be present, but cannot be separated from dust heated by recent distributed star formation with the present data. As an additional literature context, \citet{key-Appleton2017} presented Spitzer/IRS spectra extracted across the nuclei of NGC~7318b and NGC~7318a (corresponding to HCG\,92b and HCG\,92d, respectively). They noted likely 10\,$\mu$m silicate emission in both nuclei, which they briefly discussed as possible evidence for low-luminosity AGN viewed above the accretion disk. The same nuclear spectra also show PAH emission, with the feature appearing stronger in NGC~7318b than in NGC~7318a. This comparison suggests that at least some level of nuclear star formation may still be present in NGC~7318b, even though recent CO observations indicate a marked lack of cold gas in the main bodies of NGC~7318a/b. Therefore, while our broadband SED analysis does not allow us to isolate the nuclear component directly, the existing Spitzer spectroscopy supports a picture in which weak AGN-related hot dust and modest nuclear star formation can coexist with the more spatially distributed dust-heating components discussed above.

\subsubsection{HCG\,92e (NGC\,7317)}
The SED of HCG\,92e is shown in the bottom left panel of Figure~\ref{92e_fig.1}. Although no mid-infrared spectroscopic data are available, the SED shows a relatively weak far-infrared peak and a declining continuum toward shorter wavelengths, indicating a low level of dust emission and minimal ongoing star formation. The absence of a prominent warm dust component in the 10--30\,$\mu$m range further supports the interpretation that this system is likely dominated by the evolved stellar population with little recent star formation activity. Compared to other HCG\,92 members, HCG\,92e appears to be among the most quiescent members, with an infrared luminosity consistent with that of an early-type, gas-poor galaxy.

\subsubsection{SQ-A}
Table~\ref{92sqa_table} presents a summary of the near- to far-infrared photometry for the ISOCAM infrared source SQ-A. The SED of SQ-A is shown in the bottom right panel of Figure~\ref{92sqa_fig.1}. Although no spectroscopic data are available, the SED exhibits a pronounced far-infrared peak and a rising continuum in the mid-infrared, indicative of substantial warm dust emission. Given that SQ-A is not a galaxy but rather an infrared-bright region within the Stephan\textquoteright s Quintet intragroup medium, this emission is likely powered by intense, localized star formation triggered by large-scale shocks and/or interactions between group members. The high infrared luminosity and relatively warm far-infrared colors indicate that SQ-A is among the most active star-forming regions in Stephan\textquoteright s Quintet. Previous studies estimated ${\rm SFR}\simeq1.45\,M_\odot\,{\rm yr}^{-1}$ for SQ-A based on extinction-corrected H$\alpha$ and derived $L_{\rm IR}(5$--$1000\,\micron)\simeq7\times10^{42}\,{\rm erg\,s}^{-1}$ \citep{key-Xu2003}. Far-infrared SED analyses further suggest dust temperatures of the order of $\sim22$--24\,K in the region including SQ-A, where the PACS data are extracted \citep{key-Appleton2013}. More recently, \citet{key-Xu2025} argued that SQ-A is best understood as a collision-triggered starburst in the intragroup medium, caused by a high-speed collision between distinct gas systems. This interpretation is also consistent with the detailed multi-wavelength view of the shocked intergalactic environment presented by \citet{key-Appleton2023}. In this paper, we therefore emphasize the nature of SQ-A as an intragroup starburst region associated with the shocked intergalactic environment, without attempting a definitive classification of its TDG nature.

\begin{table*}[htbp!]
\centering
\caption{Near- to far-infrared photometry for SQ-A/B.}
\label{92sqa_table}\label{92sqb_table}
\begin{adjustbox}{max width=\textwidth, max totalheight=\textheight, keepaspectratio}
\begin{tabular}{l l c c c}
\hline
Source & Band & $\lambda$ [$\mu$m] & Flux $\pm$ Error [mJy] & Area [$\times 10^{3}$ arcsec$^{2}$] \\
\hline
SQ-A    & AKARI/IRC N3 & 3.2  & $5.98 \pm 0.15$ & 2.97 \\
        & AKARI/IRC N4 & 4.1  & $3.92 \pm 0.13$ & 2.80 \\
        & AKARI/IRC S7 & 7.0  & $14.3 \pm 0.3$ & 2.71 \\
        & AKARI/IRC S11 & 11.0 & $14.9 \pm 0.4$ & 2.71 \\
        & Spitzer/IRAC ch.1 & 3.6 & $5.38 \pm 0.16$ & 2.96 \\
        & Spitzer/IRAC ch.2 & 4.5 & $3.71 \pm 0.11$ & 2.91 \\
        & Spitzer/IRAC ch.3 & 5.8 & $8.58 \pm 0.26$ & 2.91 \\
        & Spitzer/IRAC ch.4 & 7.9 & $17.7 \pm 0.5$ & 2.94 \\
        & Spitzer/MIPS ch.1 & 24.0 & $11.0 \pm 1.7$ & 3.00 \\
        & Herschel/PACS green & 100 & $(3.72 \pm 0.10)\times 10^{2}$ & 3.00 \\
        & Herschel/PACS red   & 160 & $(6.68 \pm 0.29)\times 10^{2}$ & 3.00 \\
        & Herschel/SPIRE PSW  & 250 & $(4.25 \pm 0.25)\times 10^{2}$ & 3.03 \\
        & Herschel/SPIRE PMW  & 350 & $(1.84 \pm 0.14)\times 10^{2}$ & 3.02 \\
        & Herschel/SPIRE PLW  & 500 & $(1.10 \pm 0.09)\times 10^{2}$ & 3.05 \\
\hline
SQ-B    & AKARI/IRC N3 & 3.2  & $3.12 \pm 0.12$ & 2.55 \\
        & AKARI/IRC N4 & 4.1  & $2.01 \pm 0.10$ & 2.54 \\
        & AKARI/IRC S7 & 7.0  & $5.42 \pm 0.18$ & 2.51 \\
        & AKARI/IRC S11 & 11.0 & $5.88 \pm 0.19$ & 2.50 \\
        & Spitzer/IRAC ch.1 & 3.6 & $2.94 \pm 0.13$ & 2.52 \\
        & Spitzer/IRAC ch.2 & 4.5 & $2.14 \pm 0.10$ & 2.49 \\
        & Spitzer/IRAC ch.3 & 5.8 & $3.89 \pm 0.17$ & 2.52 \\
        & Spitzer/IRAC ch.4 & 7.9 & $8.31 \pm 0.29$ & 2.50 \\
        & Spitzer/MIPS ch.1 & 24.0 & $5.27 \pm 1.02$ & 2.56 \\
        & Herschel/PACS green & 100 & $(1.56 \pm 0.05)\times 10^{2}$ & 2.53 \\
        & Herschel/PACS red   & 160 & $(2.88 \pm 0.12)\times 10^{2}$ & 2.55 \\
        & Herschel/SPIRE PSW  & 250 & $(1.84 \pm 0.12)\times 10^{2}$ & 2.56 \\
        & Herschel/SPIRE PMW  & 350 & $87.4 \pm 6.4$ & 2.54 \\
        & Herschel/SPIRE PLW  & 500 & $48.9 \pm 4.8$ & 2.52 \\
\hline
\end{tabular}
\end{adjustbox}
\end{table*}

\subsubsection{SQ-B}
\begin{figure}[htbp!]
\begin{center}
\includegraphics[width=\linewidth]{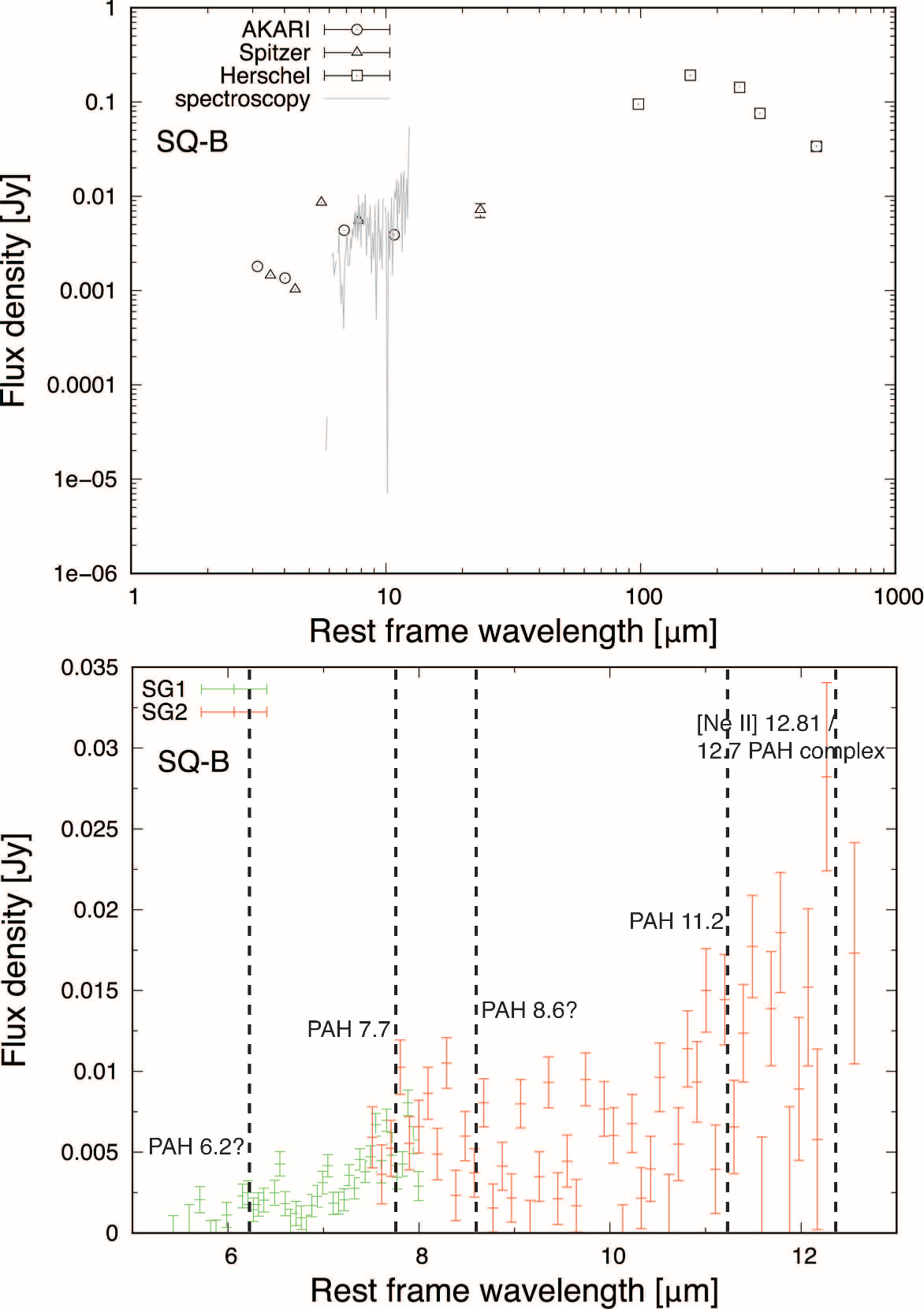}
\end{center}
\captionof{figure}{(Top): Same as Figure~\ref{92a_fig.1} (top), but for SQ-B (Table~\ref{92sqb_table}). (Bottom): AKARI/IRC slitless spectrum of SQ-B, showing the PAH features and the 12.81\,$\mu$m feature, likely including [Ne\,\textsc{ii}] emission blended with the 12.7\,$\mu$m PAH band. {Alt text: Two panel figure for SQ-B. The top panel is a spectral energy distribution with photometric points across near to far infrared wavelengths with an overlaid spectrum. The bottom panel shows the rest frame spectrum.}}
\label{92sqb_fig.1}\label{92sqb_fig.2}
\end{figure}

Table~\ref{92sqb_table} summarizes the near- to far-infrared photometry for the ISOCAM infrared source SQ-B. The SED of SQ-B is shown in the top panel of Figure~\ref{92sqb_fig.1}. The near- to mid-infrared spectrum of SQ-B, extracted from AKARI/IRC slitless spectroscopic observations, is shown in the bottom panel of Figure~\ref{92sqb_fig.2}. Spitzer/IRS Short-Low spectroscopy of SQ-B is presented in the SAINTS study by \citet{key-Higdon2010}, providing a mid-infrared (5--14~$\mu$m) view of the ISM in this tidal star-forming knot.  

In the wavelength range overlapping with IRS-SL ($\gtrsim$5~$\mu$m), the AKARI/IRC spectrum reproduces the same prominent PAH bands and the 12.81\,$\mu$m feature, likely including [Ne\,\textsc{ii}] blended with the 12.7\,$\mu$m PAH band, indicating that the overall spectral characteristics are consistent within the expected calibration and aperture differences. We detect PAH band features at 6.2~$\mu$m, 7.7~$\mu$m, 8.6~$\mu$m, and 11.2~$\mu$m. A simple modified blackbody fit to the far-infrared SED yields a temperature of submicrometer-sized dust of $T_d = 18$~K, assuming an emissivity power-law index of $\beta = 1$, and an integrated far-infrared flux of $L_{\rm FIR} = 3.7 \times 10^{-15}$~W\,m$^{-2}$. The strength of the PAH feature at 11.2~$\mu$m relative to the total far-infrared luminosity, $L_{\rm PAH\,11.2}/L_{\rm FIR}$, is approximately 0.006---significantly lower than typical values in the Galactic interstellar medium \citep[e.g., $L_{\rm PAH\,11.2}/L_{\rm FIR} = 0.02$--$0.04$;][]{key-Sakon2004}. Throughout this paper, we refer to this band as the 11.2~$\mu$m PAH feature (often labeled 11.3~$\mu$m in the literature); given the spectral resolution, we do not attempt to distinguish minor shifts within this band. Furthermore, the PAH 7.7~$\mu$m to 11.2~$\mu$m luminosity ratio in SQ-B is $\sim$ 3, which lies at the lower end of the range for actively star-forming regions.  Since the 11.2~$\mu$m band is supposed to arise from neutral PAHs, while the 7.7~$\mu$m mostly comes from ionized PAHs, the low ratio is consistent with relatively ``neutral'' PAH populations reported for SQ-B from Spitzer/IRS analyses \citep[e.g.,][]{key-Higdon2010}. While SQ-B is a well-known intergalactic star-forming region and a tidal dwarf galaxy candidate in the tidal debris of Stephan\textquoteright s Quintet \citep[e.g., an extinction-free ${\rm SFR}\simeq 0.5~M_\odot~{\rm yr^{-1}}$;][]{key-Lisenfeld2004}, the relatively low $L_{\rm PAH\,11.2}/L_{\rm FIR}$ ratio, modest PAH 7.7/11.2 ratio, and the subdued far-infrared luminosity suggest that its current star formation is not in a starburst-like phase. Instead, these diagnostics indicate that SQ-B is presently undergoing weak and/or fading star formation compared to typical starburst galaxies \citep{key-Brandl2006}, consistent with its interpretation as a tidal star-forming knot/TDG candidate \citep{key-Lisenfeld2004}.

\subsubsection{Dust SED Analysis of HCG\,92 Galaxies and Debris}
\label{92_sedmodel}
As in the analysis of HCG\,56, we examine the near- to far-infrared SEDs of the member galaxies in HCG\,92 using both the AC composition model \citep{key-Galliano2011} and the THEMIS composition model \citep{key-Jones2013}. For HCG\,92c, we account for the contribution of the AGN by incorporating emission from AGN-related templates provided by \citet{key-Siebenmorgen2015}. Among the member galaxies of HCG\,92, only HCG\,92c shows a preference for an additional screen-like extinction component to reproduce its observed SED. The best-fit values are $A^{\rm screen}_{V} = 2.74^{+2.81}_{-2.74}$\,mag for the AC dust composition (consistent with zero within $1\sigma$) and $A^{\rm screen}_{V} = 8.50 \pm 4.33$\,mag for the THEMIS composition. This may be phenomenologically related to the absorption feature detected at an observed-frame wavelength of $\sim$ 4.27~$\mu$m in HCG\,92c, although the physical origin of this feature is uncertain. We therefore treat the screen extinction component as a phenomenological attenuation term in the SED fitting, rather than as evidence for a specific absorbing medium.

\begin{figure*}[htbp!]
\begin{adjustbox}{max width=\textwidth, max totalheight=\textheight, keepaspectratio, center}
\includegraphics{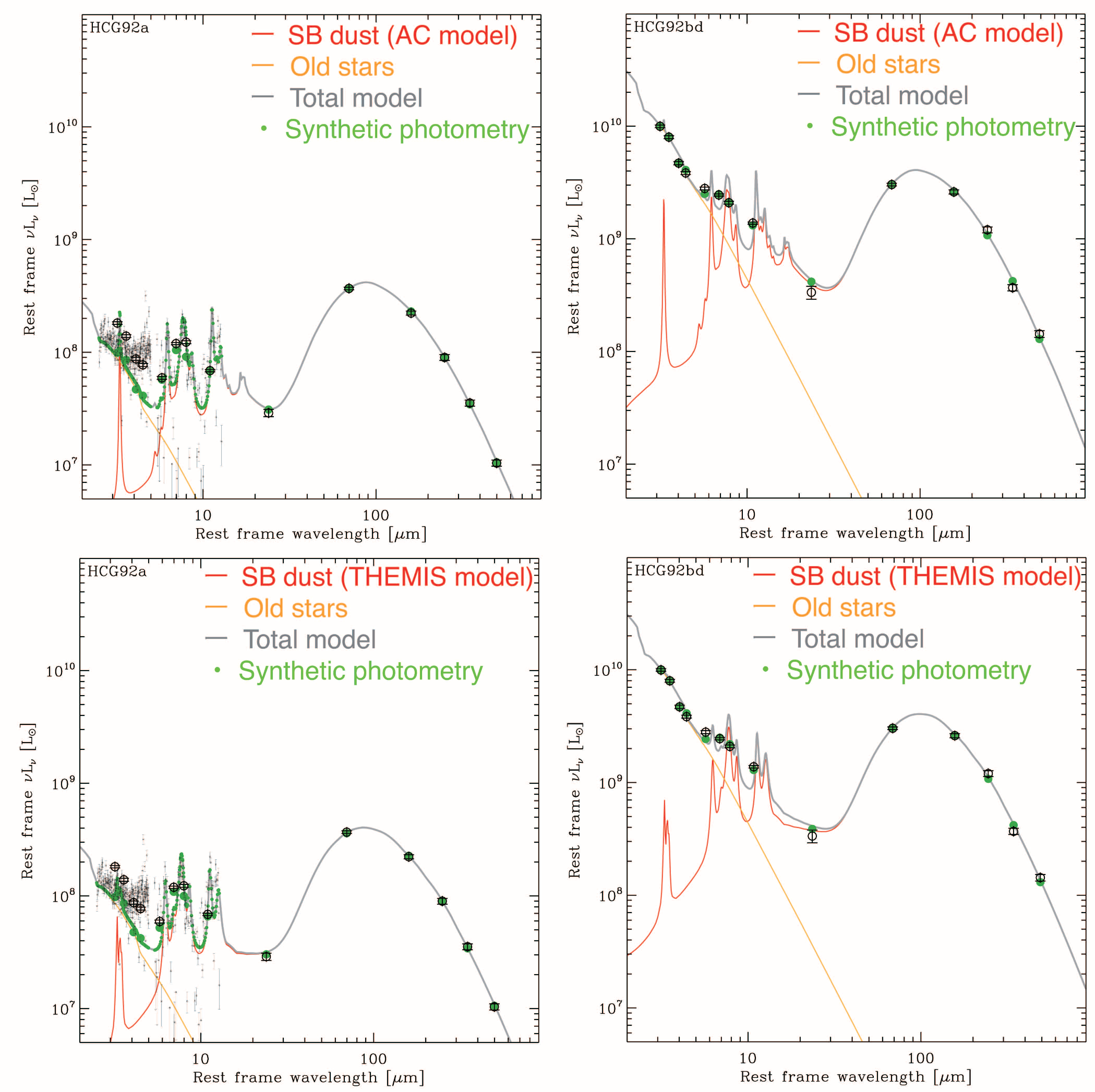}
\end{adjustbox}
\caption{(Top left/right): Best-fit SED models of HCG\,92a and HCG\,92b+d using the AC dust model \citep{key-Galliano2011}.\\(Bottom left/right): Same as top panels but using the THEMIS dust model \citep{key-Jones2013}. \\Orange, red, gray, and green follow the notation in Figure~\ref{56a_sed_fig.1}. {Alt text: Four panel model fit figure for HCG 92a and the combined HCG 92b and HCG 92d. The top row shows best fit spectral energy distribution models using one dust model. The bottom row shows the same targets using the alternative dust model.}}
\label{92_sed_fig.1}\label{92_sed_fig.2}\label{92_sed_fig.3}\label{92_sed_fig.4}
\end{figure*}

\begin{figure*}[htbp!]
\begin{adjustbox}{max width=\textwidth, max totalheight=\textheight, keepaspectratio, center}
\includegraphics{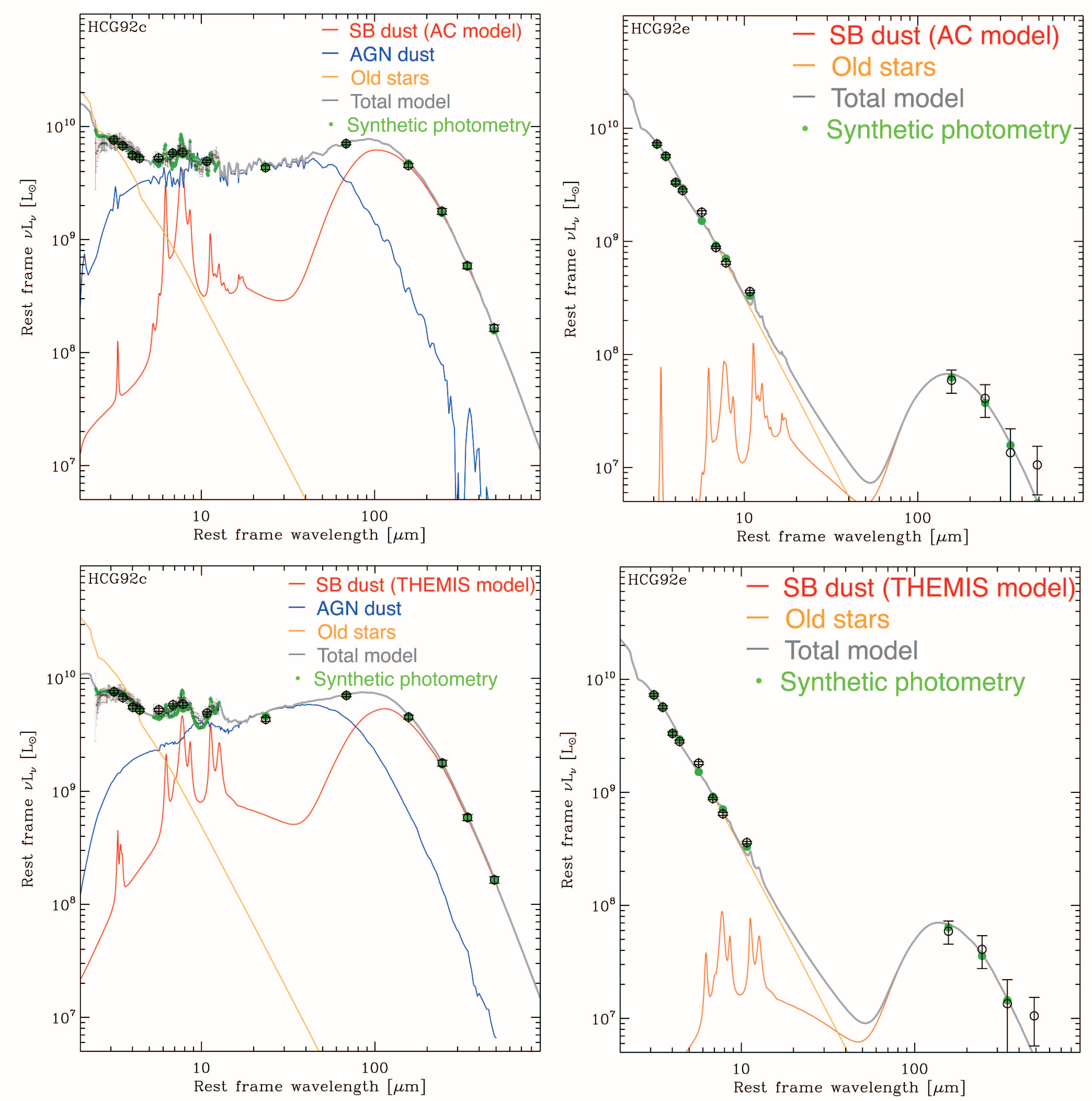}
\end{adjustbox}
\caption{Top left: SED fitting of HCG\,92c using the AC model combined with AGN templates \citep{key-Siebenmorgen2015}.\\(Top right): SED of HCG\,92e using the AC model with a uniformly illuminated dust geometry.\\(Bottom left/right): Same as top panels, but with the THEMIS dust model.\\Notation follows Figure~\ref{92_sed_fig.1}. {Alt text: Four panel model fit figure for HCG 92c and HCG 92e. The top row shows best fit spectral energy distribution models using one dust model. The bottom row shows the same targets using the alternative dust model. HCG 92c includes an additional active galactic nucleus dust component.}}
\label{92_sed_fig.5}\label{92_sed_fig.6}\label{92_sed_fig.7}\label{92_sed_fig.8}
\end{figure*}

\begin{figure*}[htbp!]
\begin{adjustbox}{max width=\textwidth, max totalheight=\textheight, keepaspectratio, center}
\includegraphics{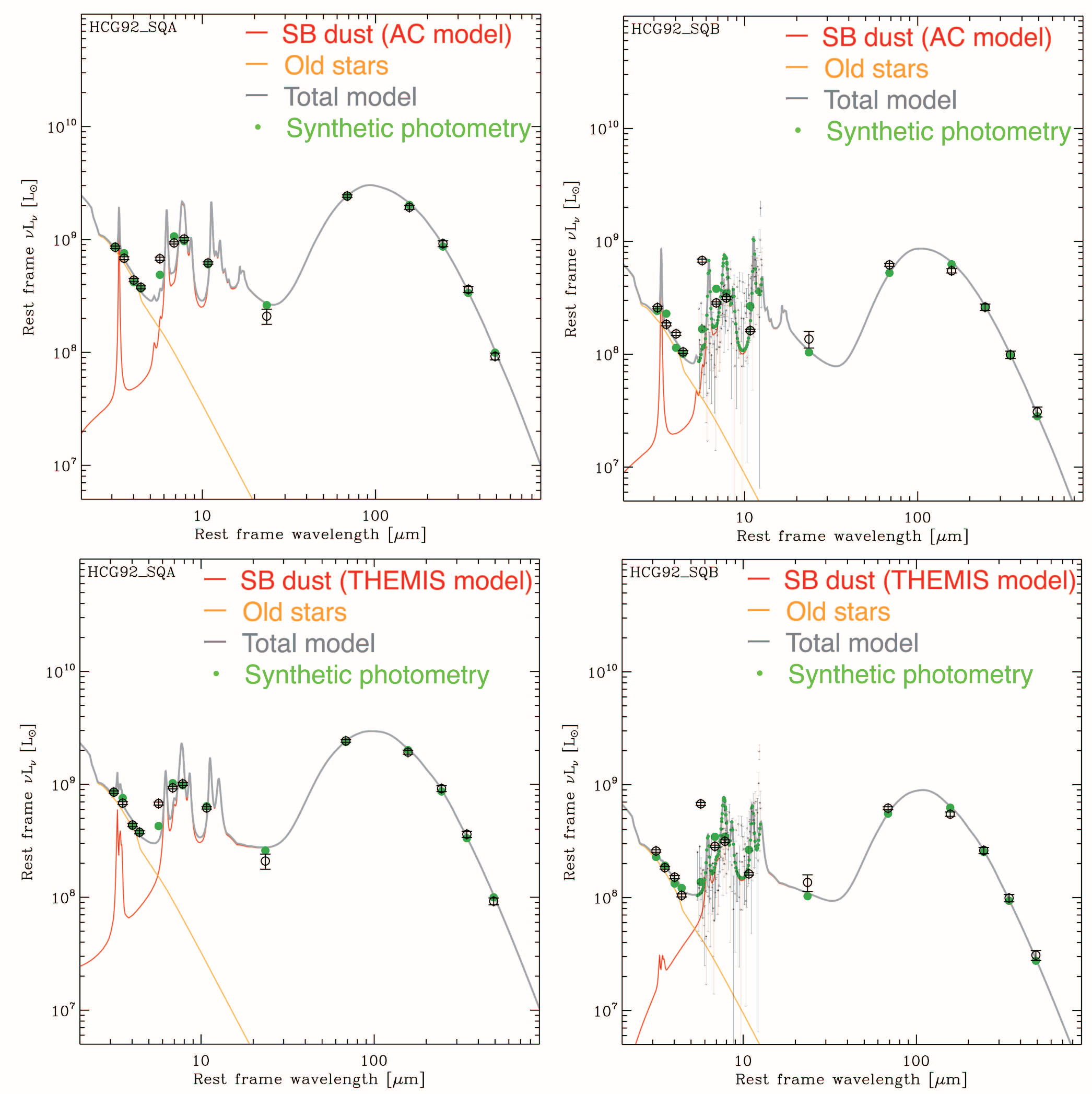}
\end{adjustbox}
\caption{(Top left/right): Best-fit SEDs of SQ-A and SQ-B using the AC model.\\(Bottom left/right): Same as top panels but using the THEMIS model.\\Notation is the same as in Figure~\ref{92_sed_fig.1}. {Alt text: Four panel model fit figure for SQ-A and SQ-B. The top row shows best fit spectral energy distribution models using one dust model. The bottom row shows the same targets using the alternative dust model.}}
\label{92_sed_fig.9}\label{92_sed_fig.10}\label{92_sed_fig.11}\label{92_sed_fig.12}
\end{figure*}

For HCG\,92a and 92b+d, we set $M_{\rm dust}$, $U_{\rm min}$, $U_{\rm max}$, $\alpha$, $q_{\rm PAH}$, $f_{+}$, and $M_{\star}$ as free parameters. The best-fit results are presented in Table~\ref{92_sed_table}. In the case of HCG\,92e, the faintness of the MIR data limits our ability to fit the SED using our non-uniformly illuminated dust model (see Table~\ref{92e_table}, Figure~\ref{92e_fig.1}). Consequently, we adopted an illuminated dust model with $M_{\rm dust}$, $U$, $q_{\rm PAH}$, $f_{+}$, and $M_{\star}$ as free parameters. The corresponding best-fit results are also shown in Table~\ref{92_sed_table}. For HCG\,92c, we included AGN-related parameters---$L^{\rm AGN}_{\star}$, $R^{\rm AGN}$, $V^{\rm AGN}_{c}$, $A^{\rm AGN}_{c}$, $A^{\rm AGN}_{d}$, and $\theta^{\rm AGN}$---and a foreground extinction parameter $A^{\rm screen}_{V}$ in addition to the dust-related parameters ($M_{\rm dust}$, $U_{\rm min}$, $U_{\rm max}$, $\alpha$, $q_{\rm PAH}$, $f_{+}$, and $M_{\star}$), treating them all as free parameters. The inclusion of this screen component is also motivated by the overall attenuated near- to mid-infrared continuum shape of HCG\,92c and by the absorption feature detected at $\sim$ 4.27~$\mu$m in the AKARI/IRC/NG spectrum (Section~3.2.3). However, because the physical origin of this feature is uncertain, we do not use it to identify a specific absorbing medium. For SQ-A and SQ-B, the SED fitting is performed in the same manner as for other HCG\,92 member galaxies, except for HCG\,92c and 92e. The best-fit parameters, including $M_{\rm dust}$, $U_{\rm min}$, $U_{\rm max}$, $\alpha$, $q_{\rm PAH}$, $f_{+}$, and $M_{\star}$, are listed in Table~\ref{92_sed_table2}. As illustrated in Figures~\ref{92_sed_fig.1}--\ref{92_sed_fig.12}, the near- to mid-infrared SEDs of HCG\,92a, 92b+d, 92c, 92e, SQ-A, and SQ-B are well reproduced by these models, and the choice between the two dust models generally has minor effects on the best-fit parameter values.

\begin{table*}[htbp!]
\begin{adjustbox}{max width=\textwidth, max totalheight=\textheight, keepaspectratio, center}
\begin{threeparttable}
\caption{Best fit parameters of HCG\,92 members and debris.\tnote{$\dag$}}
\label{92_sed_table}\label{92_sed_table2}
\begin{tabular}{ccc}
\hline\hline
&AC COMPOSITION&THEMIS COMPOSITION\\ \hline
\multicolumn{3}{c}{\bf {HCG\,92a}}\\ \hline
ln ($M_{\rm dust}$ [$M_{\odot}$])&14.8$\pm$0.240&14.7$\pm$0.260\\
ln ($U_{\rm min}$ [2.2$\times$10$^{-5}$Wm$^{-2}$])&-4.26$^{+0.810}_{-0.340}$&-3.93$^{+0.790}_{-0.680}$\\
ln ($U_{\rm max}-U_{\rm min}$ [2.2$\times$10$^{-5}$Wm$^{-2}$])&3.92$\pm$0.240&3.96$\pm$0.200\\
$\alpha$&1.40$\pm$0.0800&1.48$\pm$0.0640\\
$q_{\rm PAH}$&($5.24\pm0.160$)$\times$10$^{-2}$&($7.47\pm0.190$)$\times$10$^{-2}$\\
$f_{+}$&($7.80\pm1.80$)$\times$10$^{-2}$&($3.36\pm0.300$)$\times$10$^{-1}$\\
$M_{\star}$ [$M_{\odot}$]&(2.14$\pm$0.0230)$\times$10$^{9}$&(2.09$\pm$0.0240)$\times$10$^{9}$\\
ln ($\langle U\rangle$ [2.2$\times$10$^{-5}$Wm$^{-2}$])&0.277$\pm$0.244&0.118$\pm$0.265\\ \hline
\multicolumn{3}{c}{\bf {HCG\,92b+d}}\\ \hline
ln ($M_{\rm dust}$ [$M_{\odot}$])&17.2$\pm$0.250&17.2$\pm$0.240\\
ln ($U_{\rm min}$ [2.2$\times$10$^{-5}$Wm$^{-2}$])&-4.61$^{+0.890}_{-0}$&-4.61$^{+1.01}_{-0}$\\
ln ($U_{\rm max}-U_{\rm min}$ [2.2$\times$10$^{-5}$Wm$^{-2}$])&3.16$\pm$0.161&3.09$\pm$0.373\\
$\alpha$&1.30$\pm$0.0910&1.35$\pm$0.173\\
$q_{\rm PAH}$&($7.84\pm0.270$)$\times$10$^{-2}$&($1.03\pm0.0400$)$\times$10$^{-1}$\\
$f_{+}$&($4.30\pm4.19$)$\times$10$^{-2}$&($2.47\pm0.772$)$\times$10$^{-1}$\\
$M_{\star}$ [$M_{\odot}$]&(2.38$\pm$0.0390)$\times$10$^{11}$&(2.38$\pm$0.0340)$\times$10$^{11}$\\
ln ($\langle U\rangle$ [2.2$\times$10$^{-5}$Wm$^{-2}$])&0.0803$\pm$0.258&-0.154$\pm$0.250\\ \hline
\multicolumn{3}{c}{\bf {HCG\,92c}}\\ \hline
ln ($M_{\rm dust}$ [$M_{\odot}$])&16.5$\pm$0.150&16.6$\pm$0.200\\
ln ($U_{\rm min}$ [2.2$\times$10$^{-5}$Wm$^{-2}$])&-0.330$\pm$0.661&-0.570$\pm$0.240\\
ln ($U_{\rm max}-U_{\rm min}$ [2.2$\times$10$^{-5}$Wm$^{-2}$])&2.30$\pm$0.0140&2.09$\pm$0.176\\
$\alpha$&1.23$\pm$0.126&1.63$\pm$0.319\\
$q_{\rm PAH}$&($4.50\pm0.830$)$\times$10$^{-2}$&($1.30\pm0.600$)$\times$10$^{-1}$\\
$f_{+}$&1.00$\pm$0.101&($7.18^{+36.0}_{-7.18}$)$\times$10$^{-2}$\\
$L^{\rm AGN}_{\star}$ [$L_{\odot}$]&(5.68$\pm$0.805)$\times$10$^{10}$&(4.21$\pm$0.921)$\times$10$^{10}$\\
$L^{\rm AGN}_{\rm IR}$ [$L_{\odot}$]&(4.45$\pm$0.663)$\times$10$^{11}$&(4.46$\pm$1.17)$\times$10$^{11}$\\
$R^{\rm AGN}$ [$\mathit pc$]&($1.00\pm0.278$)$\times$10$^{3}$&($1.19\pm0.284$)$\times$10$^{3}$\\
$V^{\rm AGN}_{c}$ [\%]&62.1$\pm$18.0&$4.64^{+10.8}_{-3.14}$\\
$A^{\rm AGN}_{c}$&$5.53^{+13.2}_{-5.53}$&(1.86$\pm$0.902)$\times$10$^{2}$\\
$A^{\rm AGN}_{d}$&1.00$\times$10$^{3}$&($1.00\pm0.304$)$\times$10$^{3}$\\
$\theta ^{\rm AGN}$ [degrees]&20.8$^{+5.88}_{-1.84}$&27.1$\pm$3.86\\
$M_{\star}$ [$M_{\odot}$]&(1.59$\pm$0.329)$\times$10$^{11}$&(2.65$\pm$0.569)$\times$10$^{11}$\\
$A^{\rm screen}_{V}$&$2.74^{+2.81}_{-2.74}$&8.50$\pm$4.33\\
ln ($\langle U\rangle$ [2.2$\times$10$^{-5}$Wm$^{-2}$])&1.18$\pm$0.228&0.718$\pm$0.282\\ \hline
\multicolumn{3}{c}{\bf {HCG\,92e}}\\ \hline
ln ($M_{\rm dust}$ [$M_{\odot}$])&13.6$\pm$0.530&13.4$\pm$0.450\\
ln ($U$ [2.2$\times$10$^{-5}$Wm$^{-2}$])&-0.202$\pm$0.318&-0.202$\pm$0.275\\
$q_{\rm PAH}$&($1.61\pm0.637$)$\times$10$^{-1}$&($1.86\pm0.579$)$\times$10$^{-1}$\\
$f_{+}$&0&0\\
$M_{\star}$ [$M_{\odot}$]&(1.74$\pm$0.0200)$\times$10$^{11}$&(1.74$\pm$0.0270)$\times$10$^{11}$\\ \hline
\multicolumn{3}{c}{\bf {SQ-A}}\\ \hline
ln ($M_{\rm dust}$ [$M_{\odot}$])&16.5$\pm$0.190&16.5$\pm$0.170\\
ln ($U_{\rm min}$ [2.2$\times$10$^{-5}$Wm$^{-2}$])&-1.99$\pm$0.627&-2.13$\pm$0.558\\
ln ($U_{\rm max}-U_{\rm min}$ [2.2$\times$10$^{-5}$Wm$^{-2}$])&4.01$\pm$0.390&4.08$\pm$0.380\\
$\alpha$&1.68$\pm$0.135&1.74$\pm$0.106\\
$q_{\rm PAH}$&($6.48\pm0.170$)$\times$10$^{-2}$&($9.37\pm0.210$)$\times$10$^{-2}$\\
$f_{+}$&($2.20\pm0.326$)$\times$10$^{-1}$&($3.08\pm0.380$)$\times$10$^{-1}$\\
$M_{\star}$ [$M_{\odot}$]&(1.87$\pm$0.0310)$\times$10$^{10}$&(1.77$\pm$0.0300)$\times$10$^{10}$\\
ln ($\langle U\rangle$ [2.2$\times$10$^{-5}$Wm$^{-2}$])&0.547$\pm$0.198&0.316$\pm$0.177\\ \hline
\multicolumn{3}{c}{\bf {SQ-B}}\\ \hline
ln ($M_{\rm dust}$ [$M_{\odot}$])&15.4$\pm$0.250&15.4$\pm$0.120\\
ln ($U_{\rm min}$ [2.2$\times$10$^{-5}$Wm$^{-2}$])&-4.61$^{+1.23}_{-0}$&-4.61\\
ln ($U_{\rm max}-U_{\rm min}$ [2.2$\times$10$^{-5}$Wm$^{-2}$])&2.30&2.30\\
$\alpha$&1.00$^{+0.0160}_{-0}$&1.08$\pm$0.0470\\
$q_{\rm PAH}$&($1.08\pm0.121$)$\times$10$^{-1}$&($1.26\pm0.140$)$\times$10$^{-1}$\\
$f_{+}$&0$^{+0.0230}_{-0}$&0\\
$M_{\star}$ [$M_{\odot}$]&(4.71$\pm$0.266)$\times$10$^{9}$&(5.12$\pm$0.0240)$\times$10$^{9}$\\
ln ($\langle U\rangle$ [2.2$\times$10$^{-5}$Wm$^{-2}$])&0.370$\pm$0.249&0.177$\pm$0.125\\ \hline
\end{tabular}
\begin{tablenotes}
\item[$\dag$] The values in the second and third columns are the results obtained from the AC composition model \citep{key-Galliano2011} and the THEMIS composition model \citep{key-Jones2013}, respectively. The $L^{\rm AGN}_{\rm IR}$ and $\langle U\rangle$ are calculated from the best fit parameters.
\end{tablenotes}
\end{threeparttable}
\end{adjustbox}
\end{table*}

\section{Star Formation Rates of HCG\,56 and HCG\,92 Galaxies}
We estimate the star formation rates (SFRs) for each member galaxy in HCG\,56 and HCG\,92 based on our results. Two different methods are employed to estimate the SFR: (i) the intensities of the PAH features at 6.2, 7.7, and 11.2\,$\mu$m in the mid-infrared spectra, and (ii) the far-infrared (FIR) luminosity (8--1000\,$\mu$m) based on our dust models. We use the calibration equations provided in \citet{key-Shipley2016}, which allow us to apply the same initial mass function \citep[Kroupa IMF;][]{key-Kroupa2002} for direct comparison. It is important to note that the PAH-based SFR calibration of \citet{key-Shipley2016} is derived by directly relating the measured PAH band (e.g., 6.2, 7.7, and 11.2~$\mu$m) luminosities, obtained with Spitzer/IRS, to independently estimated SFRs based on recombination and fine-structure lines (H$\alpha$, Pa$\alpha$, and [Ne\,\textsc{ii}]+[Ne\,\textsc{iii}]) in a sample of star-forming galaxies. In other words, their PAH--SFR relations are obtained without invoking FIR--SFR calibration as an intermediate step. Therefore, the lower PAH-based SFRs compared to those derived from FIR luminosities in some of our targets do not necessarily imply that the FIR emission contains a substantial non-star-forming (e.g., cirrus-like) component; rather, these differences may reflect environmental effects that reduce PAH emission relative to the ionizing photon rate---such as PAH destruction in harsh radiation fields, suppression in AGN environments, or deviations from the star-forming galaxy conditions assumed in the \citet{key-Shipley2016} calibration. In this context, it should also be noted that the \citet{key-Shipley2016} calibration is derived from samples of star-forming galaxies and is primarily applicable to such systems. Because the PAH emission does not necessarily originate exclusively from massive star-forming regions, but can also arise from PAHs in the diffuse ISM stochastically heated by softer interstellar radiation fields (e.g., \citealt{key-Draine2007,key-Peeters2004}), a non-negligible fraction of the PAH luminosity may persist even when the current ($\lesssim 10$~Myr) SFR is very low. Therefore, applying the \citet{key-Shipley2016} PAH--SFR calibration to quiescent/early-type systems can yield spurious non-zero SFRs; in such cases, we treat PAH-based SFRs as upper limits and exclude them from the following discussion.  In particular, HCG\,92e---likely an early-type galaxy---exhibits a very low SFR and no obvious indicators of ongoing star formation in our dataset. For completeness, the SFR derived from its far-infrared luminosity is reported in Table~\ref{table_SFR}, but it is not included in the following discussion and comparison with other HCG\,92 members.

\subsection{The SFR Derived from PAH Bands}
The PAH emission bands at 6.2, 7.7, and 11.2\,$\mu$m are extracted and  summarized in Table~\ref{PAH_table}. Among these, the 7.7\,$\mu$m PAH integrated luminosity is selected to estimate the SFR, as it is typically the strongest of the observed PAH features. The SFR is calculated using the following equation \citep[see Eq.~(12) in][]{key-Shipley2016}:

\begin{center}
\begin{eqnarray}
  {\rm SFR_{7.7 {\mu}m}} {\rm [M_{\odot}/yr]} = 10^{-42.38} \times (L_{\rm PAH 7.7 {\mu}m \rm[erg/s]}).
\end{eqnarray}
 \end{center}

\begin{table*}[htbp!]
\caption{Measured PAH luminosity.}
\begin{adjustbox}{max width=\textwidth, max totalheight=\textheight, keepaspectratio, center}
\begin{tabular}{cccc} \hline
Source & $\mathit{L}_{6.2}$ & $\mathit{L}_{7.7}$ & $\mathit{L}_{11.2}$\\ \hline \hline
HCG\,56a & $(1.8^{+0.2}_{-0.2})$ $\times$ $10^8\mathit{L}_{\odot}$ & $(0.81^{+0.07}_{-0.04})$ $\times$ $10^9\mathit{L}_{\odot}$ & $(3.5^{+1.6}_{-1.5})$ $\times$ $10^8\mathit{L}_{\odot}$ \\
HCG\,56b & $(0.44^{+0.10}_{-0.06})$ $\times$ $10^8\mathit{L}_{\odot}$ & $(1.05^{+0.31}_{-0.15})$ $\times$ $10^8\mathit{L}_{\odot}$ & $\lesssim$ $3.5$ $\times$ $10^8\mathit{L}_{\odot}$ \\
HCG 56c & $\lesssim$ $2.9$ $\times$ $10^7\mathit{L}_{\odot}$ & $\lesssim$ $3.1$ $\times$ $10^7\mathit{L}_{\odot}$ &  $\lesssim$ $2.3$ $\times$ $10^7\mathit{L}_{\odot}$ \\
HCG\,56d & $(1.3^{+0.0}_{-0.1})$ $\times$ $10^8\mathit{L}_{\odot}$ &  $(4.7^{+0.2}_{-0.3})$ $\times$ $10^8\mathit{L}_{\odot}$ & $(0.81^{+0.17}_{-0.07})$ $\times$ $10^8\mathit{L}_{\odot}$ \\
HCG\,56e & $2.1^{+1.3}_{-1.0}$ $\times$ $10^7\mathit{L}_{\odot}$ & $(1.3^{+0.1}_{-0.1})$ $\times$ $10^8\mathit{L}_{\odot}$ & $(3.5^{+1.0}_{-1.3)}$ $\times$ $10^7\mathit{L}_{\odot}$ \\
HCG\,92a & $(0.96^{+0.06}_{-0.05})$ $\times$ $10^7\mathit{L}_{\odot}$ & $(1.8^{+0.2}_{-0.1})$ $\times$ $10^7\mathit{L}_{\odot}$ & $(1.8^{+0.2}_{-0.1})$ $\times$ $10^7\mathit{L}_{\odot}$ \\
HCG\,92c & $\lesssim$ $2.1$ $\times$ $10^8\mathit{L}_{\odot}$ & $(1.01^{+0.74}_{-0.2})$ $\times$ $10^8\mathit{L}_{\odot}$ & $\lesssim$ $1.2$ $\times$ $10^8\mathit{L}_{\odot}$ \\
SQ-B & $0.58^{+0.19}_{-0.17}$ $\times$ $10^7\mathit{L}_{\odot}$ & $(2.9^{+0.5}_{-0.6})$ $\times$ $10^7\mathit{L}_{\odot}$ & $\lesssim$ $2.7$ $\times$ $10^7\mathit{L}_{\odot}$ \\ \hline
\end{tabular}
\end{adjustbox}
\label{PAH_table}
\end{table*}

\subsection{The SFR Derived from the FIR Luminosity}
SED models are constructed for all galaxies with available photometric data. We fit two different dust models---the AC model \citep{key-Galliano2011} and the THEMIS model \citep{key-Jones2013}---to investigate how dust properties affect the results. For galaxies hosting active galactic nuclei (AGN), specifically HCG\,56b and HCG\,92c, we incorporated AGN templates from \citet{key-Siebenmorgen2015} into the dust models using a linear combination. We then integrated the luminosity from 8 to 1000\,$\mu$m ($L_{\rm 8-1000\,\mu m}$) for each SED model and converted it into the SFR using the following equation \citep[see Eq.~(20) in][]{key-Shipley2016}. For AGN-hosting galaxies, the SFR is calculated using only the starburst component, explicitly excluding the contribution from the AGN component.

\begin{center}
\begin{eqnarray}
  {\rm SFR_{FIR}} {\rm [M_{\odot}/yr]} =  2.54 \times10^{-44} L_{8-1000 {\rm {\mu}m}} \rm[erg/s].
\end{eqnarray}
\end{center}

\subsection{Comparison of the SFRs}
Table~\ref{table_SFR} summarizes the SFRs derived from the PAH 7.7\,$\mu$m feature, from the total infrared luminosities based on the AC and THEMIS dust models, and from the MAGPHYS analyses in the literature. In this section, we first compare these SFR indicators internally and then discuss how our results relate to previous studies.

\begin{figure*}[htbp!]
\begin{adjustbox}{max width=\textwidth, max totalheight=\textheight, keepaspectratio, center}
\includegraphics{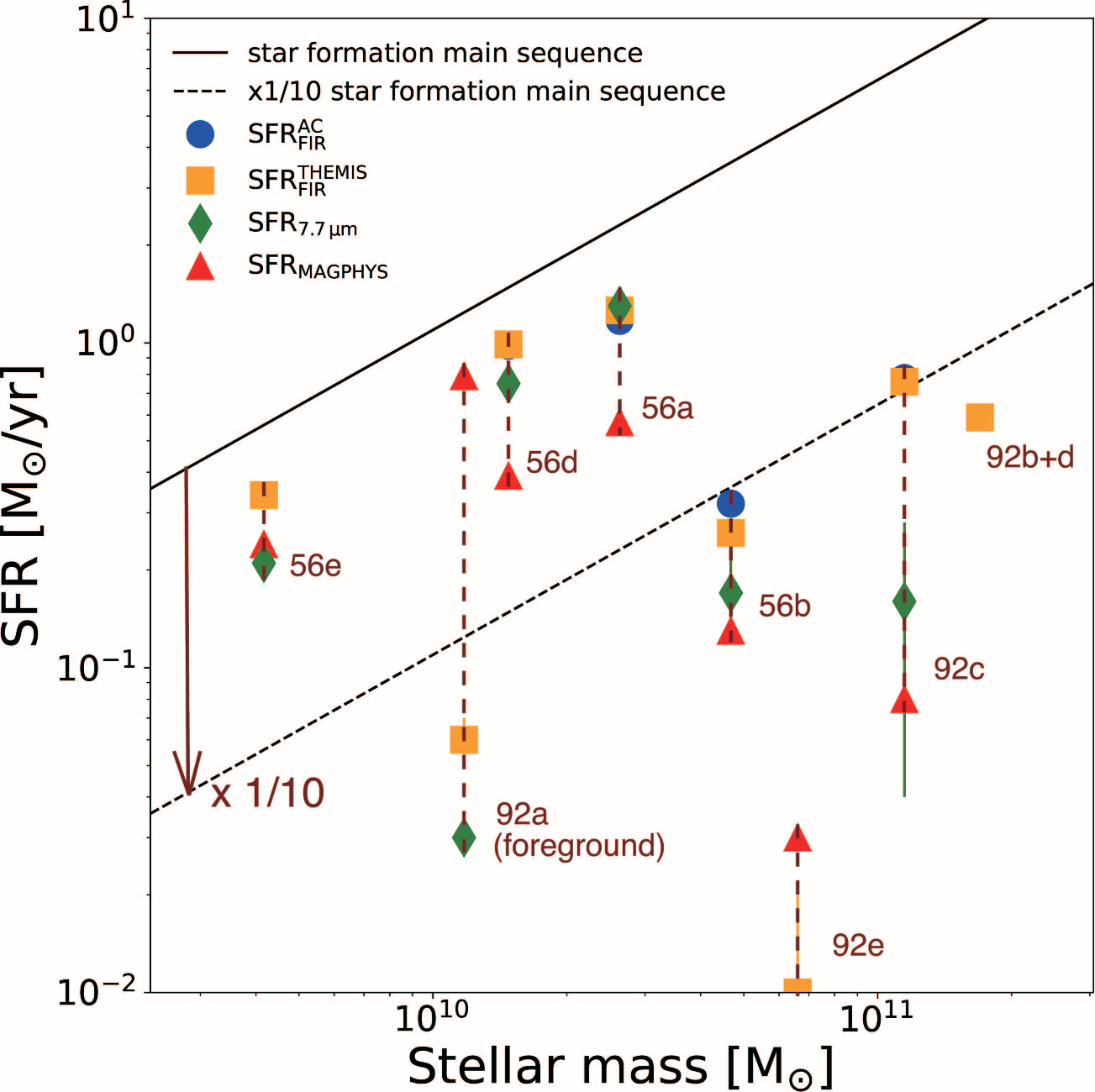}
\end{adjustbox}
\captionof{figure}{Star formation rate (SFR) versus stellar mass (M$_\star$) relation. The solid and dashed lines indicate the star-forming main sequence and its tenth-percentile boundary at $z \sim 0.1$, derived from SDSS galaxies \citep{key-Brinchmann2004}. The blue circles represent SFRs derived from the total infrared luminosity based on the AC dust model \citep{key-Galliano2011}, the orange squares from the THEMIS dust model \citep{key-Jones2013}, the green diamonds from the PAH 7.7\,$\mu$m band luminosity \citep{key-Shipley2016}, and the red triangles from MAGPHYS \citep{key-Bitsakis2011,key-Bitsakis2014}. The stellar masses are taken from \citet{key-Bitsakis2011,key-Bitsakis2014}. The labels indicate the individual HCG\,56/92 member galaxies studied in this work. {Alt text: A plot of star formation rate versus stellar mass. The horizontal axis is stellar mass and the vertical axis is star formation rate. A solid line marks the star forming main sequence. Multiple markers per galaxy compare different star formation estimates.}}
\label{SFR_mainsequence_fig.1}
\end{figure*}

Our results indicate that the SFR derived from the 7.7\,$\mu$m PAH feature is generally lower than that obtained using the AC and THEMIS dust models. This discrepancy may arise from the possible destruction of PAHs in harsh environments, such as inside H\,\textsc{ii} regions \citep{key-Peeters2004} and/or in the vicinity of weak AGNs \citep{key-Smith2007}, resulting in an underestimation of the SFR when inferred from the PAH band strengths compared to the FIR luminosity. We also compare the derived SFRs with the star formation main sequence (SFMS) of nearby galaxies \citep[$z \sim 0.1$;][]{key-Brinchmann2004}. We determine the normalization of the SFMS by combining the observed cosmic star formation rate density (SFRD) with the local galaxy stellar mass function. We adopt the SFRD
\[
\mathrm{SFRD} = 1.915 \times 10^{-2} \ M_\odot \ \mathrm{yr}^{-1} \ \mathrm{Mpc}^{-3}
\]
from \citet{key-Brinchmann2004}, assuming $h_{70} = 1$. The slope of the SFMS is fixed to $a = 0.77$, following \citet{key-Elbaz2007}. For the stellar mass distribution, we use the local ($z \approx 0$) double Schechter function from \citet{key-Baldry2012}, converted to $h = 0.7$:
\begin{equation}
\begin{split}
\phi(M) &= \ln 10 \ e^{-10^{M-M^\star}} \\
&\times \left[ \phi_1^\star 10^{(M-M^\star)(\alpha_1+1)}
+ \phi_2^\star 10^{(M-M^\star)(\alpha_2+1)} \right],
\end{split}
\end{equation}
where $M = \log_{10}(M_\star/M_\odot)$, $M^\star = 10.66$, $\alpha_1 = -0.35$, $\alpha_2 = -1.47$, $\phi_1^\star = 3.96\times 10^{-3} \ \mathrm{Mpc^{-3}}$, and $\phi_2^\star = 0.79\times 10^{-3} \ \mathrm{Mpc^{-3}}$. We assume the SFMS relation
\begin{equation}
\mathrm{SFR}(M_\star) = 10^{a \log_{10}(M_\star/M_\odot) + b},
\end{equation}
and determine the intercept $b$ by solving
\begin{equation}
\int_{M_{\mathrm{min}}}^{M_{\mathrm{max}}} \mathrm{SFR}(M_\star)\,\phi(M_\star)\,dM_\star = \mathrm{SFRD},
\end{equation}
with $M_{\mathrm{min}} = 10^8 \ M_\odot$ and $M_{\mathrm{max}} = 10^{12} \ M_\odot$. This yields $b \approx -7.662$, and thus the SFMS is expressed as
\begin{equation}
\log_{10} \mathrm{SFR} \ [M_\odot \ \mathrm{yr}^{-1}] \approx 0.77 \ \log_{10}(M_\star / M_\odot) - 7.66.
\end{equation}
As shown in Fig.~\ref{SFR_mainsequence_fig.1}, the galaxies in our sample are generally located below the SFMS, indicating suppressed star formation compared to typical field galaxies.

We next compare our results with the MAGPHYS-based studies of \citet{key-Bitsakis2011,key-Bitsakis2014}, which derived stellar masses and SFRs for HCG galaxies from broad-band photometry. Their main conclusion---that most HCG galaxies lie below the local SFMS---is broadly consistent with our results. The main difference concerns AGN-host galaxies, especially HCG\,56b. In \citet{key-Bitsakis2011}, HCG\,56b appeared as an extreme outlier with a very high SFR, whereas in \citet{key-Bitsakis2014} this value was revised downward after Herschel/PACS and SPIRE data were incorporated in their data. This trend is consistent with the interpretation that improved FIR coverage leads to a better constraint on the cold dust emission and total infrared luminosity. In addition, our spectroscopy-constrained analysis further reduces the ambiguity in the mid-infrared decomposition, because the PAH band strengths and the continuum shape are directly constrained by the AKARI/IRC spectra rather than inferred solely from broad-band photometry.  Comparison for each galaxy is summarized in Table~\ref{table_prevcomp}.

In this context, our results for HCG\,56b and HCG\,92c suggest that the most important improvement in the present study is not a redefinition of the stellar masses, but a more robust estimate of the star-forming contribution to the infrared output in AGN host galaxies. For such galaxies, SFRs based on a single infrared flux point or broad-band mid-infrared photometry can be overestimated when AGN-heated dust contributes significantly to the continuum. The PAH-based SFRs and the literature MAGPHYS values are therefore useful as relatively conservative star-formation tracers, while the FIR-based estimates obtained with the AC and THEMIS dust models provide an upper limit when some AGN-related degeneracy may still remain. We note that HCG\,92a shows a larger offset between the PAH-based SFR and the other estimates than the other star-forming galaxies. Since HCG\,92a is a foreground galaxy unrelated to the Stephan\textquoteright s Quintet interaction, this discrepancy is more likely attributable to observational systematics rather than to interaction-driven dust heating. In particular, the spectrum of HCG\,92a was obtained with slit spectroscopy and scaled to match the imaging photometric points, implying a modest but non-negligible aperture mismatch. Because PAH emission can be spatially extended, some fraction of it may fall outside the slit, biasing the PAH-based SFR low compared to the FIR- and SED-based estimates.

For HCG\,92, it is also important to compare our results with the detailed infrared study of Stephan\textquoteright s Quintet by \citet{key-Natale2010}. Their analysis was organized by physically distinct components of the system, such as NGC\,7319, SQ-A, SQ-B, the shock region, and extended emission, whereas our study is based primarily on galaxy-centered apertures, supplemented by the star-forming regions SQ-A and SQ-B. While the comparison is  informative, it is not strictly one-to-one. Nevertheless, it provides an important bridge between the earlier Spitzer-based view of the dust-emitting structures in Stephan\textquoteright s Quintet and the present spectroscopy-constrained galaxy-scale SED analysis. In particular, SQ-A, SQ-B, and NGC\,7319 (HCG\,92c) allow a qualitative comparison of the derived star-forming activity, while differences for the total Stephan\textquoteright s Quintet system or for the shock-related components should be interpreted with caution because of the different object definitions and aperture choices. For clarity, we summarize the comparison with previous works, including \citet{key-Bitsakis2011,key-Bitsakis2014} and \citet{key-Natale2010}, in Table~\ref{table_prevcomp}. \citet{key-Natale2010} is particularly valuable for characterizing the large-scale dust-emitting structures of Stephan\textquoteright s Quintet, including the shock-related and extended components, whereas our study is optimized for galaxy-centered apertures and AGN/starburst decomposition. 

Taken together, these comparisons indicate that the main advantage of the present work lies in the combination of improved FIR coverage and direct spectroscopic constraints on the PAH strengths and mid-infrared continuum shape. This combination is especially important for AGN-host galaxies, where broad-band SED fitting alone can suffer from degeneracies between star-forming and AGN-heated dust components. In this sense, our results are best regarded as a refinement of previous infrared studies of HCG\,56 and HCG\,92: these preserve the overall picture that most member galaxies lie below the local SFMS, while providing a more robust estimate of the star-forming contribution to the infrared emission, particularly in systems hosting an AGN.

\begin{table*}[htbp!]
\caption{Star formation rates derived from different methods}
\label{table_SFR}
\begin{adjustbox}{max width=\textwidth, max totalheight=\textheight, keepaspectratio, center}
\begin{threeparttable}
\begin{tabular}{ccccccc}
\hline
&SFR$_{\rm{7.7{\mu}m}}$ [M$_{\odot}$/yr]\tnote{a}
&SFR$_{\rm{FIR}}^{\rm{AC}}$ [M$_{\odot}$/yr]\tnote{b}
&SFR$_{\rm{FIR}}^{\rm{THEMIS}}$ [M$_{\odot}$/yr]\tnote{b}
&SFR$_{\rm{MAGPHYS}}$ [M$_{\odot}$/yr]\tnote{c}
&$\log M_{\star}$ [M$_{\odot}$]\tnote{d}
&starburst component [\%]\tnote{e} \\
\hline\hline
HCG\,56a & 1.3$\pm$0.1 & 1.17$\pm$0.06 & 1.26$\pm$0.06 & 0.57 & 10.42 & 100 \\
HCG\,56b & 0.17$^{+0.05}_{-0.03}$ & 0.32$\pm$0.02 & 0.26$\pm$0.01 & 0.13 & 10.67 & 13.1 (AC) \& 10.5 (THEMIS) \\
HCG\,56d & 0.75$^{+0.03}_{-0.05}$ & 0.98$\pm$0.05 & 0.99$\pm$0.05 & 0.39 & 10.17 & 100 \\
HCG\,56e & 0.21$^{+0.01}_{-0.02}$ & 0.34$\pm$0.02 & 0.34$\pm$0.02 & 0.24 & 9.62 & 100 \\
HCG\,92a & 0.029$^{+0.003}_{-0.002}$ & 0.06$\pm$0.01 & 0.06$\pm$0.01 & 0.79 & 10.07 & 100 \\
HCG\,92b+d & -- & 0.59$\pm$0.03 & 0.59$\pm$0.03 & -- & 11.23 & 100 \\
HCG\,92c & 0.16$^{+0.08}_{-0.03}$ & 0.78$\pm$0.04 & 0.76$\pm$0.04 & 0.08 & 11.06 & 42.3 (AC) \& 39.2 (THEMIS) \\
HCG\,92e & -- & 0.01$\pm$0.01 & 0.01$\pm$0.01 & 0.03 & 10.82 & 100 \\
SQ-A & -- & 0.46$\pm$0.02 & 0.46$\pm$0.02 & -- & -- & 100 \\
SQ-B & 0.046$^{+0.008}_{-0.009}$ & 0.13$\pm$0.01 & 0.13$\pm$0.01 & -- & -- & 100 \\
\hline
\end{tabular}
\begin{tablenotes}
\item[(a)] Star formation rate derived from PAH 7.7 $\mu$m band luminosity \citep{key-Shipley2016}.
\item[(b)] Star formation rate calculated from 8--1000 $\mu$m luminosity \citep{key-Shipley2016}.
\item[(c)] Star formation rate derived from MAGPHYS fit \citep{key-Bitsakis2011,key-Bitsakis2014}.
\item[(d)] Stellar mass derived from MAGPHYS. The values are retrieved from \citet{key-Bitsakis2011,key-Bitsakis2014}.
\item[(e)] Calculated by $L_{\rm SB}$/$L_{\rm SB+AGN}$. If the SED model does not take into account the effect of AGN, the value becomes 100.
\end{tablenotes}
\end{threeparttable}
\end{adjustbox}
\end{table*}

\begin{table*}
\begin{adjustbox}{max width=\textwidth, max totalheight=\textheight, keepaspectratio, center}
\begin{threeparttable}
\caption{Comparison of the present work with previous infrared studies}
\label{table_prevcomp}
\centering
\footnotesize
\begin{tabular}{lccccc>{\raggedright\arraybackslash}p{6.2cm}}
\hline
This work &
$\log M_{\star}^{\rm lit}$ &
SFR$_{\rm prev}$ &
SFR$_{\rm this}$ &
$\log {\rm sSFR}_{\rm prev}$ &
$\log {\rm sSFR}_{\rm this}$ &
Remarks \\
 &
$[M_\odot]$ &
$[M_\odot\,{\rm yr}^{-1}]$ &
$[M_\odot\,{\rm yr}^{-1}]$ &
$[{\rm yr}^{-1}]$ &
$[{\rm yr}^{-1}]$ &
 \\
\hline
\multicolumn{7}{@{}l}{\emph{(A) Galaxy-based comparison with broad-band MAGPHYS studies of Bitsakis et al.\ (2011, 2014)}} \\
\hline
HCG 56a & 10.42 & 0.57 & 1.17 & -10.66 & -10.35 & Broad agreement; our spectroscopy-constrained fit gives a moderately higher SFR. \\
HCG 56b & 10.67 & 0.13 & 0.32 & -11.56 & -11.16 & AGN host; MIR spectroscopy helps separate AGN-heated continuum from the star-forming component. \\
HCG 56d & 10.17 & 0.39 & 0.98 & -10.58 & -10.18 & Star-forming-dominated, though a weak nuclear contribution cannot be excluded; the present SFR is higher than the MAGPHYS value, but both studies indicate non-quiescent activity. \\
HCG 56e & 9.62 & 0.24 & 0.34 & -10.24 & -10.09 & Reasonable agreement within methodology/systematics. \\
HCG 92a & 10.07 & 0.79 & 0.06 & -10.17 & -11.29 & Foreground galaxy; likely affected by aperture/systematic differences between spectroscopy and photometry. \\
HCG 92c & 11.06 & 0.08 & 0.78 & -12.16 & -11.17 & Seyfert 2 host galaxy; FIR-based SFR remains higher than MAGPHYS, while PAH/MAGPHYS provide conservative estimates. \\
HCG 92e & 10.82 & 0.03 & 0.01 & -12.34 & -12.82 & Very low-SFR system; both works indicate quiescent or weakly star-forming nature. \\
\hline
\multicolumn{7}{@{}l}{\emph{(B) Component-based comparison with the Stephan's Quintet study of Natale et al.\ (2010)}} \\
\hline
HCG 92c & 11.20 & 0.22 & 0.78 & -11.86 & -11.31 & Closest one-to-one correspondence (NGC 7319 in Natale et al.\ 2010). Their lower SFR partly reflects a different methodology and source definition; Natale et al.\ argued that much of the FIR emission in NGC 7319 may be AGN-powered rather than star-forming. \\
SQ-A & -- & 0.78 & 0.46 & -- & -- & Directly comparable extranuclear star-forming region. Natale et al.\ treated SQ-A as one of the most active star-forming regions in the system. \\
SQ-B & -- & 0.28 & 0.13 & -- & -- & Directly comparable extranuclear star-forming region in the tidal debris. \\
\hline
\end{tabular}

\begin{tablenotes}
\item The column SFR$_{\rm this}$ lists the representative star-formation rate adopted from the present work, taken to be the FIR-based SFR derived with the AC dust model.
\item The specific star-formation rates are calculated as ${\rm sSFR}={\rm SFR}/M_\star$.
\item The comparison with Natale et al.\ (2010) is informative but not strictly one-to-one, because their study decomposed Stephan's Quintet into physically motivated large-scale components, whereas the present work is based primarily on galaxy-centered apertures plus SQ-A and SQ-B.
\item For the Natale et al.\ (2010) comparison, stellar masses are available for NGC\,7319 but not for SQ-A or SQ-B; accordingly, the sSFR values are given only where a published stellar mass is available.
\end{tablenotes}
\end{threeparttable}
\end{adjustbox}
\end{table*}

\section{Summary}
We analyze the near- to far-infrared properties of member galaxies in Hickson Compact Groups (HCGs) 56 and 92. These compact groups are characterized by strong gravitational interactions, providing an ideal environment to study the effects of galaxy interactions on star formation, active galactic nuclei (AGN), and overall evolutionary processes. AKARI/IRC spectroscopic observations (primarily slitless, supplemented by an additional slit pointing for HCG\,92a) are conducted for the member galaxies of HCG\,56 and HCG\,92, enabling examination of their spectral properties across the 2--14\,$\mu$m range.

Clear PAH bands are detected in HCG\,56a and HCG\,92a, which are consistent with normal and/or star-forming galaxies. HCG\,56d also shows strong PAH emission indicative of substantial star formation. Taken at face value, its mid-infrared spectrum is more naturally interpreted as star-forming-dominated than as AGN-dominated, although some uncertainty remains because a weak nuclear contribution cannot be completely excluded. HCG\,56e likewise exhibits significant star-formation-related PAH emission, but its moderately red mid-infrared continuum suggests that it is more safely interpreted as a composite system than as a purely normal star-forming galaxy. In contrast, the near- and mid-infrared spectrum of HCG\,56b shows strong continuum emission with no discernible PAH bands or silicate absorption, indicating the presence of an AGN---a characteristic commonly associated with Seyfert~1 galaxies. HCG\,92c, on the other hand, exhibits strong continuum emission along with the 9.7\,$\mu$m silicate absorption feature, consistent with its classification as a Seyfert~2 galaxy.

Our analysis shows that the star formation rates (SFRs) of the member galaxies in HCG\,56 and HCG\,92 do not lie above the local star formation main sequence. Therefore, we find no evidence in the present analysis for interaction-induced enhancement of star formation in these systems. Instead, most members lie below the local SFMS, indicating that star formation is globally suppressed relative to typical field galaxies.

In particular, the spectroscopy provides (i) PAH-based SFRs that trace star formation more directly and (ii) direct constraints on the mid-infrared continuum shape, which helps identify AGN-heated dust and reduces degeneracies in the IR decomposition. Using these spectroscopy-constrained diagnostics, we find that most members of HCG\,56 and HCG\,92 lie below the local SFMS, with the clearest suppression in the AGN hosts HCG\,56b and HCG\,92c (Fig.~\ref{SFR_mainsequence_fig.1}). This framework also clarifies the differences with previous MAGPHYS-based studies (e.g., \citealt{key-Bitsakis2011,key-Bitsakis2014}). For HCG\,56b, our fits yield a substantially lower SFR than \citet{key-Bitsakis2011}, which we attribute to the fact that, without spectroscopic constraints, a portion of the AGN-heated IR continuum can be absorbed into the star-forming dust component. The revised values in \citet{key-Bitsakis2014}, based on improved FIR coverage, are consequently closer to our results. Overall, the MAGPHYS-based SFRs in the literature are broadly consistent with our PAH-based estimates for most galaxies, whereas the largest discrepancy is between these conservative SF tracers (PAH/MAGPHYS) and FIR-luminosity-based SFRs (including our AC/THEMIS fits) for AGN hosts, where FIR-driven fits can still return somewhat higher SFRs due to remaining AGN--SF degeneracies and/or aperture/photometry differences.

Our AKARI/IRC spectroscopic data, obtained primarily in slitless mode and supplemented by an additional slit pointing for HCG\,92a, provide new spectra for HCG\,56a, HCG\,92a, and HCG\,92c. These datasets enable more accurate estimation of SFRs and deeper investigation of AGN activity within the member galaxies. Future analyses of additional AKARI/IRC slitless spectroscopic datasets from other HCGs (e.g., HCG\,13, 31, 62, and 79) will allow us to assess whether galaxies in interacting systems exhibit distinct interstellar medium (ISM) properties and evolutionary pathways compared to field galaxies. This study provides the first spectroscopically constrained infrared SED analysis of HCG\,56 and HCG\,92 incorporating both PAH- and FIR-based SFR diagnostics and explicit AGN--starburst decomposition. By demonstrating that all member galaxies---including AGN hosts---lie below the local SFMS, we present quantitative evidence for widespread star formation suppression in these compact group environments.

\begin{ack}
This research is based on observations with AKARI, a JAXA project with the participation of ESA. This work is based in part on observations made with the Spitzer Space Telescope, which was operated by the Jet Propulsion Laboratory, California Institute of Technology under a contract with NASA. This work is based in part on observations made with Herschel, an ESA space observatory with science instruments provided by European-led Principal Investigator consortia and with important participation from NASA. We thank Takuma Yoshida for his contribution at the initial stage of this research. We thank the reviewer Philip N. Appleton for his careful reading of the manuscript and for constructive comments that helped improve the clarity and scope of this paper. TO acknowledges support from JSPS and CNRS under the Japan--France Research Cooperative Program, JSPS Bilateral Program Grant No. 120219939, and JSPS KAKENHI Grant-in-Aid for Scientific Research (C) Grant No. JP24K07087. IS acknowledges support from JSPS KAKENHI Grant-in-Aid for Scientific Research (B) Grant Nos. JP22H01261, JP23K22532, and JP26K00738, JSPS KAKENHI Grant-in-Aid for Scientific Research (C) Grant No. JP20K04028, and NINS Astrobiology Center Program Research Grant No. AB0813. FG acknowledges support from the French National Research Agency under the contracts WIDENING (ANR-23-ESDIR-0004) and REDEEMING (ANR-24-CE31-2530), as well as from the Actions Th{\'e}matiques ``Physique et Chimie du Milieu Interstellaire'' (PCMI) of CNRS/INSU, with INC and INP, and ``Cosmologie et Galaxies'' (ATCG) of CNRS/INSU, with INP and IN2P3, both programs being co-funded by CEA and CNES.
\end{ack}

\begin{appendix}
\section{Spectral Decomposition of the HCG\,56 Galaxies}
\begin{figure}[htbp!]
\begin{center}
\includegraphics[width=\linewidth]{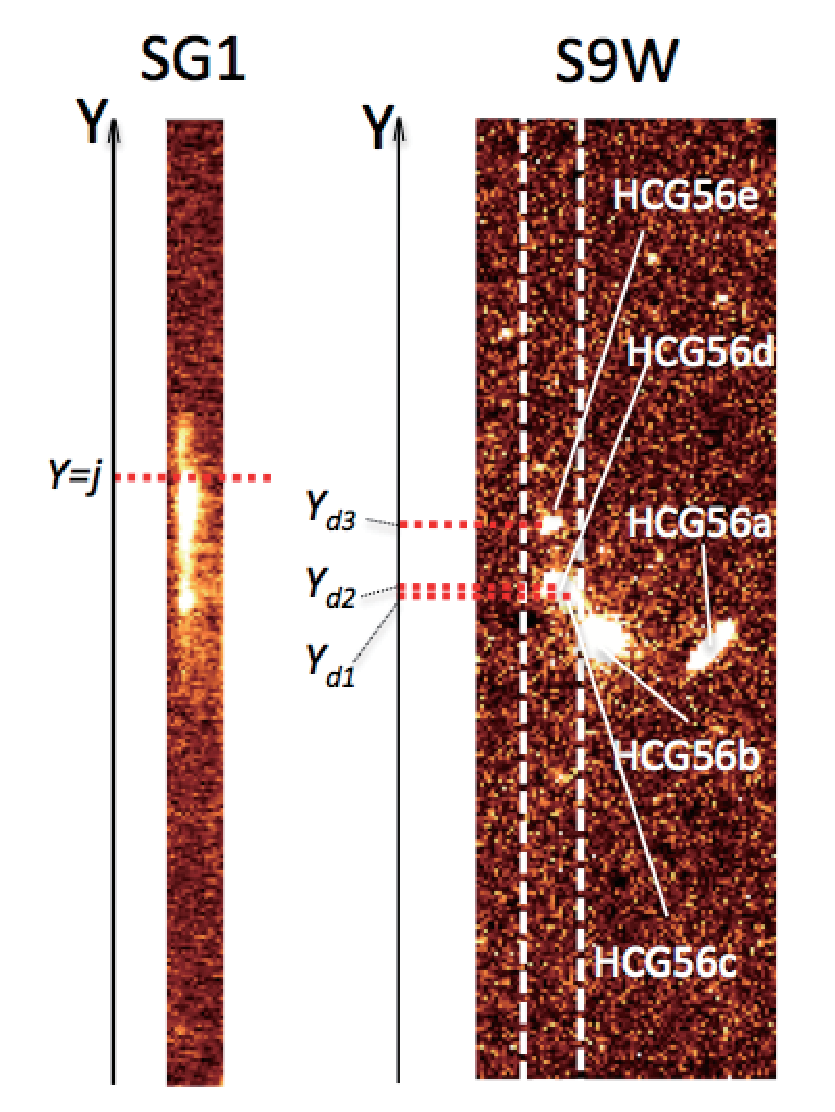}
\end{center}
\captionof{figure}{Data reduction method for slitless spectroscopy of HCG\,56 galaxies.\\(left): AKARI/IRC slitless spectroscopic image of HCG\,56c, HCG\,56d, and HCG\,56e.\\ (right): Reference image of HCG\,56 taken with the S9W (9~$\mu$m) band during the same pointed observation. The direct light positions, $Y_{d_i} (i=1, 2, 3)$, for HCG\,56c, HCG\,56d and HCG\,56e, respectively, are shown in the image. {Alt text: Two panel illustration of slitless spectroscopy reduction for HCG 56. The left panel shows a slitless spectroscopic image containing dispersed spectra from nearby sources. The right panel shows a direct reference image with marked source positions used for wavelength reference.}}
\label{decomposition_fig.1}
\end{figure}

For the data reduction of slitless spectroscopy it is necessary to first determine the wavelength reference position.  This can be done using the position of the object in the reference image (Figure~\ref{decomposition_fig.1} left).  Also, because of the nature of slitless, objects located close in the direction perpendicular to the dispersion have a chance to overlap although they are separated in the cross-dispersion direction.  For the case of HCG\,56c, 56d, and 56e, their spectra overlap in the present AKARI observations (see Figure~\ref{decomposition_fig.1} right).  Therefore, we need to separate each contribution appropriately. The relation between the pixel position ($Y$) and the wavelength ($\lambda$), expressed as $\lambda = \Lambda_{\mathrm{SG1/SG2}}(Y - Y_{d_i})$, is determined for HCG\,56c, HCG\,56d, and HCG\,56e based on the position of each galaxy, $Y_{d_i}$ ($i=1,2,3$), in the S9W image. Because the mid-infrared emission from HCG\,56c, HCG\,56d, and HCG\,56e is unresolved in our S9W data, we model the signal count profile along the $Y = j$ cross-dispersion cut, $f^{Y=j}_{\mathrm{SG1/SG2}}(X)$, as a linear combination of three components:  
$l^{Y=j}_{\mathrm{SG1/SG2}}(X)|_{\mathrm{HCG\,56c}}$,  
$l^{Y=j}_{\mathrm{SG1/SG2}}(X)|_{\mathrm{HCG\,56d}}$, and  
$l^{Y=j}_{\mathrm{SG1/SG2}}(X)|_{\mathrm{HCG\,56e}}$.
Each component corresponds to the spectral signal profile of HCG\,56c, HCG\,56d, and HCG\,56e, respectively, along $Y = j$. These profiles are modeled as those of point sources in the cross-dispersion direction at $\lambda = \Lambda_{\mathrm{SG1/SG2}}(j - Y_{d_i})$. From preliminary analyses, we find that the spectrum profile of a point source observed with AKARI/IRC/MIR-S SG1 and SG2 along the cross-dispersion direction at wavelength $\lambda$ is well represented by a summation of two Lorentzian functions. The functional form and parameters are described in \citet{key-Sakon2008} and adopted in the present analysis. In their calibration, the typical uncertainties are $\pm0.2$\,pixel in the peak positions of the 0-th order light, $\pm0.13$--$0.32$\% in its relative intensity to the 1st order light, and $\pm0.5$\,pixel in the aperture mask width, with slit efficiencies determined within $\pm0.06$--$0.09$\%. It is important to note that the dispersion characteristics of AKARI/IRC/NIR NP cause the spectrum profile of a point source to vary with its position on the detector array in the cross-dispersion direction. It is difficult to separate overlapping spectra unambiguously.  Therefore, the near-infrared spectra of HCG\,56c, HCG\,56d, and HCG\,56e are not provided in this paper.

\section{Infrared SED Models}
\subsection{The Dust Composition}
The SEDs are fitted by two different dust models to test the robustness of the results with respect to the adopted grain properties. In both models, the dust grains are assumed to be heated by the interstellar radiation field of \citet{key-Mathis1983}, scaled by a factor $\mathit{U}$. The value of $\mathit{U} = 1$ corresponds to the radiation field in the solar neighborhood (within $\leq$10\,kpc), with an integrated flux of $2.2 \times 10^{-5}$\,W\,m$^{-2}$ between 0.0912\,$\mu$m and 8\,$\mu$m. Note that $\mathit{U}$ is defined as a multiplicative scaling of the \citet{key-Mathis1983} ISRF; therefore, varying $\mathit{U}$ changes only the intensity, not the spectral shape of the input radiation field.

\subsubsection{The AC (Amorphous Carbon) Composition \citep{key-Galliano2011}}
This model was originally developed to fit Herschel observations of the Large Magellanic Cloud (LMC). Although the LMC has a lower metallicity than the Milky Way, the optical properties adopted here are constrained by dust emission features and are not directly dependent on the overall metallicity. Therefore, the model can be applied to systems with different metallicities as long as the dust composition and size distribution are consistent with the observed infrared spectra. It is essentially based on the BARE-GR-S model of \citet{key-Zubko2004}, in which the graphite component of the standard model is replaced by ACAR\footnote{\citet{key-Zubko1996} defined the ACAR sample as the result of arc discharge between amorphous carbon electrodes in an argon atmosphere at 10 mbar.} amorphous carbon grains \citep{key-Zubko1996}, resulting in a bare silicate + amorphous carbon (ACR) + PAH mixture. This dust mixture is more emissive than the standard silicate and graphite dust model \citep{key-Draine2007} by a factor of approximately 2--3 in the submillimeter range. This enhancement in emissivity has also been observed in the Milky Way \citep{key-Planck2016}. In the Milky Way, the PAH mass fraction is approximately 4.6\%, denoted by $q_{\rm PAH}$.

\subsubsection{The THEMIS (The Heterogeneous dust Evolution Model at the Institut d’Astrophysique Spatiale) Composition \citep{key-Jones2013}}
This model was originally developed to reproduce Planck observations of the Milky Way. It has two key features: (i) aromatic material coats both silicate and carbon grains, and (ii) it does not include PAHs as an independent component explicitly. Instead, the model incorporates small amorphous carbon particles, a-C(:H), which contain both aromatic and aliphatic bonds. The submillimeter opacity of this model is comparable to that of the AC model. For simplicity, we adopt the notation $q_{\rm PAH}$, as defined by \citet{key-Galliano2011}, to represent the mass fraction of a-C(:H) grains with radii smaller than 1.5\,nm. These small grains are the primary contributors to the PAH features in this model. Although larger grains also contain aromatic structures, they are not hot enough to emit these features. In the Milky Way, the fraction of such small a-C(:H) grains is 18.5\%. The mass fraction of small a-C(:H) is considerably higher than that of PAHs because a-C(:H) grains have fewer aromatic C-H bonds per hydrogen atom compared to typical PAH molecules ($q_{\rm PAH}^{\rm AC} \simeq 0.45 \times q_{\rm PAH}^{\rm THEMIS}$; \citealt{key-Galliano2021}).

\subsection{The AGN Template}
Some of the galaxies in our sample exhibit strong mid- to far-infrared emission, indicating the presence of AGNs. Therefore, we incorporate the AGN templates developed by \citet{key-Siebenmorgen2015}. The model assumes that photons are emitted within a clumpy medium and/or a homogeneous disk. Due to the dense environment near the AGN, fluffy grains \citep{key-Krugel1994} are adopted as the dust population. These grains exhibit higher submillimeter emissivity compared to those found in the diffuse interstellar medium (ISM). The templates are computed using a fully self-consistent three-dimensional radiative transfer code that treats the interaction between dust and radiation. The free parameters in the model include the inner radius $R$, the clump volume filling factor $V_c$, the clump $V$-band optical depth $A_c$, the disk $V$-band optical depth $A_d$, and the viewing angle $\theta$. In addition, we vary the luminosity of the central source, $L_\star$, which differs from the infrared luminosity of the dusty torus due to a significant escape fraction of the radiation.

\subsection{The SED Fitting}
To account for the diversity of physical conditions within the galaxies, we fit the SEDs using a range of radiation field intensities. Following \citet{key-Dale2001}, we assumed that the dust is exposed to a distribution of radiation field strength described by
\begin{equation}
{\rm d}M_{\rm dust} \propto U^{-\alpha} {\rm d}U \quad \text{between } U_{\rm min} \text{ and } U_{\rm max}.
\label{Mdust_eq.1}
\end{equation}
The parameters $U_{\rm min}$, $U_{\rm max}$, and $\alpha$ are treated as free parameters of the ISM topology and stellar distribution within the galaxies.
To model the near-infrared emission, we added an old stellar population template (P\'EGASE; \citealt{key-Fioc1997}) to the SED. This component is scaled by varying the total stellar mass $M_\star$. The model SED is expressed as
\begin{equation}
L_\nu^{\mathrm{model}} = M_\star \, l_\nu^{\mathrm{star}} \;+\; M_{\mathrm{dust}} \, l_\nu^{\mathrm{dust}}(U_{\min}, \ldots),
\end{equation}
where $M_\star$ is the mass of the old stellar population, scaling the stellar continuum template $l_\nu^{\mathrm{star}}$. Note that the stellar light component is introduced to subtract its contribution in the near-infrared. Therefore, the derived $M_\star$ should not be interpreted as an accurate stellar mass estimate.

We search the best-fit parameters by minimizing $\chi^2$ using the Levenberg--Marquardt algorithm \citep{key-Markwardt2009}, and we performed Monte Carlo propagation of uncertainties for all parameters. This strategy is applied for all sources except HCG\,56b, c, d, and e. For sources HCG\,56b, 56c+d, and 56e, we adopt the following strategy: we simultaneously fit the SEDs of the three sets of observations up to the PACS wavebands, and fit the sum of the three modeled SEDs to the SPIRE bands of the combined three sources. Namely, the $\chi^2$ to be minimized is defined as:
\begin{eqnarray}
  \chi^2 & = &
    \chi^2_{\le160{\mu} m}(\mbox{56b}) + \chi^2_{\le160{\mu} m}(\mbox{56c+d})
    + \chi^2_{\le160{\mu} m}(\mbox{56e}) \nonumber\\
    & &
    + \chi^2_{>160{\mu} m}(\mbox{56 b+c+d+e}),
\end{eqnarray}
where $\chi^2_{\le160{\mu} m}$ denotes the contribution from bands up to 160\,$\mu$m (AKARI/IRC2 to Herschel/PACS3) for each source separately, and $\chi^2_{>160 {\mu} m}$ is computed for the SPIRE bands (Herschel/SPIRE1 to SPIRE3) using the summed flux of all three sources.

\end{appendix}

\end{document}